\RequirePackage{snapshot}
\documentclass[nofootinbib,english]{revtex4-1}
\usepackage[T1]{fontenc}
\usepackage[latin9]{inputenc}
\usepackage{color}
\usepackage{amsmath}
\usepackage{amssymb}
\usepackage{graphicx}

\usepackage[normalem]{ulem }
\usepackage{float}
\usepackage{subfig}
\usepackage{soul}
\usepackage{hyperref}
\usepackage{xcolor}
\hypersetup{
    colorlinks,
    linkcolor={red!50!black},
    citecolor={blue!50!black},
    urlcolor={blue!80!black}
}

\definecolor{nred}{RGB}{224,0,0}
\definecolor{nblue}  {RGB}{28,130,185}
\definecolor{dgreen} {RGB}{24,111,19}
\definecolor{norange}{RGB}{230,120,20}

\begin{document}

\title{\textcolor{black}{Incompleteness} of the Large $N$ Analysis of the O($N$) Models: Nonperturbative Cuspy Fixed Points and their Nontrivial Homotopy at finite $N$}
\author{S. Yabunaka}\email{yabunaka123@gmail.com}
\affiliation{\textcolor{black}{Advanced Science Research Center, Japan Atomic Energy Agency, Tokai, 319-1195, Japan}}
\author{C. Fleming} \email{fleming@lptmc.jussieu.fr}
\affiliation{Sorbonne Universit\'e, CNRS, Laboratoire de Physique Th\'eorique de la Mati\`ere Condens\'ee,  F-75005, Paris, France}
\author{B. Delamotte} \email{bertrand.delamotte@sorbonne-universite.fr}
\affiliation{Sorbonne Universit\'e, CNRS, Laboratoire de Physique Th\'eorique de la Mati\`ere Condens\'ee,  F-75005, Paris, France}

\date{\today}

\begin{abstract}
We summarize the usual implementations of the large $N$ limit of $O(N)$ models and show in detail why and how they can miss some physically important fixed points  when they become  singular in the limit $N\to\infty$. Using Wilson's renormalization group in its functional nonperturbative versions, we show how the singularities build up as $N$ increases. 
In the Wilson-Polchinski version of the nonperturbative renormalization group, we show that the singularities are cusps, which become boundary layers for finite but large values of $N$. The corresponding fixed points  being never close to the Gaussian, are out of reach of the usual perturbative approaches. We find four new fixed points  and study them  in all dimensions
 and for all $N>0$  and show that they play an important role for the tricritical physics of $O(N)$ models. Finally, we show that some of these fixed points  are bi-valued when they are considered as functions of $d$ and $N$  thus revealing  important and nontrivial homotopy structures.
 The  Bardeen-Moshe-Bander phenomenon that occurs at $N=\infty$ and $d=3$ is shown to play a crucial role for the internal consistency of all our results.
\end{abstract}
\maketitle

\section{Introduction}

The O($N$) models are probably the simplest and best studied scalar field theories. In their euclidean version, their hamiltonian reads:
\begin{equation}
    H=\int d^d x \left( \frac{1}{2}(\nabla\boldsymbol\varphi)^2 +\frac{1}{2} r\boldsymbol\varphi^2 +\frac{g_4}{4!}(\boldsymbol\varphi^2)^2 +\frac{g_{6}}{6!}(\boldsymbol\varphi^2)^3+\cdots\right)
    \label{hamiltonienO(N)}
\end{equation}
 where $\boldsymbol\varphi$ is a $N$-component field.
It is widely believed that their physics is almost fully understood in all dimensions at least at a qualitative and semi-quantitative level. This belief relies on the fact that all experimental, theoretical and numerical methods yield the same physical picture. The best known analytical methods are the $\epsilon$- \cite{4-epsilon, 2-epsilon} and $1/N$-expansions \cite{BrezinWallace, Zinn-Justin} and the high and low temperature series \cite{pelissetto2000}. More recently, the nonperturbative, also called functional, renormalization group (NPRG) \cite{berges,dupuis21}  and the conformal bootstrap \cite{El-Showk} have also been important achievements.

As for their critical behavior, the common belief is that we will perhaps never have an exact solution of these models in three dimensions but that this is not very important  because we already know everything. By using the NPRG, we show in this article that this statement is wrong and that plenty of unknown and unexpected features show up in these models and not only in three dimensions. 

The interest of our study is five-fold. 

First, we show in the following that there exists new and nontrivial fixed points (FPs) of the renormalization group (RG) in the 3$d$ O($N$) models that could drive the 
multicritical physics of these models.
\textcolor{black}{In particular, we find for large values of $N$ a new twice unstable FP  contrary to the common belief that the upper critical dimension of tricriticality  is 3 for O($N$) models because it is described by the massless $({\boldsymbol\varphi}^2)^3$ theory which is renormalizable for $d\le3$ \cite{Yabunaka-Delamotte-PRL2017,Yabunaka-Delamotte-PRL2018}. }

Second, we show why these new multicritical FPs were not found previously in any 
approach. In particular, we show that the usual large $N$ limit is inappropriate to find all the relevant FPs of the O($N$) model at $N=\infty$ even though it is widely believed  that a complete and exact solution for all possible O($N$) FPs  at $N=\infty$ is available (see \cite{Zinn-Justin} for many interesting and classical references on this subject and \cite{ref-large-N,dattanasio,katsis}). We also show how to generalize the large $N$ limit so as to find the other physically relevant  FPs at $N=\infty$.

Third, we show that the existence of these new FPs is intimately related to the existence of FPs with a cusp at $N=\infty$, that is, of FPs whose dimensionless effective potential $\tilde U(\tilde\phi)$  is singular and typically shows a cusp at some value of the field $\tilde\phi$ \cite{Yabunaka-Delamotte-PRL2018}. This feature is interesting because FPs exhibiting a cusp in their effective potential have already been found but mainly   in systems \textcolor{black}{ showing supersymmetry,} that have disorder or that are out of equilibrium \textcolor{black}{\cite{tisser08,tisser11,Gredat,Doussal1,Doussal2, Doussal3,Fisher1,Fisher2,Wiese1,Wipf}} but not in systems as simple as the O($N$) model.

Fourth,  a subset of these new fixed points  leads to two nontrivial homotopy structures. They can be understood as follows: A FP potential $\tilde U$ is of course a function of the field $\tilde\phi$ but is also a function of $d$ and $N$: 
$\tilde U=\tilde U(\tilde\phi,d,N)$. We find in certain regions of the $(d,N)$ plane that two distinct FPs  having the same number of infrared unstable eigendirections exist for a given couple of values of $d$ and $N$: They are both twice unstable for instance. 
We then show that these FPs are bi-valued in the $(d,N)$ plane, that is, when continuously followed along certain closed paths of this plane they are exchanged after one cycle along these paths and  two cycles  are needed to retrieve the FP we started with. These homotopy structures are intimately related  to "imperfect" or "cusp" bifurcations that we briefly review in Appendix \ref{sec: bifurcation_section}. This bi-valued nature of the new FPs is shown to be necessary to get a fully consistent picture of the new FPs in the $(d,N)$ plane including $N=\infty$.

\textcolor{black}{Fifth, it is known at $N=\infty$ and $d=3$ that there exists a line of tricritical FPs where the $\tau\left(\vec{\phi}^2\right)^3$ coupling of the O($N$) model is exactly marginal and where $\tau$ is bounded by a maximal value $\tau_{\textrm{BMB}}$ \cite{Bardeen-Moshe-Bander,David,david1985study}. We refer to the existence of this line as the Bardeen-Moshe-Bander (BMB) phenomenon in the following.  However it has remained unclear what  the BMB phenomenon implies for the multicritical physics of $O(N)$ models at finite $N$. We show how the BMB phenomenon is related to the new multicritical FPs mentioned above thus solving the long-standing problem of its \textcolor{black}{counterpart and impact}  at finite $N$. }

Finally, it is interesting to notice that the recourse to functional RG is mandatory in our study because the usual perturbative approaches based on a Taylor expansion of the action cannot deal with singular effective potentials and/or with the non conventional scaling in $N$ that is required to find the new FPs when $N\to\infty$.

An important output of our study is to solve a paradox of the large $N$ limit of the O($N$) models. The paradox can be stated in the following way. On one hand, the only nontrivial FP that was found in generic dimensions $d\in]2,4[$ at $N=\infty$ in the usual large $N$ limit of the O($N$) models was the Wilson-Fisher (WF) critical point: There are no nontrivial multicritical FPs at $N=\infty$ in generic dimensions. On the other hand, for all finite values of $N$, each time the dimension $d$ crosses one of the critical dimensions $d_c(p)=2+2/p$, a new, $p$ times unstable, multi-critical FP bifurcates from the gaussian FP G and becomes nontrivial for $d<d_c(p)$. For each value of $p$ and for all $N$ they are therefore found perturbatively  in the $\epsilon=d_c(p)-d$ expansion of the massless $(\boldsymbol\varphi^2)^{p+1}$ theory \cite{Izykson}. The paradox comes from the fact that it is {\it a priori} impossible to reconcile these two well-established facts: either the FPs found perturbatively in $d=d_c(p)-\epsilon$ survive at fixed and finite $\epsilon$ when $N\to\infty$ and the question is: ``Why aren't they  found at $N=\infty$ in the usual large $N$ limit?'', or they disappear at finite $\epsilon$ when $N$ is increased and this must occur by collision with other FPs. In this latter case, the question is: ``What are these other  FPs with which the perturbative multicritical FPs collide at large $N$?''. We show in the following that the two possibilities mentioned above  are realized: existence of FPs at $N=\infty$ not found in the usual large $N$ limit and collision of perturbative multicritical FPs with other  FPs that are nonperturbative. This solves completely the above paradox at the price of revisiting the large $N$ limit of the O$(N)$ model and finding nonperturbative FPs. 

In the present paper, we study in detail  the usual, that is, perturbative tricritical FP and find all the necessary FPs to understand the fate of this tricritical FP when both $d$ and $N$ are varied. 

Since our study deeply questions the large $N$ limit of the O($N$) models, let us first revisit its usual implementations.

\section{The usual large $N$ approaches}
\label{usual-large-N}
\begin{figure}
\includegraphics[scale=0.3]{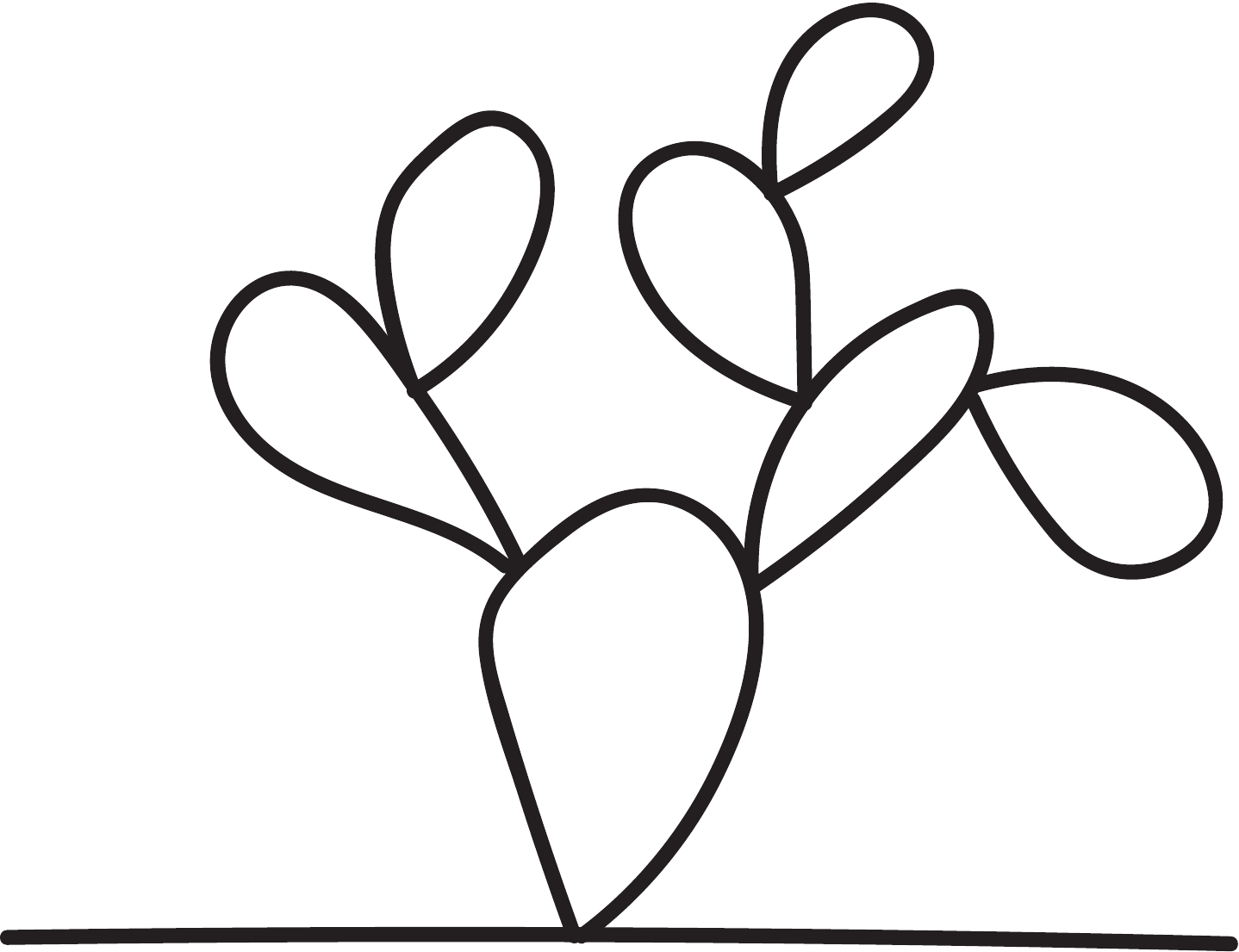}\caption{A cactus diagram. }
\label{cactus}
\end{figure}

\begin{figure}
\includegraphics[scale=0.4]{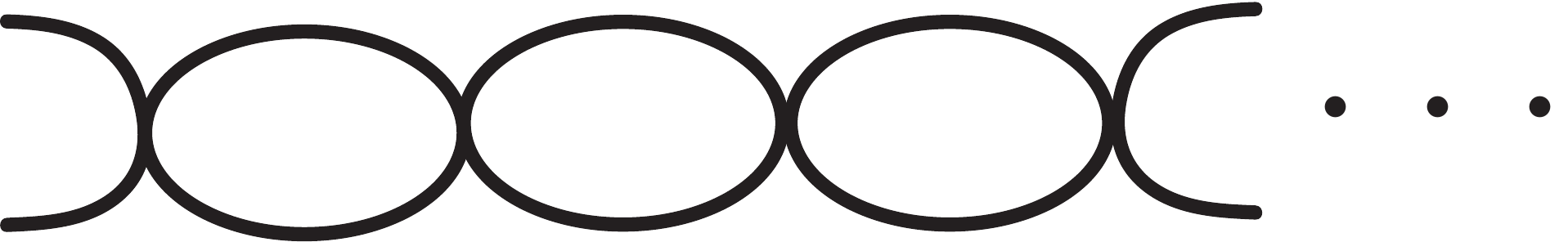}\caption{A chain of bubble diagrams. }
\label{bubble}
\end{figure}

There are three principal ways of deriving the large $N$ limit of a vector model. The first is to consider the Feynman graphs of the usual weak-coupling perturbative expansion of any correlation function and to select the graphs that involve the leading behavior in $N$ \cite{BrezinWallace}. The explicit dependence in $N$ of a graph comes from its combinatoric factors that themselves are the results of the contraction of the tensors attached to the vertices of the graphs and to the propagators. A finite large $N$ limit is obtained when the coupling constants $g_4, g_6,\cdots$ in Eq.~(\ref{hamiltonienO(N)})  used in the perturbative expansion are tuned to scale at large $N$ as powers of $N$  in such a way that their $N$-dependence exactly compensates at each loop order the leading powers in $N$ coming from the combinatorics. For the O($N$) model  and the $(\boldsymbol\varphi^2)^2$ coupling, this consists in taking $g_4=\tilde{g}_4/N$ with $\tilde g_4\sim O(1)$  when $N\to\infty$. Once this choice of scaling of $g_4$ has been performed,
an infinite number of graphs contribute at the same order in $N$ to a given correlation function and the large $N$ limit can be analytically computed when the resummation of these graphs is possible. For instance, the leading contribution in $N$ to the 2- and 4-point functions is respectively given by the sum of cactus diagrams and by the chain of bubbles, see Figs. \ref{cactus} and \ref{bubble}. Since these diagrams can be resummed, the O($N$) $(\boldsymbol\varphi^2)^2$ model  is solvable in the limit $N\to\infty$ and all physical quantities can be obtained from these resummations. The same analysis can be applied to all $(\boldsymbol\varphi^2)^n$ couplings with the appropriate rescaling of the corresponding coupling constants.

For the $(\boldsymbol\varphi^2)^2$ model, a second method consists in disentangling the quartic term in Eq.~(\ref{hamiltonienO(N)}) by introducing a supplementary field $\lambda(x)$ that plays the role of $\boldsymbol\varphi^2(x)$ \cite{Zinn-Justin}. It is in fact better to introduce a source (a magnetic field) in the field direction 1 and to separate the field $\boldsymbol\varphi$ in $(\sigma,\pi)$ where $\pi$ is a $N-1$-component field perpendicular to the source. The resulting action being quadratic in $\pi$, the partition function can be rewritten after integration over this field as:
\begin{equation}
 Z[h]=\int D\lambda(x)\,D\sigma(x)\, e^{-S[\sigma,\lambda]+\int h(x)\sigma(x)}   
\end{equation}
with
\begin{equation}
S[\sigma,\lambda]=\int d^dx\,\Big[\frac 1 2 (\partial\sigma)^2 -\frac{3}{2g_4}\lambda^2 + \frac{3r}{g_4}\lambda
+\frac 1 2 \lambda \sigma^2\Big]+\frac{N-1}{2}{\rm Tr\, ln}(-\Delta+\lambda).
\end{equation}
The large $N$ limit is performed as above by taking  $g_4$ of order $1/N$.  The  action then scales as $N$ and  $Z[h]$ is  computed by the steepest descent method. From this calculation, all  physical quantities can be derived. They of course coincide with what is found by the first method described above.

The third method is based on the Wilsonian RG flow of the effective potential (see Section \ref{Wetterich-RG} for a detailed explanation). The idea, that will be developed at length in the following, is to design a coarse-graining procedure depending on a scale $k$ such that at this scale only the short wavelength fluctuation modes of the field $\boldsymbol\varphi(q)$, that is  the modes $\boldsymbol\varphi(\vert q\vert>k)$, are integrated over \cite{wetterich93b,Ellwanger,Morris94}. The dimensionless effective potential $\tilde U_k(\tilde\phi)$ for these "rapid modes", that is, for the modes that have already been integrated out, follows an RG evolution equation as $k$ is decreased from the UV cut-off $\Lambda$  to $k=0$. This flow equation is in general not closed which means that it belongs to an infinite hierarchy of coupled flow equations. The Local Potential Approximation (LPA) consists in neglecting in the $k$-dependent effective action all contributions but the effective potential $\tilde U_k$ and the bare kinetic term. Its evolution equation becomes then closed and reads:

\begin{equation}
   \begin{split}
 \partial_{t}\tilde U_{t}(\tilde\phi)=-d\,\tilde U_{t}(\tilde\phi)+\frac{1}{2}(d-2)\tilde\phi\, \tilde U_{t}'(\tilde\phi)+\left(N-1\right)\frac{\tilde\phi}{\tilde\phi+\tilde U_{t}'(\tilde\phi)}+\frac{1}{1+\tilde U_{t}''(\tilde\phi)}\hspace{0.5cm}
\end{split}
\label{flow-LPA}
\end{equation}
where $t=\ln (k/\Lambda)$ and $\tilde\phi$ is the dimensionless modulus of $\boldsymbol\phi$ with $\boldsymbol\phi=\langle\boldsymbol\varphi\rangle$. As such, this equation is not exact when $N<\infty$ and is only an approximation of the exact flow equation for $\tilde U$ (see below) \cite{wetterich91,Ellwanger,Morris94}. However, when $N\to\infty$ and when the last term of the right hand side of Eq.~(\ref{flow-LPA}) -- corresponding to the contribution of the longitudinal mode -- is assumed to be negligible  because its prefactor is of order 1, Eq. (\ref{flow-LPA})  can be shown to become exact \cite{dattanasio}. In this limit, it reads:
 \begin{equation}
\partial_t \tilde U_t(\tilde\phi)=-d\,\tilde U_t+\frac{1}{2}(d-2)\tilde\phi\,\tilde U'_t+ N\frac{\tilde\phi}{\tilde\phi+ \tilde U_{t}'}.
\label{flow-LPA-dimensionless-N-infinity}
\end{equation}

  A finite large $N$ limit of the FP potential $\tilde U$ satisfying $\partial_t \tilde U(\tilde\phi)=0$, is obtained by assuming that $\tilde\phi$ scales as $\sqrt N$ and $\tilde U$ as $N$ which entails that the FP coupling $g$ in front of the $\tilde\phi^4$ term in $\tilde U$ scales as $1/N$ in agreement with the first two methods explained above.
 In the  $N\to\infty$ limit, the rescaling by a factor $N$ of $\tilde U_t$ and by $\sqrt N$ of $\tilde \phi$  eliminates all explicit dependence on $N$ in  Eq.~(\ref{flow-LPA-dimensionless-N-infinity}), which allows us to study its solutions in this limit. We therefore define:
 \begin{equation}
\bar\phi= \frac{\tilde\phi}{\sqrt N}\ \ ,\ \ \bar{U}_t(\bar\phi)= \frac{\tilde{U}_t(\tilde\phi)}{ N}.
\label{rescaling-N}
\end{equation}
In terms of these variables, the FP equation following from  (\ref{flow-LPA})  reads:
\begin{equation}
\partial_t \bar U_t(\bar\phi)=0=-d\,\bar U_t+\frac{1}{2}(d-2)\bar\phi\,\bar U'_t+\left(1-\frac{1}{N}\right) \frac{\bar\phi}{\bar\phi+ \bar U_{t}'}+\frac 1 N \frac{1}{1+ \bar U_{t}''}
\label{flow-LPA-dimensionless-rescaled-first}
\end{equation}
and in the limit $N\to\infty$, Eq.~(\ref{flow-LPA-dimensionless-rescaled-first}) becomes:
\begin{equation}
\partial_t \bar U_t(\bar\phi)=0=-d\,\bar U_t+\frac{1}{2}(d-2)\bar\phi\,\bar U'_t+ \frac{\bar\phi}{\bar\phi+ \bar U_{t}'}.
\label{flow-LPA-dimensionless-N-infinity-rescaled}
\end{equation}
Notice that a generic solution of the above FP equations, either Eq. \eqref{flow-LPA-dimensionless-rescaled-first} at finite $N$ or Eq. \eqref{flow-LPA-dimensionless-N-infinity-rescaled} at infinite $N$, \textcolor{black}{blows up} at a finite value of the field $\bar\phi$. The physical FP solutions are of course those that are globally defined on the interval $\bar\phi\in [0,\infty[$. \textcolor{black}{ Generically, only a finite number of such solutions exist and they are the FPs that drive the critical or multi-critical physics of the model}. 
All regular solutions of the FP equation following from Eq.~(\ref{flow-LPA-dimensionless-N-infinity-rescaled}) are  exactly known \cite{dattanasio,Tetradis-Litim}. Only one  nontrivial solution exists in generic dimensions $d\in\, ]2,4[$ and it is nothing but the limit when $N\to\infty$ of the Wilson-Fisher FP found at criticality at finite $N$. All physical quantities such as critical exponents can be computed from this solution. In the limit $N\rightarrow\infty$, they of course agree with the results found in the first two methods explained above. 

When the contribution of the longitudinal propagator in Eq.  (\ref{flow-LPA-dimensionless-rescaled-first}), that is, its last term,  is indeed negligible, all FPs can be obtained from Eq.  (\ref{flow-LPA-dimensionless-N-infinity-rescaled})  in the large $N$ limit. However, as we show in the following, this assumption is not always valid and as a consequence, this equation  does not yield the full set of physically relevant FPs of the O($N$) models in the large $N$ limit contrary to what is commonly believed. At first sight, this looks surprising because these methods look exact: They have  indeed been considered so in the literature up to now. The origin of the problem is nevertheless very clear on Eqs.~(\ref{flow-LPA-dimensionless-rescaled-first}) and (\ref{flow-LPA-dimensionless-N-infinity-rescaled}): The last term of 
Eq.~(\ref{flow-LPA-dimensionless-rescaled-first}) has been neglected in 
Eq.~(\ref{flow-LPA-dimensionless-N-infinity-rescaled}) whereas we show in the following that on a whole interval of $\bar\phi$, $(1+\bar{U}_t'')^{-1}$ can be of order   $N$ at large $N$ for some FPs. Thus, this term which is not taken into account in any of the three approaches described above, can in fact be for some FPs of the same order as the other terms, at least for some values of $\bar\phi$. Notice that this means that the rescaling of both the field and the potential that looks so natural in the three usual methods can be inappropriate because the rescaling which is necessary to find some FPs depends actually on the value of the field $\bar\phi$: In some intervals of values of $\bar\phi$, it is  the usual rescaling by a factor $\sqrt N$ which is necessary, while another rescaling on another interval of values of $\bar\phi$ will be necessary. The notion of boundary layer is the appropriate framework to implement a rescaling that can depend on which  interval in field space is considered. We explain it in detail in the following. 

Obviously, the NPRG which is in essence functional in $\bar\phi$,  is the appropriate framework to study these new FPs. In fact, we believe that they cannot be found with the first method based on graphs because the loop-expansion corresponds to an expansion of the bare action around a definite value of the field while nonanalyticities of some FP potentials show up for nontrivial values of $\bar\phi$.

Notice that the LPA that we extensively use in the following, is only exact at $N=\infty$ for the usual FPs and does not contain all  $1/N$ corrections. This is why we have checked the stability of our approach by incorporating higher order terms in the effective action -- technically, all terms involving two derivatives.  Known perturbative results obtained at high orders of the $\epsilon=3-d$ expansion are also shown to be fully consistent with ours when they overlap.

We now explain in detail  the NPRG approach to the $N\to\infty$ limit.

\section{The Nonperturbative Renormalization Group}

We consider below a scalar field theory with O$(N)$ symmetry  where momentum modes beyond the scale $\Lambda$ are cutoff. If the system studied is the effective large distance continuous version of a lattice model, $\Lambda$ is typically the inverse of the lattice spacing.

\subsection{Ellwanger-Morris-Wetterich  approach to the NPRG}
\label{Wetterich-RG}
The  NPRG is based on the idea of integrating  step by step the fluctuations existing on all momentum scale between the ultraviolet cut-off scale $\Lambda$ and 0 \cite{PhysRevB.4.3174}.  In the Ellwanger-Morris-Wetterich  version of the NPRG \cite{Ellwanger,Morris94,wetterich93b}, with the usual $(\boldsymbol\varphi^2)^2$ model is associated a one-parameter family of models with Hamiltonians $H_k=H+ \Delta H_k$ and partition functions ${\cal Z}_k$, where  $k$ is a momentum scale. In $H_k$, $\Delta H_k$ is chosen such that the slow fluctuations of the original model described by $H$ are frozen. Thus, only the rapid fluctuations, those with wave numbers $\vert q\vert > k$, are summed over in the partition function ${\cal Z}_k$. The slow modes ($\vert q\vert < k$) need therefore to be decoupled in ${\cal Z }_k$ and this is achieved by making them shortly correlated, that is, by giving them a mass of order $k$. We thus take for $\Delta H_k$ a quadratic (mass-like) term, which is nonvanishing only for the slow modes:
\begin{equation}
 {\cal Z}_k[\boldsymbol{J}]= \int D\boldsymbol\varphi_i \exp(-H[\boldsymbol\varphi]-\Delta H_k[\boldsymbol\varphi]+ \boldsymbol{J}\cdot\boldsymbol\varphi)
 \label{partition}
\end{equation}
with $\Delta H_k[\boldsymbol\varphi]=\frac{1}{2}\int_q R_k(q^2) \boldsymbol\varphi_i(q)\boldsymbol\varphi_i(-q)$ and $\boldsymbol{J}\cdot\boldsymbol\varphi=\int_x J_i(x) \boldsymbol\varphi_i(x)$. In the following, we use two kinds of regulator functions depending on the approximation scheme we are using. For the local potential approximation (see below for the definition), we use
\begin{equation}
R_k(q^2)=(k^2-q^2) \theta(k^2-q^2)
\label{theta}
\end{equation}
where $\theta(x)$ is the Heaviside function. For the second order of the derivative expansion (see below) we use
\begin{equation}
R_k(q^2)=\bar{Z}_k \frac{q^2}{e^{q^2/k^2}-1}
\label{exonential}
\end{equation}
where $\bar{Z}_k$ is a running field renormalization factor to be defined more precisely below. This running renormalization factor defines a running anomalous dimension according to $ \eta_k = -\partial_k \log\bar{Z}_k$. $\eta_k$ becomes the true anomalous dimension at criticality when $k\to 0$, that is, at the fixed point. Notice that there is no field renormalization at the LPA level, that is,  $\bar Z_k=1$, which implies that the rescaling of $\boldsymbol\phi( x)$ is performed according to its canonical dimension only and the anomalous dimension at the FP is vanishing. This approximation is justified here because we will be mainly interested in the large $N$ limit of the O($N$) model where the anomalous dimension  decays sufficiently fast with $N$ to neglect it.

All our results should in principle be independent of the precise shape of the regulator $R_k(q^2)$. However, once approximations are performed, a residual and spurious dependence on $R_k(q^2)$ shows up that decreases with the order of the derivative expansion\cite{balog19}.

The NPRG is based on the $k$-dependent  Gibbs free energy $\Gamma_k[\boldsymbol\phi]$ where $\boldsymbol\phi(x)=\langle\boldsymbol\varphi(x)\rangle$ ($\Gamma_k$ is also called the one-particle-irreducible generating functional). It
is defined as  the (slightly modified) Legendre transform of $\log  {\cal Z}_k[\boldsymbol{J}]$:
\begin{equation}
 \Gamma_k[\boldsymbol\phi]+\log  {\cal Z}_k[\boldsymbol{J}]= \boldsymbol{J}\cdot\boldsymbol\phi-\frac 1 2 \int_q R_k(q^2) \phi_i(q)\phi_i(-q)
 \label{def-gammak}
 \end{equation}
with $\int_q=\int d^dq/(2\pi)^d$. The properties of the regulator explained above implies that :
\begin{equation}
   \begin{array}{llll}
  R_{k=\Lambda}(q)\simeq \Lambda^2\ \  &\Rightarrow\ \ &\Gamma_{k=\Lambda}\simeq H \ \ &{\text{and all fluctuations are frozen}} \\
 R_{k=0}(q)=0\ \  &\Rightarrow\ \ &\Gamma_{k=0}=\Gamma\ \ &{\text{and all fluctuations are integrated over.}}
\end{array}
\label{properties-Gammak}
\end{equation}
Thus, at a generic scale $k$, $\Gamma_k$ is the Gibbs free energy of a model where only the fluctuations between $\Lambda$ and typically $k$ have been integrated over. Decreasing $k$ therefore means integrating over more and more fluctuations.
The exact flow of $\Gamma_k[\boldsymbol\phi]$ follows from 
Eqs.~(\ref{partition}) and (\ref{def-gammak}). It reads \cite{Morris94,wetterich93b}:
\begin{equation}
\partial_k\Gamma_k[\boldsymbol\phi]= \frac 1 2 \int_q \partial_k R_k(q){\rm Tr} \left(\Gamma_k^{(2)}(q,-q;\boldsymbol\phi) + R_k(q) \right)^{-1} 
\label{flot-wetterich}
\end{equation}
where $\Gamma_k^{(2)}(q,-q;\boldsymbol\phi)$ is the Fourier transform of the second functional derivative of $\Gamma_k[\boldsymbol\phi]$ with respect to $\boldsymbol\phi(x)$ and $\boldsymbol\phi(y)$, and ${\rm Tr}$ stands for a trace over group indices \cite{wetterich93b}.

Integrating Eq.~(\ref{flot-wetterich}) together with the initial condition at $k=\Lambda$ given in Eq.~(\ref{properties-Gammak}) would amount to solving exactly the model.
However, this equation is in general impossible to solve exactly and the recourse to approximations mandatory.
The derivative expansion is the approximation scheme which is both simple and well suited for our purpose. It has been used with much success in many instances \cite{tarjus04,dupuis21}. It leads for instance to very accurate results for the determination of the critical exponents and competes with the best results obtained by other methods including the conformal bootstrap \cite{balog19}.  It consists in expanding $\Gamma_k[\boldsymbol\phi]$ in a power series of $\nabla\boldsymbol\phi$ truncated at a finite order \cite{canet03,canet05,kloss14,delamotte04,benitez08,canet04,tissier10,tisser08,canet16,leonard15}. For instance, at second order, $\Gamma_k$ is approximated by
\begin{equation}
\Gamma_k[\boldsymbol\phi]= \int_x \left(U_k(\rho) + \frac 1 2 Z_k(\rho) (\nabla\boldsymbol\phi)^2 +\frac 1 4 Y_k(\rho)(\nabla\rho)^2+O(\nabla^4)\right)
\label{DE2}
\end{equation}
where $\rho=\boldsymbol\phi^2/2$ and $Z_k(\rho)$, $Y_k(\rho)$ and $U_k(\rho)$ are ordinary functions of $\rho$. The running field renormalization is then defined by $\bar{Z}_k=Z_k(\rho_0)$ for some arbitrary reference point $\rho_0$.  
Once this approximation has been performed, the exact flow Eq.~(\ref{flot-wetterich}) boils down to coupled differential equations for $Z_k(\rho)$, $Y_k(\rho)$ and $U_k(\rho)$. The LPA which is the approximation that is mostly used in this article consists in retaining only the potential in the ansatz for $\Gamma_k$, that is, in keeping only the (bare) derivative term already present in $H$. In the O($N$) case, it consists in approximating $\Gamma_k$ by:
\begin{equation}
\Gamma_k^{\text{LPA}}[\boldsymbol\phi]= \int_x \left(\frac 1 2  (\nabla\boldsymbol\phi)^2 +U_k(\rho) \right).
\label{LPA}
\end{equation}
The flow equation for $\Gamma_k$ becomes in the LPA a flow for the running effective potential only. For the regulator  of Eq.~(\ref{theta}), the flow of $U_k$  reads\cite{delamotte07}:
\begin{equation}
   \begin{split}
 \partial_{k} U_{k}(\phi)=\left(N-1\right)v_d \frac{\phi}{\phi+ U_{k}'(\phi)}+v_d\frac{1}{1+ U_{k}''(\phi)}
\end{split}
\label{flow-LPA-dim}
\end{equation}
 where $v_{d}^{-1}=2^{d-1}d\pi^{d/2}\Gamma(\frac{d}{2})$, $\phi_i=\langle\boldsymbol\varphi_i\rangle$ and $\phi=\sqrt{\phi_i\phi_i}$
 is the modulus of the average of $\boldsymbol\varphi$. Notice that in the following, $U_k$ is considered for convenience as a function of $\phi=\sqrt{2\rho}$ and the derivative $U_{k}'(\phi)$ is therefore the radial derivative of the potential.
 
Within the LPA, the anomalous dimension $\eta$ is neglected which is one of its main source of inaccuracy. It is however known to be one-loop exact for the $\phi^4$ theory  \cite{berges,delamottefrustrated} in $d=4-\epsilon$ and exact at $N=\infty$ at least for nonsingular FPs \cite{dattanasio}, see below. It also yields quite accurate results for the critical exponents associated with the Wilson-Fisher FP in $d=3$ and for all values of $N$. Moreover, the regulator (\ref{theta}) is known to be optimal for the LPA \cite{LitimOptimal}. 
 
 At criticality, the model is self-similar and the RG flow reaches a FP if it is expressed in terms of dimensionless quantities. We proceed as usual by rescaling fields and coordinates according to $\tilde x=k x$, $\tilde{\boldsymbol\phi}(\tilde x)=v_d^{-1/2}k^{(2-d)/2}\bar Z_k^{1/2}\boldsymbol\phi( x)$ and $\tilde U_k(\tilde\rho)=v_d^{-1}k^{-d}U_k(\rho)$. The LPA flow of the potential $\tilde U_k(\tilde\phi)$ is given by Eq.~(\ref{flow-LPA}) and becomes in the large $N$ limit the one given by Eq.~(\ref{flow-LPA-dimensionless-N-infinity}) under the assumption that the last term in Eq.~(\ref{flow-LPA}) is negligible. In this limit, it can be shown that if this last term is indeed negligible then the LPA equation becomes exact in the large $N$ limit because the coupling between $\tilde U_k(\tilde\phi)$ and the other functions such as $Z_k(\rho)$ and $Y_k(\rho)$ is subleading.
 The usual field and potential rescalings is given in Eq.~(\ref{rescaling-N}) and the flow thus reads:
\begin{equation}
\partial_t \bar U_t(\bar\phi)=-d\,\bar U_t+\frac{1}{2}(d-2)\bar\phi\,\bar U'_t+ \left(1-\frac{1}{N}\right) \frac{\bar\phi}{\bar\phi+ \bar U_{t}'}+\frac 1 N \frac{1}{1+ \bar U_{t}''}.
\label{flow-LPA-dimensionless-rescaled}
\end{equation}
As shown in Eq. (\ref{properties-Gammak}), the potential $U_{k=\Lambda}$ is the bare potential given by the hamiltonian of the model. This yields the boundary condition at $t=0$ necessary to solve unambiguously Eq. (\ref{flow-LPA-dimensionless-rescaled}).
In the large $N$ limit and again under the hypothesis that the last term is negligible in this limit, it becomes Eq.~(\ref{flow-LPA-dimensionless-N-infinity-rescaled}).

\subsection{Wilson-Polchinski approach}
Another formulation of the NPRG is  the approach {\it\`a la} Wilson-Polchinski (W-P)  based on  running effective Hamiltonians instead of running Gibbs free energies \cite{Polchinski}, see Appendix \ref{sec:WP} for  details. What is important for what follows is that, as pointed out by Morris \cite{Morris}, there is an exact mapping between LPAs in W-P and in  Ellwanger-Morris-Wetterich versions of the RG when  the cutoff in Eq.~(\ref{theta}) is used (the LPA is universal in the W-P version, that is, is independent of the function $R_k$). Here we denote the potential part of the effective Hamiltonian in W-P approach  by $V(\varrho)$ and $\tilde V(\tilde\varrho)$ its dimensionless analog which is a function of the dimensionless field $\tilde\varrho$. Then, the mapping is given by:
\begin{equation}
   \tilde V(\tilde\varrho)=\tilde U(\tilde\rho)+(\tilde\phi_i-\tilde\Phi_i)^2/2 \ \ \text{and}\ \  \tilde\phi_i-\tilde\Phi_i=-\tilde\Phi_i \tilde V'(\tilde\varrho)=-\tilde\phi_i\tilde U'(\tilde\rho)
   \label{transformation-WP-Wetterich}
\end{equation}
 with $\tilde\varrho=\tilde\Phi_i\tilde\Phi_i/2$ \cite{Morris}. We then perform the same rescaling as in Eq.~(\ref{rescaling-N}): 
  \begin{equation}
\bar\varrho= \frac{\tilde\varrho}{ N}\ \ ,\ \ \bar{V}_t(\bar\varrho)= \frac{\tilde{V}_t(\tilde\varrho)}{ N}.
\label{rescaling-N-WP}
\end{equation}
  With these changes of variables, the LPA FP equation (\ref{flow-LPA-dimensionless-rescaled}) in Ellwanger-Morris-Wetterich version of the RG is transformed into the LPA in the W-P parametrization \cite{Polchinski,Hasenfratz,Comellas} which is given by: 
\begin{equation}
 0=1-d\,\bar V+(d-2)\bar\varrho \bar V'+2\bar\varrho{\bar V'}{}^2-\bar V'-\frac{2}{N}\bar\varrho\,\bar{V}''.
\label{flow-LPA-WP-essai}
\end{equation}

\section{The large $N$ limit revisited: new fixed points, boundary layers and cusps}
\label{section-largeN}

We  show below that the usual large $N$ approaches  described in section \ref{usual-large-N} are too restrictive and miss several FPs that are relevant to the multicritical physics of $O(N)$ models. As already explained in Section \ref{usual-large-N}, the problem is that some  FP potentials that are physically relevant show singularities when $N=\infty$. It turns out that the analysis of these singularities is by far simpler in W-P than in Ellwanger-Morris-Wetterich version of the RG and we therefore switch to the former from now on.

\subsection{The usual large $N$ limit in the functional RG}
\label{usual-large-N-functional}

The LPA FP equation on the potential is given in the Ellwanger-Morris-Wetterich version of the RG by Eq.
(\ref{flow-LPA-dimensionless-rescaled-first}) when the regulator of Eq.~(\ref{theta})  is used. It becomes Eq.~(\ref{flow-LPA-WP-essai}) in the W-P version of the RG.

In the usual large $N$ approach, the FP potential $\bar U(\bar\rho)$  is assumed to be smooth for all values of the field when $N\to\infty$. Thus, the last term in  Eq.
(\ref{flow-LPA-dimensionless-rescaled-first}) is neglected since its prefactor is of order $1/N$. The resulting equation  has been shown to be exact in this limit \cite{dattanasio} which means that although the free energy $\Gamma_k[\boldsymbol\phi]$ involves terms other than $U_k(\rho)$, see Eq.~(\ref{DE2}), the exact flow equation for the potential at $N=\infty$ is given by  Eq.~(\ref{flow-LPA-dimensionless-N-infinity-rescaled}). 
 Its analogue in the W-P version of the NPRG is also obtained by neglecting the term with a $1/N$ prefactor but this time in Eq. (\ref{flow-LPA-WP-essai}). It reads:
\begin{equation}
 0=1-d\,\bar V+(d-2)\bar\varrho \bar V'+2\bar\varrho{\bar V'}{}^2-\bar V'.
\label{flow-LPA-WP-N-infini}
\end{equation}

Apart from the gaussian FP G, this equation is known to have two smooth solutions defined for all $\bar\varrho \ge 0$ in generic dimensions $2<d<4$: the WF FP solution, whose  analytic expression has been derived in \cite{Kubyshin}  and the high temperature FP  $\bar{V}(\bar{\varrho})=\bar{\varrho}$ \cite{Zumbach}.

\begin{figure}[t]
\includegraphics[scale=0.017]{Vnew-717.pdf}\includegraphics[width=0.42\linewidth]{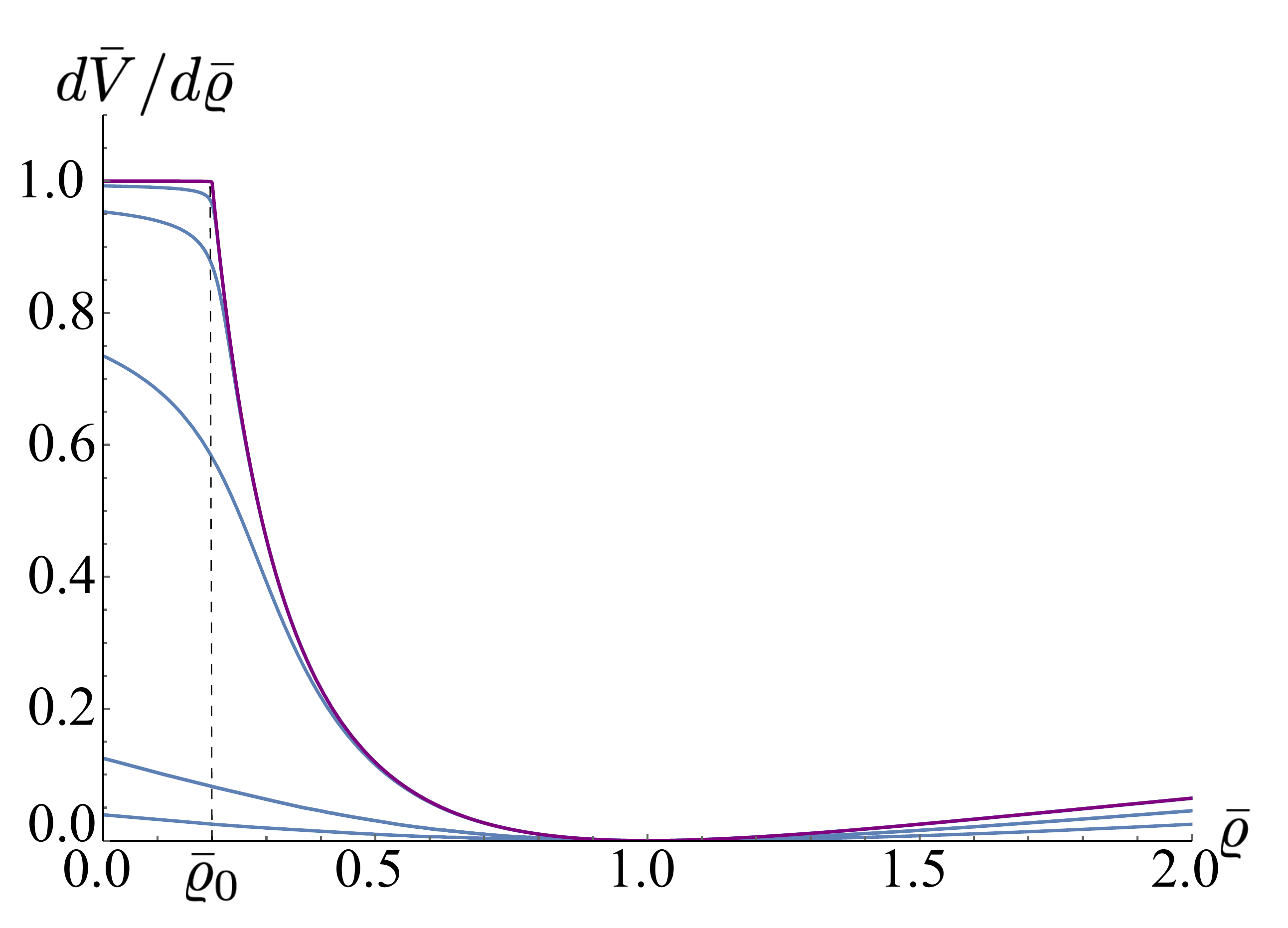}
\caption{$d=3$ and $N=\infty$: (Left) Some tricritical potentials of the  ${\cal A}(\tau)$ FPs along the BMB line (blue)  together with the Wilson-Fisher FP potential (red) that are solutions of Eq. (\ref{flow-LPA-WP-essai}). The gaussian FP G corresponds to the horizontal line and all FP potentials along the BMB line are continuous deformations of G. The BMB FP is the end-point of the BMB line (purple): It involves a linear part for $\varrho\in\,[0,\varrho_0]$ and shows a discontinuity in its second derivative at $\bar\varrho_0$, see the right panel. All these potentials are given by Eq.~(\ref{solrho}) (in the Wilson-Polchinski version of the LPA flow). (Right) Derivatives of the ${\cal A}(\tau)$ FP potentials of the BMB line (blue). The BMB FP potential (purple) shows a discontinuity in its second derivative at $\bar\varrho=\bar\varrho_0$.}
\label{V(rho)}
\end{figure}

Notice that the dimensions $d_c=2+2/p$ with $p\in\mathbb{N}^*$ are exceptional. In these dimensions, the $(\boldsymbol\varphi^2)^{p+1} $ term becomes marginal and at $N=\infty$, takes place the BMB phenomenon that has no counterpart in generic dimensions \cite{Bardeen-Moshe-Bander,Omid,Mati2017}. For example, for $p=2$, that is, $d=3$, there exists a line of tricritical FPs at $N=\infty$ called the BMB line of FPs. The complete set of regular solutions of Eq.~(\ref{flow-LPA-WP-N-infini}) in $d=3$ are given by the following implicit expression \cite{Litim2018}:
\begin{equation}
\bar{\varrho}_{\pm}=1+\frac{\bar{V}'\left(\frac{5}{2}-\bar{V}'\right)}{\left(1-\bar{V}'\right)^{2}}+\frac{\frac{3}{2}\arcsin\sqrt{\bar{V}'}\pm \sqrt{2/\tau}}{\left(\bar{V}'\right)^{-1/2}\left(1-\bar{V}'\right)^{5/2}}
\label{solrho}
\end{equation}
 where $\bar{\varrho}_{+}\left(\bar{V}'\right)$ and $\bar{\varrho}_{-}\left(\bar{V}'\right)$ correspond to the two branches $\bar{\varrho}>1$ and $\bar{\varrho}<1$ respectively, and $\tau$ is an integration constant.  They consist in (i) the gaussian FP G obtained for $\tau=0$ for which $\bar V'(\bar\varrho)=0$, (ii) a set of well-defined solutions $\bar V(\bar\varrho)$ indexed by $\tau\in[0,\tau_{\text{ BMB}}=32/(3\pi)^2]$ which correspond to the BMB line of FPs and denoted here by ${\cal A}(\tau)$, with the BMB FP obtained at $\tau=\tau_{\text{ BMB}}$  being the endpoint of the line   \cite{Bardeen-Moshe-Bander,david1985study,David,Litim2018,Mati2017}, (iii) an isolated solution associated with $\sqrt{2/\tau}=0$ which corresponds to the Wilson-Fisher FP (an analytic continuation is needed when $V'<0$). Notice that for $\tau>\tau_{\text{ BMB}}$  the solutions of Eq.~(\ref{solrho}) are not defined on the whole interval $\bar\varrho\in[0,\infty[$ \cite{Litim2018}.
 The potentials of the ${\cal A}(\tau)$ FPs with $\tau<\tau_{\text{ BMB}}$ are regular for all values of the field. Approaching $\tau_{\text{BMB}}$, the FP potential approaches a limiting shape which shows a singularity: It is made of a linear part $\bar V(\bar\varrho)=\bar\varrho $ starting at  $\bar\varrho=0$ up to the point $\bar\varrho_0$ where this straight line crosses the nontrivial part of the potential, see Fig. \ref{V(rho)}.  
 The BMB FP potential is obtained as this limiting shape. Notice that the linear part of the BMB FP potential existing for $\bar\varrho<\bar\varrho_0$, see Figs. \ref{V(rho)} can be replaced by a smooth analytic continuation of the other part of the potential, that is, the part corresponding to $\bar\varrho>\bar\varrho_0$, without having any physical consequence. This can be most easily realized by going to the Ellwanger-Morris-Wetterich version of the flow where this linear part is entirely mapped onto the point $\bar\phi=0$ \cite{Litim2018}. The RG flow is given in Appendix \ref{sec:Phase diagram} where the BMB line together with the WF FP are provided.

\subsection{Explicit construction of the new FPs SWF$_2$ and SG$_3$ at $N=\infty$ in $d>3$ in the Wilson-Polchinski approach}\label{FPd>3}

The problem with the usual large $N$ approach is particularly clear on Eqs.~(\ref{flow-LPA-WP-essai}) and (\ref{flow-LPA-WP-N-infini}):  It is well known from singular perturbation theory that when the small expansion-parameter, that is,  $1/N$ in our case, multiplies the term of highest derivative, that is, $\bar{V}''$, it is not legitimate in general to neglect this term in the limit where the small parameter goes to zero \cite{Holmes}. Thus, the limit $N\to\infty$ has to be taken with care because  a boundary layer that becomes an isolated singularity in the large $N$ limit can exist, see below.  One possibility to derive the correct large $N$ limit allowing for potentials with isolated singularities is to connect two smooth large $N$ solutions across  an isolated point $\bar\varrho=\bar\varrho_0$. We show in the following the detail of this procedure and afterwards that the new FPs  thus obtained at $N=\infty$ have a finite $N$ counterpart.

\begin{figure}[t]
\includegraphics[scale=0.45]{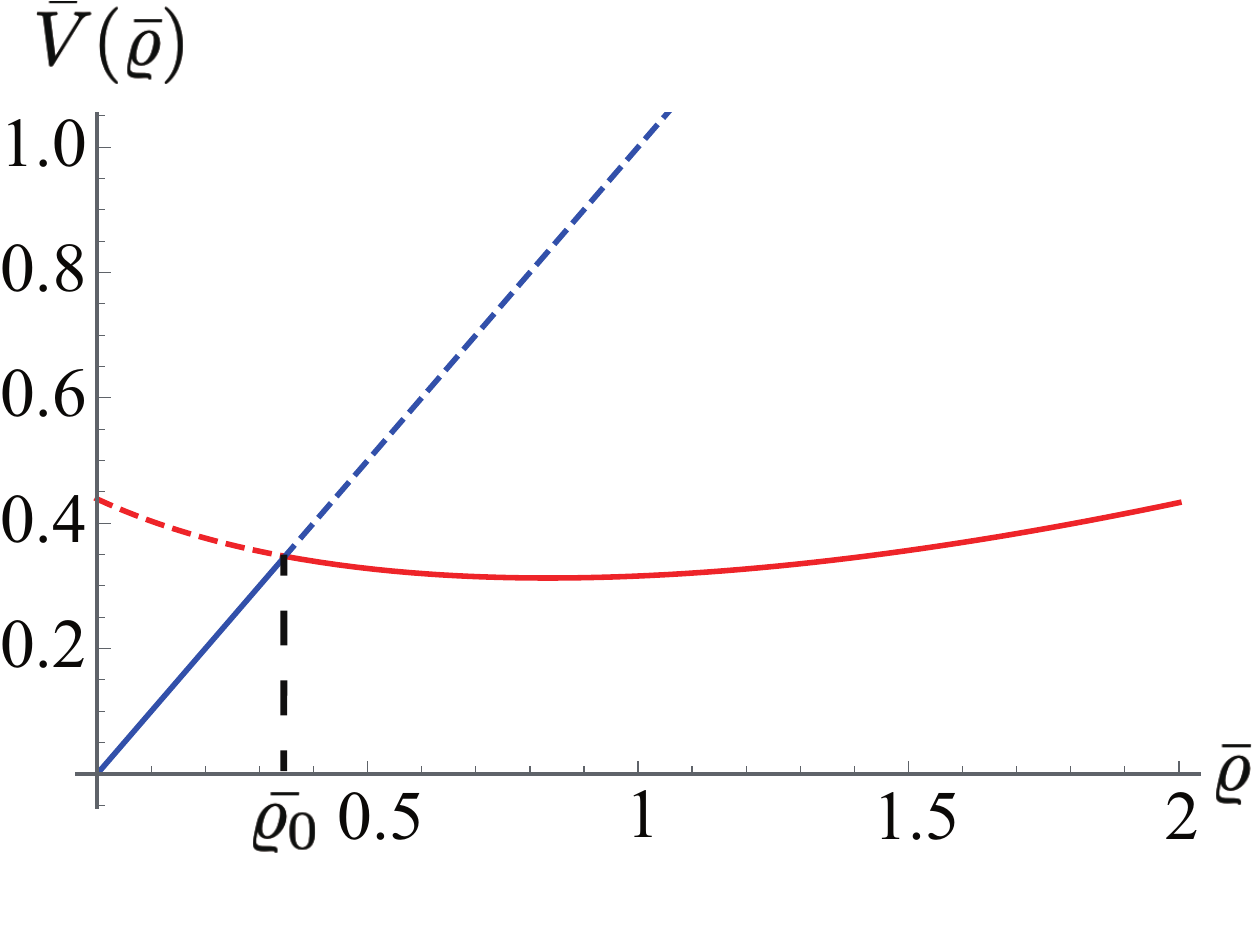}
\caption{ The  SWF$_2$ FP of Eq.~(\ref{flow-LPA-WP-essai}) in $d=3.2$ at $N=\infty$. It is shown  as a solid line and is made of two parts that match at $\bar{\varrho}_0=0.347$. For $\bar{\varrho}<\bar{\varrho}_0$, it is the linear solution $\bar{V}(\bar{\varrho})=\bar{\varrho}$ in blue and for $\bar{\varrho}>\bar{\varrho}_0$, it is identical to the WF FP in red. }
\label{matching}
\end{figure}

We now show how to construct FP potentials  at $N=\infty$ showing an isolated singularity in the W-P version of the RG. Two regular FP solutions of Eq.~(\ref{flow-LPA-WP-N-infini}) are displayed in Fig. \ref{matching}: One is the WF FP and the other one is the high temperature FP $\bar{V}(\bar{\varrho})=\bar{\varrho}$. Since these two solutions cross at a point $\bar\varrho=\bar\varrho_0$ , it is easy to find another continuous solution of the FP equation (\ref{flow-LPA-WP-N-infini}): We take for $\bar\varrho<\bar\varrho_0$ the linear solution $\bar{V}(\bar{\varrho})=\bar{\varrho}$ and for $\bar\varrho>\bar\varrho_0$ the WF solution. This potential has therefore the usual large field behavior of the WF FP, see Fig. \ref{matching}. Although continuous, this potential is not differentiable at $\bar\varrho_0$. Notice that since the WF potential is exactly known at $N=\infty$, $\bar\varrho_0$ can be computed with an arbitrary accuracy. We call this FP SWF$_2$, where the $S$ means singular. [This FP was called $C_2$ in \cite{Yabunaka-Delamotte-PRL2017,Yabunaka-Delamotte-PRL2018}]. We have computed the number of IR unstable directions of the RG flow around this FP and found that it is twice unstable, hence the index 2. Notice that this FP has one more unstable direction than its regular counterpart, the WF FP. This is a general phenomenon shown in Appendix \ref{sec: eigenfunctions-sing-FP}. \textcolor{black}{Note also that strictly speaking, although twice unstable, SWF$_2$ is not necessarily tricritical because a model is said to be tricritical if in parameter space it lies at the intersection of a first and of a second order hypersurface, a question we have not studied here.  }

\begin{figure}[t]
\includegraphics[scale=0.287]{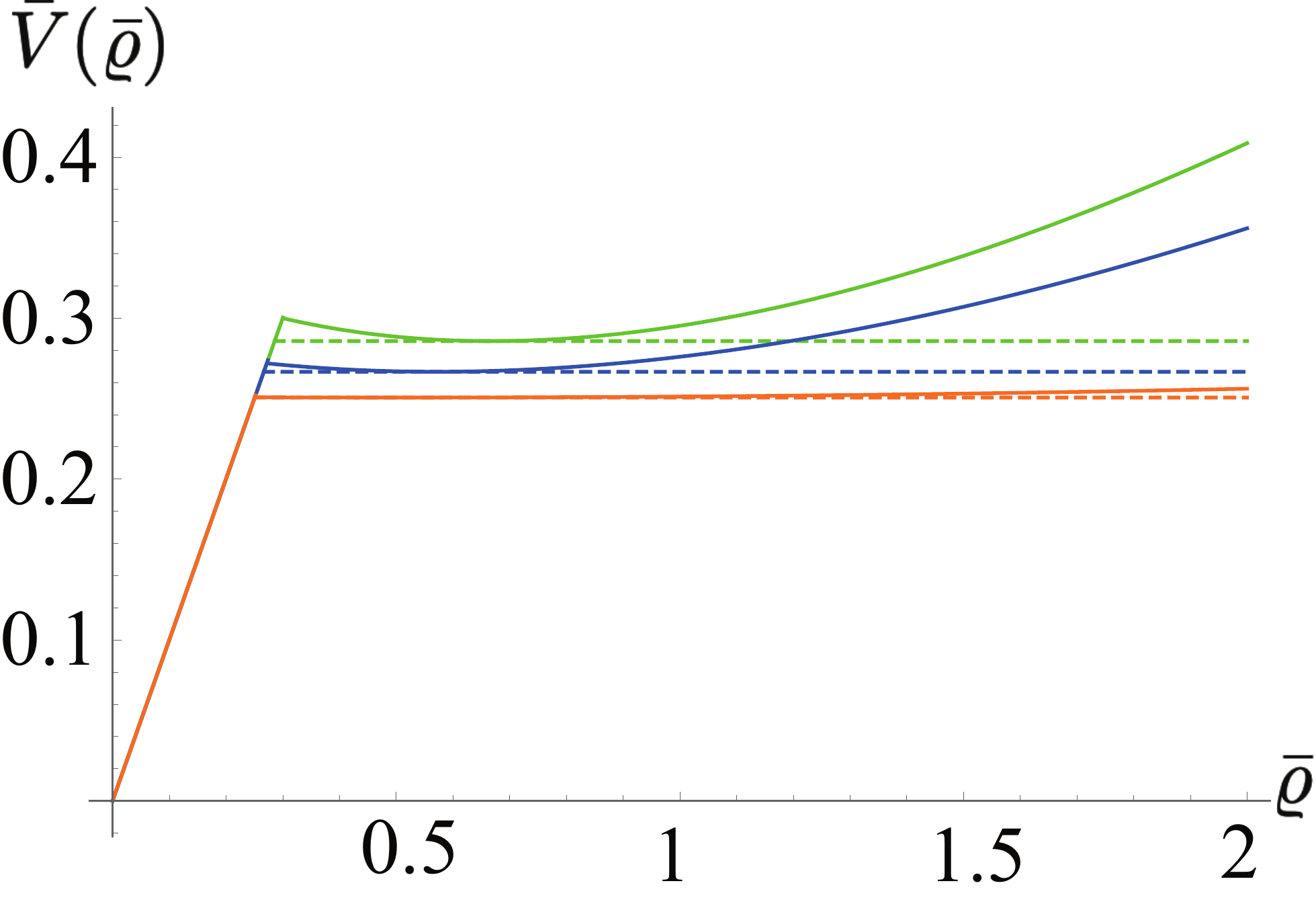}
\caption{The SWF$_2$ (thick lines) and SG$_3$  (dashed lines) FPs of Eq.~(\ref{flow-LPA-WP-essai}) in $d=3.5$ (green), $d=3.75$ (blue) and $d=3.99$ (orange) at $N=\infty$. At large field, that is,  $\bar{\varrho}>\bar{\varrho}_0$, SWF$_2$ coincides with the  WF FP solution and thus becomes flat when  $d\to4^-$. In this limit, it coincides with SG$_3$ which shows that they appear together at $N=\infty$ in $d=4^-$.}
\label{2solutions}
\end{figure}

In Fig. \ref{2solutions}, we show how SWF$_2$ changes with $d$ at $N=\infty$. On this figure, we can see that $\bar\varrho_0$ approaches $1/4$ when $d\to 4^-$ and that the potential becomes flat for $\bar\varrho>\bar\varrho_0 $ when $d\to4^-$, since the WF  FP approaches the Gaussian FP in this limit.  Notice however that even in this double limit where $N\to\infty$ and $d\to 4$, SWF$_2$ does not become gaussian because it is not flat for all values of the field.

Another FP potential showing an isolated singularity can be  constructed at $N=\infty$. It is obtained by taking the trivial solution $\bar{V}(\bar{\varrho})=\bar{\varrho}$ for $\bar\varrho<1/d$ and  $\bar{V}(\bar{\varrho})=1/d$ for $\bar\varrho>1/d$ which corresponds to the Gaussian FP. We call this FP SG$_3$ because it is the singular counterpart of the gaussian FP G and because it is three times unstable. Notice that it was called $C_3$ in \cite{Yabunaka-Delamotte-PRL2017,Yabunaka-Delamotte-PRL2018}. We can see that at $N=\infty$, SWF$_2$ and SG$_3$ become identical in the limit $d\to4^{-}$, which shows that they appear as a pair of FPs just below $d=4$ (see Section \ref{sectionV}  for a discussion of the line $N_c'(d)$ where these FPs appear at finite $N$). 

Let us now study these FPs at finite $N$. We start by describing how the singularity of SWF$_2$ and SG$_3$ builds up as $N\to\infty$.

\subsection{The boundary layer analysis}
\label{BL}

Having determined SWF$_2$ and SG$_3$ at $N=\infty$, the natural question now is to know whether they are the limits of FPs existing at finite $N$. At finite $N$, these FPs are expected to be regular and therefore this question requires to know how the singularity of their potential builds up as $N$ increases. The notion of boundary layer yields the relevant framework to tackle this problem. We give in Appendix \ref{sec:Toyboundarylayer} a toy model of the formation of a boundary layer for a simple differential equation that shares many features with Eq.~(\ref{flow-LPA-WP-essai}).

\begin{figure}[t]
\includegraphics[scale=0.45]{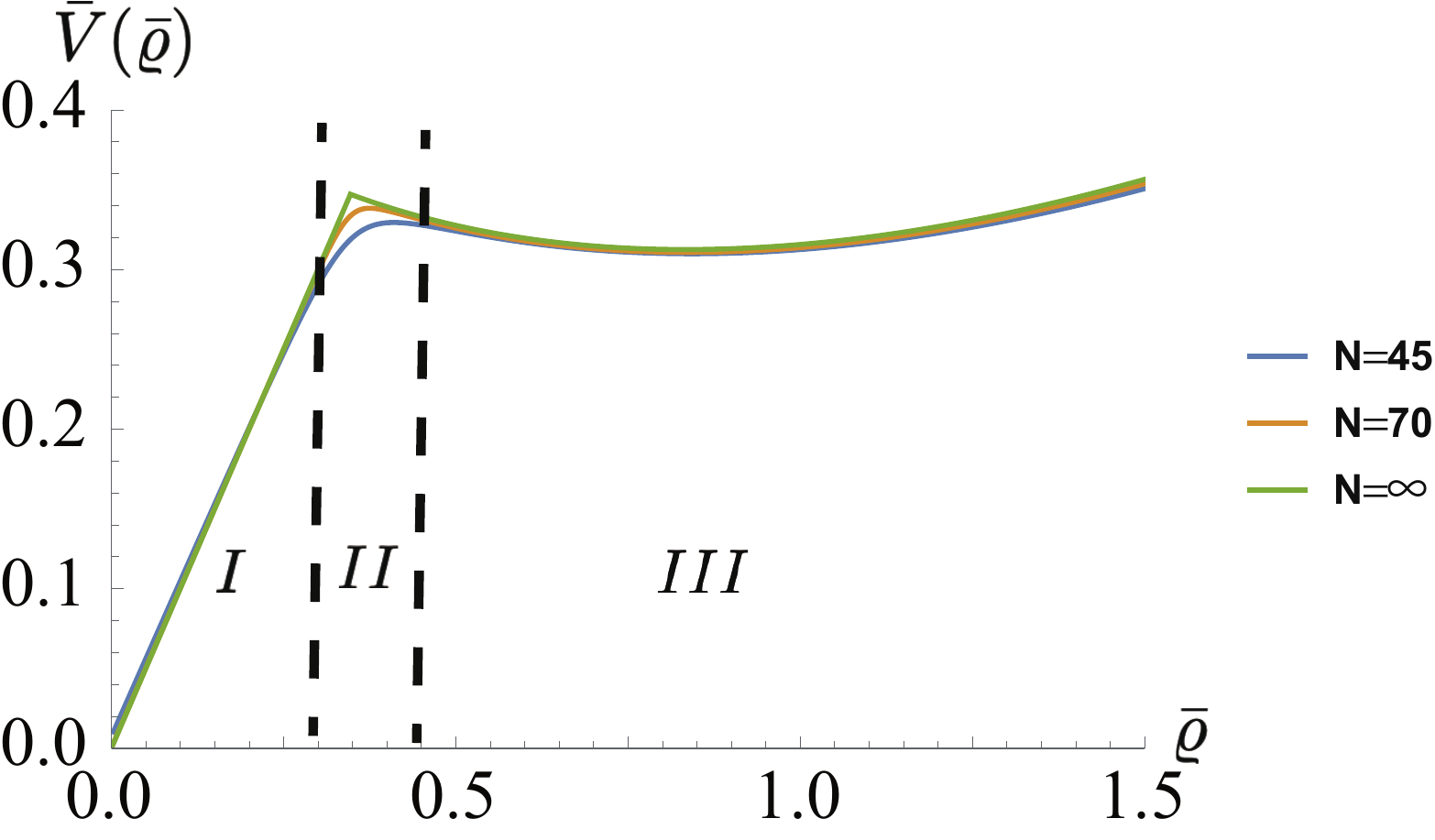}
\caption{$d=3.2$ and Wilson-Polchinski version of the RG:  Potential of the FP SWF$_2$ obtained by numerically integrating  Eq.~(\ref{flow-LPA-WP-essai}) at $N=45$ and 70   showing the boundary layer and the $N=\infty$ limit. The regions I, II and III correspond respectively to the trivial linear solution $\bar{V}(\bar{\varrho})=\bar{\varrho}$, the boundary layer and the WF FP. Region II shrinks to zero in the limit $N\to\infty.$ }
\label{c2cusp}
\end{figure}

Let us therefore  describe the detail of the boundary layer analysis of Eq.~(\ref{flow-LPA-WP-essai}) for  SWF$_2$ in a generic dimension $d>3$. We assume that $\bar{V}'(\bar{\varrho})$ remains of order 1 close to the matching point $\bar\varrho_0$, see Fig. \ref{matching} and that it changes  from $\bar{V}'(\bar{\varrho}_0^-)=1$, for the trivial solution $\bar{V}(\bar{\varrho})=\bar{\varrho}$, to $\bar{V}'(\bar{\varrho}_0^+)$
 for the WF FP. This occurs  at finite and large $N$ across a thin boundary layer located around $\bar\varrho_0$, whose width is of order  $1/N$ so that $\bar {V}''$ scales as $N$. Inside this boundary layer, we introduce a scaled coordinate $\tilde{\varrho}=N (\bar{\varrho}-\bar{\varrho}_0)$ and denote  $\bar{V}'(\bar{\varrho})$ by $F(\tilde\varrho)$. Then, from Eq.~(\ref{flow-LPA-WP-essai}) follows the  differential equation, valid inside this boundary layer  at  leading order in $1/N$:
\begin{equation}
 0=1-d\,\bar V(\bar{\varrho}_0)+(d-2)\bar\varrho_0 F+2\bar\varrho_0{F}{}^2-F-{2}\bar\varrho_0\,F'
\label{flow-LPA-WP-essai22}
\end{equation}
where $\bar\varrho$ has been replaced by $\bar{\varrho}_0$ in $-d\,\bar V(\bar{\varrho})$, $\bar\varrho F$, $\bar\varrho F^2$ and $\bar\varrho F'$. The primes in Eq.~(\ref{flow-LPA-WP-essai22}) stand for derivatives with respect to the scaled variable $\tilde\varrho$. The solution  of this differential equation reads
\begin{eqnarray}
F(\bar{\varrho})=V_1 - V_2 \tanh(V_2 \tilde\varrho),
\end{eqnarray}
where we have defined  $V_1=1/2+\bar{V}'(\bar{\varrho}_0^+)/2$ and $V_2=1/2-\bar{V}'(\bar{\varrho}_0^+)/2$. This boundary layer solution smoothly  connects the two values $\bar{V}'(\bar{\varrho}_0^-)$ and $\bar{V}'(\bar{\varrho}_0^+)$  across the boundary layer, as expected.

The above boundary layer analysis strongly suggests that SWF$_2$ and SG$_3$ exist at finite $N$ at least when $N$ is sufficiently large. We have confirmed their existence within the LPA by solving numerically the FP equation (\ref{flow-LPA-WP-essai}) at fixed $d$ and for various values of $N$, see Fig. \ref{c2cusp} for SWF$_2$ where the boundary layer corresponds to the region II.

\textcolor{black}{It is natural to wonder whether SWF$_2$ and SG$_3$ are artifacts of the LPA approximation. We show below that their existence is necessary for consistency reasons and that they do not exist for $d>4$ in agreement with general results about the triviality of the long distance physics of the O($N$) model in these dimensions.}

\begin{figure}[t]
\includegraphics[scale=0.27]{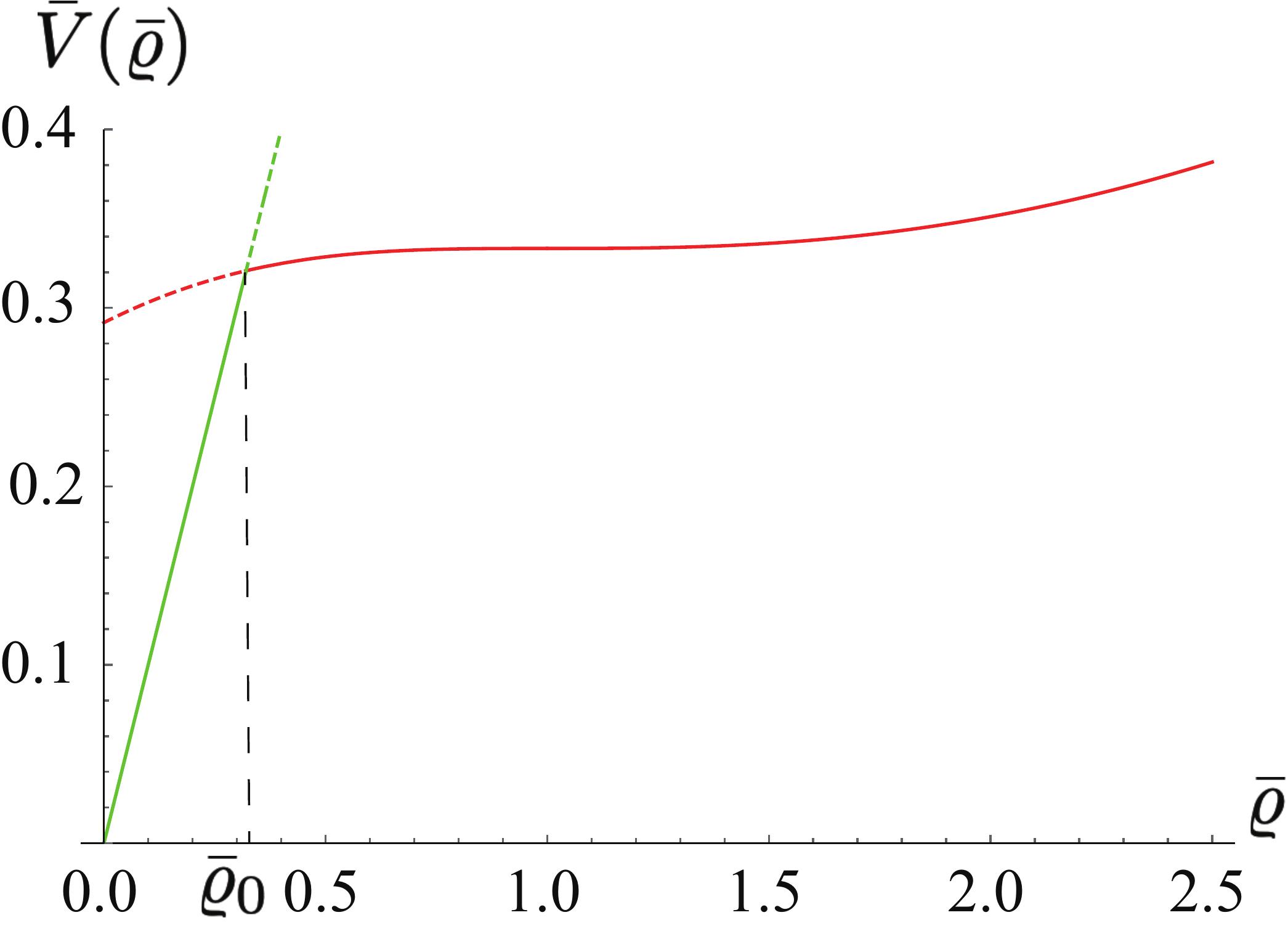}
\caption{$d=3$ and $N=\infty$: Construction of the singular counterpart $S{\cal A}$ of a regular tricritical FP ${\cal A}$ of the BMB line. The potential of ${\cal A}$ is the solution of  Eq.~(\ref{flow-LPA-WP-essai}) obtained with $\tau\simeq0.1776$ (shown in red). The potential of $S{\cal A}$ is shown  as a solid line made of two parts that match at $\bar{\varrho}_0\simeq0.32$. For $\bar{\varrho}>\bar{\varrho}_0$ it coincides with ${\cal A}(\tau)$ and for $\bar{\varrho}<\bar{\varrho}_0$ it is $\bar V(\bar{\varrho})=\bar{\varrho}$.}
\label{fig:ST}
\end{figure}

\subsection{The BMB phenomenon revisited: Explicit construction of new FPs at $N=\infty$ in $d=3$ in the Wilson-Polchinski approach}\label{FPd3}
\label{BMB-revisited}

We have recalled in Section \ref{usual-large-N-functional} that in $d=3$ and $N=\infty$, there exists a line of tricritical FPs, the BMB line, see Fig. \ref{V(rho)}. In the present section, we show that from each of these FPs $\cal{A}(\tau)$ we can construct its singular counterpart $S\cal{A}(\tau)$ showing an isolated singularity in  the same way as SWF$_2$ is the singular counterpart of the WF FP and SG$_3$ the singular counterpart of the gaussian FP G. We also show below that these FPs at $N=\infty$ and $d=3$ are also realized as limits of finite-$N$ nonperturbative multicritical FPs.

The strategy to build singular FP potentials at $N=\infty$ and $d=3$ from the potentials of the FPs of the BMB line is the same as previously for SWF$_2$ and SG$_3$ up to the difference that there is now an infinite number of regular FPs $\cal{A}(\tau)$ along the BMB line and, thus, infinitely many singular FPs $S\cal{A}(\tau)$ built out of these regular FPs. 

We first consider a regular tricritical FP $\cal{A}(\tau)$ on the BMB line, see Fig. \ref{V(rho)}, and connect the linear solution $\bar V(\bar\varrho)=\bar\varrho$ with this regular FP at a crossing point as shown in Fig. \ref{fig:ST}. In much the same way as the regular FPs along the BMB line can be obtained by continuous deformations of the gaussian FP, see Fig. \ref{V(rho)}, their singular counterparts can also be obtained as continuous deformations of the BMB FP. This FP which is the endpoint of the usual BMB line, also shows a linear part at small $\bar\varrho$, see Fig. \ref{V(rho)}. Thus, the BMB FP plays a pivotal role because it is at the same time the last point of the regular BMB line and the first singular FP of the full BMB line 
which is therefore made of a regular part, the usual BMB line, and of a second part made of the singular FPs described above. The BMB FP is thus "at the middle" of the entire BMB line.

\section{Nonperturbative Fixed points for all  $N$ and $d$ -- Tricriticality of the O($N$) models}
 \label{sectionV}

We have shown in the previous section that two singular FPs, SWF$_2$ and SG$_3$, exist at $N=\infty$ for $d\in[3,4[$ and that they are the limits when $N\to\infty$ of FPs whose potentials show a boundary layer at large $N$. We have also shown that it is possible to construct singular FPs from the BMB line of FPs. In the following sections we  study their extensions at finite $N$ and show that they play an important role for the multicritical physics of the O$(N)$ models, even in $d=3$.

The following analysis starts at $N$ large and the value of $N$ will be continuously lowered. An intricate homotopy structure among these fixed points will eventually show up for moderate values of $N$.

\subsection{Perturbative tricritical and multicritical FPs of the O($N$) model}
\label{perturbative-tricritical}

 Before tackling with the nonperturbative multicritical FPs  that are the finite $N$ counterparts of the singular FPs on the BMB line, Sec.  \ref{BMB-revisited}, let us first recall the known perturbative results about multicritical FPs of the O($N$) model.

The phase transitions in the O($N$) model can be either continuous or discontinuous. For these models, when no symmetry breaking terms are included, first order transitions are found only when terms of degree at least 6 in the fields are considered. In coupling constant space, or equivalently within a phase diagram, the boundary separating the first and second order regions corresponds to specific continuous phase transitions that are called tricritical. In absence of symmetry breaking terms, the second order region requires fine tuning a single relevant coupling, or equivalently phase diagram variable, and it is then said to be a hyperspace of codimension one. The tricritical region is at the boundary of this codimension one hypersurface and it is then naturally of codimension two and requires thus fixing two relevant couplings or equivalently phase diagram variables. The tricritical universality class can be studied perturbatively with the massless $(\boldsymbol\varphi^2)^3$ theory which is renormalizable for $d\le 3$. An $\epsilon=3-d$ expansion is therefore possible. For all finite values of $N$, the corresponding perturbative tricritical FP, that we call A$_2$, bifurcates from the gaussian FP G below $d=3$. The index 2 refers to the number of relevant eigendirections of the FP A$_2$.

As recalled in the introduction, the same holds true for all perturbative multi-critical FPs of the O($N$) model that are described by the massless $(\boldsymbol\varphi^2)^{p+1}$: They all bifurcate from the Gaussian FP G below their respective upper critical dimension which is $d_c(p)=2+2/p$ and they are $p$ times unstable in the infrared. The paradox mentioned in the introduction is that although they are all found in the $\epsilon=d_c(p)-d$ expansions for all finite values of $N$, none of them exists at $N=\infty$.

\subsection{Large $N$ analysis of the SWF$_2$ and SG$_3$ FPs: Their critical line $N_c'(d)$ }

\begin{figure}[ht]
\includegraphics[width=8 cm]{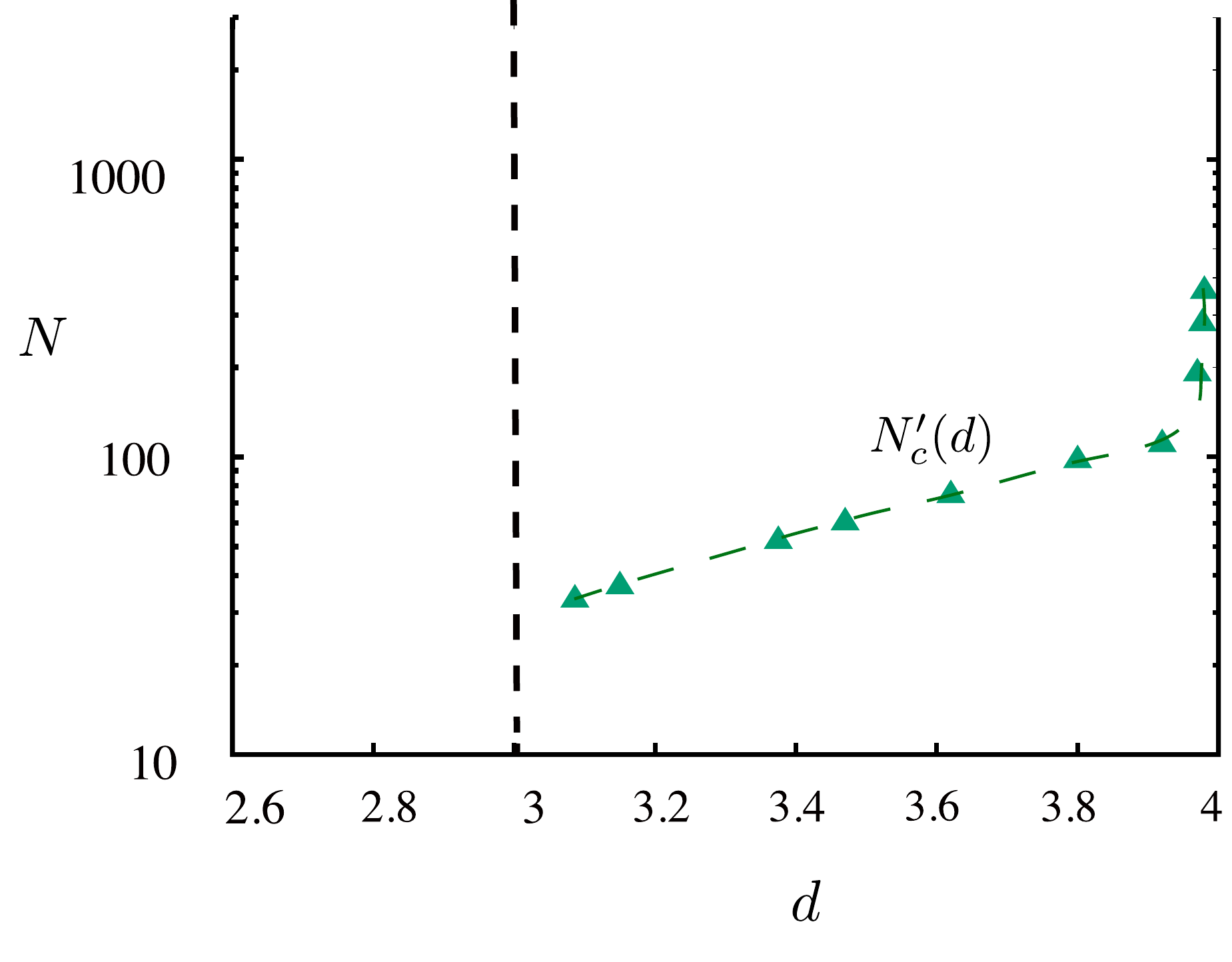}
\caption{The curve $N'_c(d)$ defined by $\textrm{SG}_3=\textrm{SWF}_2$ for $d>3$ and $N$ large. The extension of the curve below $d<3$ has been removed for clarity and is given in Fig. \ref{fig:N_c}. }
\label{Ncddg3}
\end{figure}

We start our large $N$ analysis of the new FPs with SWF$_2$ and SG$_3$. They were shown to be singular at $N=\infty$ and to appear simultaneously in $d=4^-$ where they coincide, see Fig. \ref{2solutions}. By numerically integrating the FP LPA equation at finite $N$, either (\ref{flow-LPA-dimensionless-rescaled}) or (\ref{flow-LPA-WP-essai}), we  find that these FPs always appear together along a nontrivial line in the $(d,N)$ plane and exist on the left of this curve. We call $N_{c}'(d)$ this line [or, equivalenty, $d_{c}'(N)$], see Fig. \ref{Ncddg3}. Of course, at finite $N$, their potential is regular for all values of the field and the singularity shows up only in the limit $N\to\infty$.

The line $N_c'(d)$ is asymptotic to the $d=4$ axis and can be fitted by  $N'_c(d)\simeq 71.0-23.0\log(4-d)$ at large $N$, that is, close to $d=4$, see Appendix \ref{sec:curveN'c} for more details. Notice that SWF$_2$ and SG$_3$ are nonperturbative in the sense that they never coincide with the Gaussian FP in any $(d,N)$, not even for $N=\infty$ and $d\to4^-$. This explains why they were not found perturbatively in the $\epsilon$-expansion.

\subsection{Large $N$ analysis in the vicinity of $d=3$: The finite-$N$ counterpart of the  BMB line}
\label{section-BMB}

\begin{figure}[h]
\begin{centering}
\includegraphics[scale=0.35]{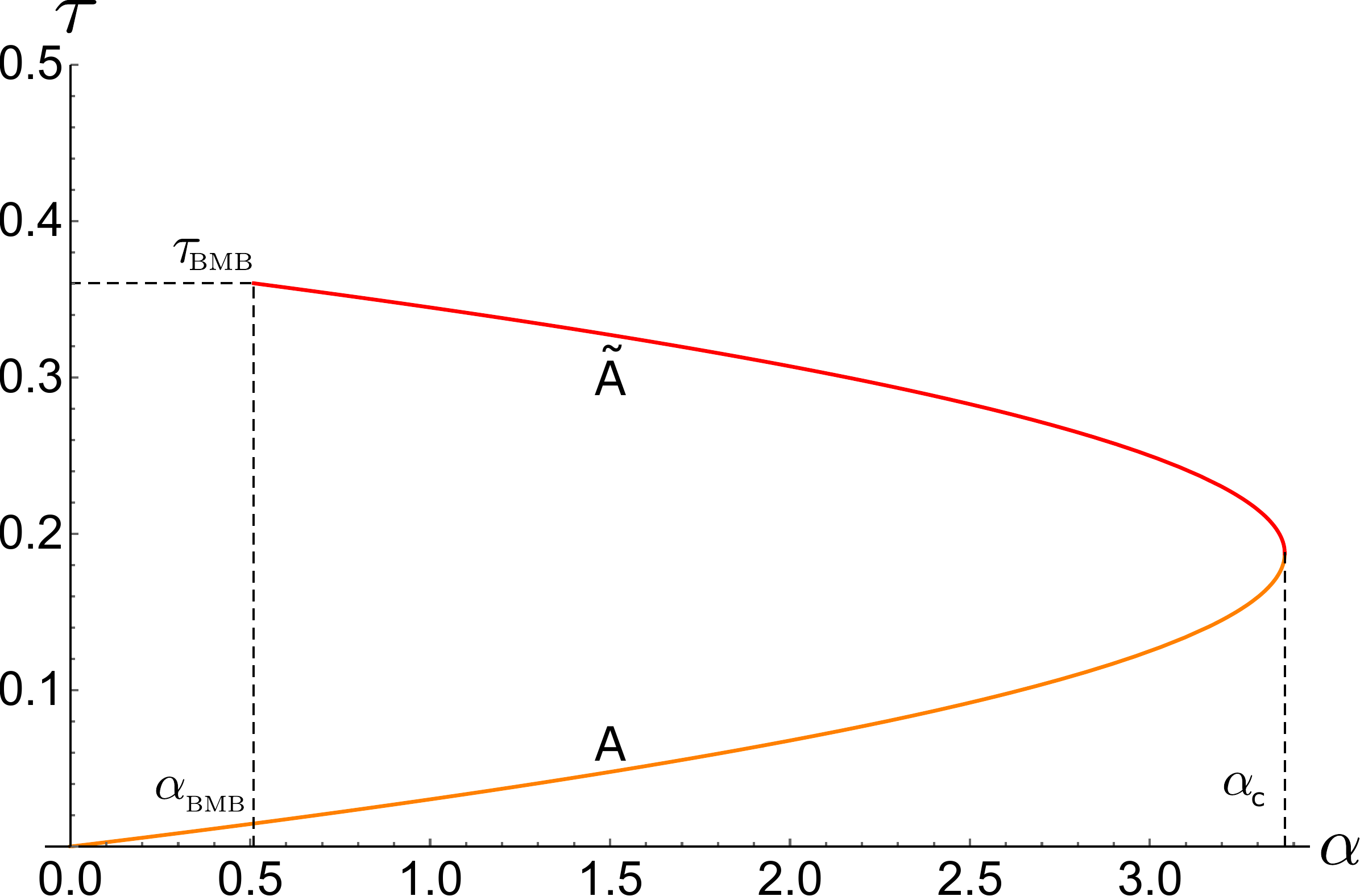}
\par\end{centering}
\caption{Relation between the parameter $\tau$ parameterizing the BMB line of FPs ${\cal A}(\tau) $, Eq.~(\ref{solrho}),  and the parameter $\alpha=(3-d)N$ within the LPA. The BMB line divides in two branches given by A$(\tau_1)$ and $\tilde{\textrm{A}}(\tau_2)$, see text. They correspond to the limits when $N\rightarrow\infty$ of two FPs existing at finite but large $N$, namely A$_2(\alpha)$ and $\tilde{\textrm{A}}_3(\alpha)$. Both branches meet at $\alpha=\alpha_{c}$ with $\alpha_{c}^{\text{LPA}}=27/8$ and the upper branch $\tilde{\textrm{A}}$ extends to the point $(\alpha_{\text{BMB}}^{\text{LPA}},\tau_{\text{BMB}}^{\text{LPA}})\simeq(0.51,0.36)$.}
\label{taualpha}
\end{figure}

In this section, we show that both the regular and the singular parts of the BMB line of FPs existing in $d=3$ at $N=\infty$ have counterparts at finite $N$. We start by the regular part of the BMB line.

Within the LPA, it can be shown \cite{BMB2020} that to each FP ${\cal A}(\tau)$ on the regular part of the BMB line, where $\tau\in\, [0,\tau_{\rm BMB}]$ is defined in Eq.~(\ref{solrho}), there is one value of $\alpha$ given by:
\begin{equation}
    \alpha(\tau)= 36\tau - 96\tau^2
\label{alpha-tau}    
\end{equation}
such that there exists a FP at finite and large  $N$ that tends to ${\cal A}(\tau)$ when $N\to\infty$ when it is followed along  the line $d=d(\alpha(\tau),N)=3-\alpha(\tau)/N$, see Appendix \ref{sec:Analytical relation}. This means that  each FP on the regular part of the BMB line has a finite $N$ counterpart with the subtlety that the  $N\to\infty$ limit of these latter FPs must be taken in a correlated way $N\to\infty$ and $d\to 3$ at fixed value of $\alpha=(3-d)N$. We now show which FPs existing at finite $N$ tend to a given FP on the BMB line.

When the relation between $\alpha$ and $\tau$ in Eq.~(\ref{alpha-tau}) can be inverted, there exists either one or two values of $\tau$ associated with one given value of $\alpha$, see Fig. \ref{taualpha}. We call them $\tau_1(\alpha)$ and $\tau_2(\alpha)$ with $\tau_1(\alpha)\le\tau_2(\alpha)$. The regular part of the BMB line is therefore made of two parts, the set of FPs ${\cal A}(\tau_1)$ that we call A$(\tau_1)$, and the set of FPs ${\cal A}(\tau_2)$ that we call $\tilde{\text A}(\tau_2)$: $\{ {\cal A}(\tau)\}=\{$A$(\tau_1)\}$ $\cup$ $\{\tilde{\text A}(\tau_2)\}$, see Fig. \ref{taualpha}.

For $\alpha=0$, that is, $d=3$, $\tau_1(0)=0$ which corresponds to the Gaussian FP G. When G is followed by continuity along the axis $\alpha=0$, that is, $d=3$, it remains Gaussian up to $N=\infty$. Thus, when $N\to\infty$ at fixed $\alpha=0$, A$_2(\alpha=0)=$\,G $\to$ A$(\tau_1=0)=$\,G.  

For small values of $\alpha$ and at finite $N$, the FP that bifurcates from G below $d=3$ is the perturbative tricritical FP A$_2$, see Section \ref{perturbative-tricritical}. By continuity, when $N\to\infty$ at small and fixed $\alpha$, A$_2(\alpha)\to$ A$(\tau_1(\alpha))$ with $\tau_1$ small. This remains true up to $\alpha_c$, see Fig. \ref{taualpha}.

\begin{figure}[t]
\includegraphics[scale=0.7]{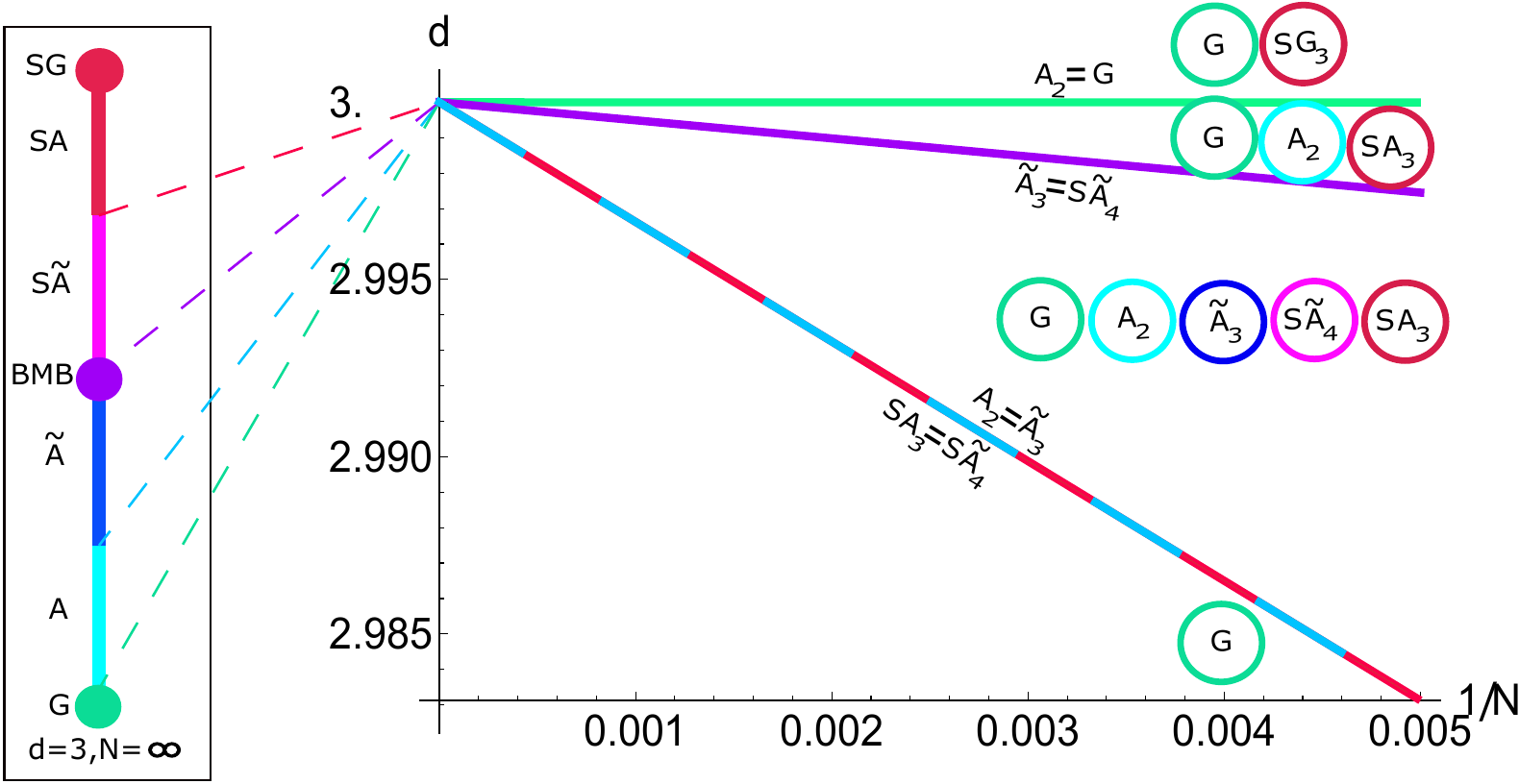}\caption{FPs existing at $N=\infty$ (left panel) and large $N$ (right panel) in $d\le3$ (the Wilson-Fisher FP is not shown although it exists everywhere). Right panel: Straight lines represent the leading order of the  critical lines $d(\alpha,N)=3-\alpha/N +O(1/N^2)$ where two FPs collapse. The horizontal line corresponds to $\alpha=0$, that is, $d=3$. In this dimension,  $A_2$ bifurcates from G and it exists for $\alpha\in[0,\alpha_{\rm c}]$, that is, $d_c(N)<d<3$. At $\alpha_c$, A$_{2}=\tilde{\textrm{A}}_{3}$ and beyond this value of $\alpha$ these two FPs do no longer exist as real FPs. The FP $\tilde{\textrm{A}}_{3}$
exists for $\alpha\in[\alpha_{\rm BMB},\alpha_c]$. At leading order in $1/N$, $\tilde{\textrm{A}}_{3}=\textrm{S}\tilde{\textrm{A}}_{4}$ for $\alpha=\alpha_{\rm BMB}$. The FP $\textrm{S}\tilde{\textrm{A}}_{4}$ collapses with SA$_{3}$ for $\alpha=\alpha_{\rm c}$ and thus, as $\tilde{\textrm{A}}_{3}$, exists for $\alpha\in[\alpha_{\rm BMB},\alpha_c]$. The line $d_c(N)$ corresponding to $\alpha=\alpha_c$ is represented as a dashed line with alternating colors: dark pink for the line where  SA$_{3}=\textrm{S}\tilde{\textrm{A}}_{4}$ and blue for the line where  A$_{2}=\tilde{\textrm{A}}_{3}$.  Notice that these lines are only superimposed  at leading order in $N$ and differ at finite $N$. The FP SA$_{3}$ can be followed above $d=3$ where it is identical to SG$_3$. Left panel: the full BMB line made of the regular FPs ${\cal A}=\{A,\tilde A\}$ between the Gaussian and the BMB FPs and of the singular FPs S$\tilde{\cal A}=\{\text{S}A,\text{S}\tilde A\}$ between the BMB and the singular Gaussian SG FPs. This line, made of four parts, corresponds to the limits when $N\to\infty$ of A$_2\to$A, $\tilde{\textrm{A}}_3\to\tilde{\textrm{A}}$, $S\tilde{\textrm{A}}_4\to S\tilde{\textrm{A}}$ and $S{\textrm{A}}_3\to S{\textrm{A}}$. The dashed lines between the right and left panels show the limits of the remarkable FPs: The Gaussian FP G on the BMB line is the limit of the Gaussian FP at finite $N$, the FP at the border of the A and $\tilde{\text A}$ domains, that is, where A$=\tilde{\textrm{A}}$ is the limit of A$_{2}=\tilde{\textrm{A}}_{3}$ along the line indexed by $\alpha_{\rm c}$, the BMB FP is the limit of $\tilde {\textrm{A}}_{3}=S\tilde{\textrm{A}}_{4}$ along the line indexed by $\alpha_{\rm BMB}$, the FP where $SA=S\tilde{\textrm{A}}$ is the limit of $SA_{3}=S\tilde{\textrm{A}}_{4}$ along the line indexed by  $\alpha_{\rm c}$ and finally SG is the limit of S$A_3$ along the line indexed by $\alpha=0$.} 
\label{master_fig}
\end{figure}

Beyond $\alpha_c$, we find by direct integration of the FP Eqs.  (\ref{flow-LPA-dimensionless-rescaled-first}) or (\ref{flow-LPA-WP-essai}) that A$_2$ does no longer exist. More precisely, for $\alpha>\alpha_c$, that is, for dimensions smaller than $d_c(\alpha,N)$, the potential of A$_2$ becomes complex and therefore unphysical. In fact, we find at $\alpha_c$ that it disappears, that is, becomes complex, by colliding with another FP which is three times unstable and that we call for this reason $\tilde{\textrm{A}}_3$. By following this latter FP towards the large values of $N$ along the lines  $d(\alpha,N)$, we find that $\tilde{\textrm{A}}_3(\alpha)\to \tilde{\text A}(\tau_2(\alpha))$ when $N\to\infty$ as expected from Eq. (\ref{alpha-tau}). The value of $\alpha_c$ found from Eq. (\ref{alpha-tau}) is $\alpha_{\rm c}^{\rm LPA}=27/8=3.375$. 

Notice that the LPA value of $\alpha_c$ given above is not exact. However, both the disappearance of A$_2$ and the exact value of $\alpha_c$ can be obtained perturbatively. The four-loop beta function of the dimensionless $(\boldsymbol\varphi^2)^3$ coupling $g_6$ rescaled according to $ g_6=\tilde{g}_6/N^2$ can be expanded  in the large $N$ limit and has been shown to involve all the leading in $N$ terms. It reads
 \cite{Osborn,Pisarski}:
  \begin{equation}
    N \beta_{{\tilde g}_6}=-2\alpha\tilde{g}_6+12\tilde{g}_6^2-\pi^2\tilde{g}_6^3/2+O(1/N)
    \label{beta-function1}
\end{equation}
where $\alpha=\epsilon N$ as usual \footnote{This $\beta$ function was computed at $d=3$ within the large $N$ expansion in \cite{Rydnell,Omid}.}.  From this $\beta$-function, valid at small $\epsilon$ and large $N$, follows the existence of two FPs:
 \begin{equation}
\tilde{g}_{6,\pm}^*=\frac{12}{\pi^2}\left(1\pm\sqrt{1-\pi^2 \alpha/36} \right).
 \label{g6}
\end{equation}
Notice that at large $N$, the anomalous dimension is given by $\eta=\tilde{g}^{*2}_6/(6 N^2)$ \cite{Osborn} and its contribution to $\beta_{{\tilde g}_6}$ is negligible at this order. This justifies why we expect the LPA to be a reasonable approximation at large $N$.

Clearly, $\tilde{g}_{6,-}^*$ corresponds to A$_2$ because it coincides with the Gaussian FP for $\alpha=0$ and identifies with the usual perturbative tricritical FP. It exists up to $\alpha=\alpha_c=36/\pi^2\simeq3.65$ which is exact at leading order in $1/N$ \footnote{Notice that this value
has been interpreted in \cite{Osborn,Pisarski} as the radius of convergence of the $\epsilon=3-d$ expansion at large $N$.}. As for the second root $\tilde{g}_{6,+}^*$, it clearly corresponds to $\tilde{\textrm{A}}_3$ because it collides with $\tilde{g}_{6,-}^*$ at $\alpha=\alpha_c$. It was however not clear from perturbation theory whether $\tilde{g}_{6,+}^*$ was a spurious root, probably because it is not Gaussian in $d=3$. In particular, from (\ref{g6}), $\tilde{g}_{6,+}^*$ seems to exist at large $N$ in all dimensions larger than $d_c(N)$ which seemed doubtful, and is actually wrong as we show below. Notice also that neither does Eq. (\ref{alpha-tau}) impose an upper bound on  $\tau$ which could suggest that $\alpha$ is unbounded from below, that is, $\tilde{A}_3$ exists in all dimensions larger than $d_c(N)$. This is due to the fact that Eq. (\ref{alpha-tau}) is obtained from an expansion about $\bar{\varrho}=1$ which is insensitive to the singularity that occurs at $\bar{\varrho}=0$ for $\tau\le\tau_{\textrm{BMB}}$. Since it is this singularity at small field that prevents having a well-defined FP potential it is not surprising that Eq. (\ref{alpha-tau}) cannot predict the existence and the value of $\alpha_{\textrm{BMB}}$. The recourse to a functional analysis of the FP, that is, of Eq. (\ref{solrho})  is therefore mandatory.

We know from Eq. (\ref{solrho}) that the BMB line has a finite extension which means that there is a maximal value $\tau_{\textrm{BMB}}=32/(3\pi)^2$ of the parameter $\tau$ that parameterizes the BMB line. Using Eq. (\ref{alpha-tau}) and Fig. \ref{taualpha}, we find that this upper bound on $\tau_2$ translates into a lower bound on $\alpha$:  $\alpha_{\textrm{BMB}}^{\textrm{LPA}}\simeq 0.51$. 

The exact value of $\alpha_{\textrm{BMB}}$ can be derived from another argument of the $N=\infty$, $d=3$ analysis. We know that the effective potentials of the FPs along the BMB line are all regular at  small values of $\tilde{g}_6$, see Fig. \ref{V(rho)}. It is only at the endpoint of the regular part of the BMB line, that is, at the BMB FP that the FP effective potential starts showing a singularity at small fields. [We recall that the linear part of the potential in the W-P version of the RG maps onto a single point in the Ellwanger-Morris-Wetterich version.] It has been shown that this occurs at $\tilde{g}_{6,+}^*=2$ \cite{David,david1985study}. Using Eq. (\ref{g6}) we conclude that the corresponding value of $\alpha$ is $\alpha_{\textrm{BMB}}=2-\pi^2\simeq 2.13$. Notice that whereas the LPA value of $\alpha_c$ differs from the exact value by about $10\%$ -- 3.375 instead of 3.65 -- the LPA value of $\alpha_{\textrm{BMB}}$ is off by a factor 4. It can be shown that this value improves significantly when going at order two of the derivative expansion \cite{Fleming}.

From our analysis of the BMB line made above, we know that it is made of a regular part and of a singular part which is nothing but a singular copy of the regular part, see Fig. \ref{master_fig}. From this point of view, the BMB FP is both the endpoint of the regular part and the starting point of the singular one. The fact that $\alpha_{\textrm{BMB}}$ is the lower bound of the values of $\alpha$ on the $\tilde{\textrm{A}}$ branch of the BMB line, see Fig. \ref{taualpha}, implies that for finite $N$ the FP $\tilde{\text A}_3$ ceases to exist above the dimension $d_\textrm{BMB}(N)=3-\alpha_{\textrm{BMB}}/N$. As usual, it is expected that this occurs by the collapse of $\tilde{\text A}_3$ with another FP. The paradox is that there is no candidate within the $\epsilon$ and $1/N$ expansions for this new FP. This paradox, which uses only known exact results from perturbation theory and the large $N$ limit at $d=3$, is crucial because it paves the way to the new cusped FPs described in this work. Indeed, we know from our analysis at $N=\infty$ and $d=3$ that there exists new FPs: the $S\cal{A}(\tau)$ FPs. It is therefore natural to assume, and we have checked it numerically, that these FPs have also  finite $N$ extensions and that $\tilde{\text A}_3$ collapses at finite $N$ with one of them. We call S$\tilde{\textrm{A}}_4$ the four times unstable FP that collapses with $\tilde{\text A}_3$ at finite $N$, that is, which is such that S$\tilde{\textrm{A}}_4 = \tilde{\text A}_3$ along the line $d_\textrm{BMB}(N)=3-\alpha_{\textrm{BMB}}/N$. Taking the limit $N\to\infty$, this last equality translates on the BMB line into $S\cal{A}(\tau_{\textrm{BMB}})=\cal{A}(\tau_{\textrm{BMB}})$ which is consistent with the fact that the BMB FP is both the last point of the regular part of the BMB line and the first of the singular part. This is similar to what occurs for A$_2$ that collapses with $\tilde{\text A}_3$ along the line $d_c(N)=3-\alpha_{ c}/N$ which corresponds at $N=\infty$ to A$(\tau_c)=\tilde{\text A}(\tau_c)$ with $\tau_c=\tau(\alpha_c)$, see Fig.~\ref{taualpha}. 

Since at large fields, that is, for $\bar\varrho>\bar\varrho_0$ in Fig. \ref{fig:ST}, the potentials of the regular $\cal{A}(\tau)$
  and singular $S\cal{A}(\tau)$ FPs are identical, the  relation (\ref{alpha-tau}) between $\alpha$ and $\tau$ holds also for the $S\cal{A}(\tau<\tau_{\textrm{BMB}})$ FPs at leading order and for sufficiently large $N$. This implies that S$\tilde{\textrm{A}}_4(\alpha)\to \text{S}\cal{A}(\tau(\alpha))$ when it is followed along the lines $d(\alpha,N)= 3 - \alpha/N$.
Thus, like $\tilde{\text A}_3$ that exists at large $N$ on a finite interval of dimensions corresponding to $\alpha\in [\alpha_c, \alpha_\textrm{BMB}]$, S$\tilde{\textrm{A}}_4$  exists on a finite interval of dimensions which, at large values of $N$, also corresponds to the interval of values $\alpha\in [\alpha_c, \alpha_\textrm{BMB}]$. It must therefore also collide with another FP for $\alpha=\alpha_c$ in the same way as $\tilde{\text A}_3$ collides with A$_2$ on the line $d_c(N)= d(\alpha_c,N)$. We call SA$_3$ the FP that collapses with S$\tilde{\textrm{A}}_4$ on the line $d_c(N)$. Similarly to S$\tilde{\textrm{A}}_4$, the large $N$ limit of the SA$_3$ FPs are singular FPs of the BMB line: We call them SA$(\tau)$ while we call S$\tilde{\textrm{A}}(\tau)$ the limits of the S$\tilde{\textrm{A}}_4(\alpha)$ FPs when $N\to\infty$ and $d\to3$. The set of singular FPs on the BMB line is thus: $\{\text{S}\cal{A}(\tau)\}=\{$S$\tilde{\textrm{A}}(\tau_2)\}\cup\{$SA$(\tau_1)\}$  with S$\tilde{\textrm{A}}_4(\alpha)\to \text{S}\tilde{A}(\tau_2)$ and SA$_3(\alpha)\to \text{SA}(\tau_1)$ which is the singular counterpart of  $\{\cal{A}(\tau)\}=$ $\{\tilde{\textrm{A}}(\tau_2)\}$ $\cup$ $\{A(\tau_1)\}$ with $\tilde{\textrm{A}}_3(\alpha)\to\tilde{\textrm{A}}(\tau_2)$ and $\textrm{A}_2(\alpha)\to\textrm{A}(\tau_1)$. 

By numerically integrating at finite $N$ the FP Eqs.  (\ref{flow-LPA-dimensionless-rescaled-first}) and (\ref{flow-LPA-WP-essai}), we have confirmed the existence of all these FPs as well as their respective limits when $N\to\infty$ and $d\to3$. 
We have in particular found that at finite and moderate values of $N$, the lines where the different FPs collapse by pairs deform and  no longer satisfy $d=3-\alpha^*/N$ with  either $\alpha^*=\alpha_c$ or $\alpha^*=\alpha_{\textrm{BMB}}$. For a reason that will become clear in the following, we call $N_{c,S}(d)$ the line where A$_2=\tilde{\text A}_3$, $N_{c,S'}'(d)$ the line where $\tilde{\text A}_3=\text{S}\tilde{\textrm{A}}_4$, $N_{c,S'}(d)$ the line where SA$_{3}=\text{S}\tilde{\textrm{A}}_4$ and $N_{c,S}'(d)$ the line where SWF$_2=\text{SG}_3$. Notice that from the discussion above $N_{c,S}(d)$ and $N_{c,S}'(d)$ are asymptotically identical at large $N$. 

Finally, it is possible to follow SA$_3$ by continuity above  $d=3$. In the range $d>3$, SA$_3$ is the same as SG$_3$. In other words, it is equivalent to call it SG$_3$ or SA$_3$. By convention, we choose to call this FP SG$_3$ for $d>3$ and SA$_3$ for $d<3$.

\begin{figure}[t]
\includegraphics[scale=0.25]{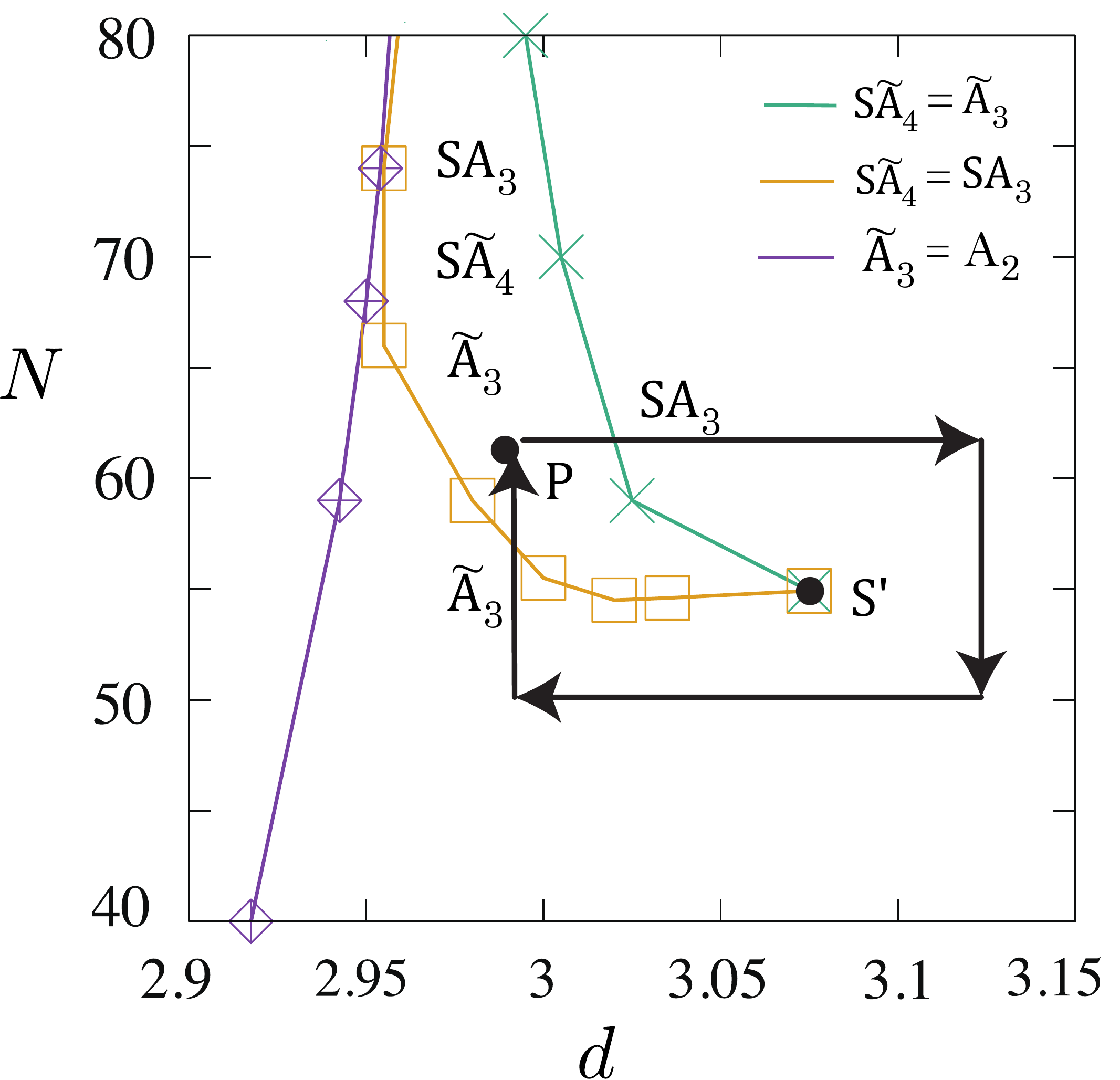}\caption{Point $S'$ and the lines $N_{c,S}(d)\left[\text{A}_2=\tilde{\text{A}}_3\right]$ (violet diamonds), $N_{c,S'}'(d)\left[\tilde{\text{A}}_3=\text{S}\tilde{\text{A}}_{4}\right]$(green crosses) and $N_{c,S'}(d)\left[\text{SA}_{3}=\text{S}\tilde{\text{A}}_{4}\right]$ (orange squares). Starting from $P$, SA$_{3}$ is followed on a clockwise closed path travelling around $S'$. When $d>3$, SA$_{3}$ and SG$_{3}$ are one and the same FP. SA$_{3}$ remains real all along the path but back to the point P, it is $\tilde{\textrm{A}}_{3}$. Following SA$_{3}$ along a path travelling twice around $S'$ it comes back to SA$_{3}$.} 
\label{SASA}
\end{figure}

\begin{figure}[t]

\includegraphics[scale=0.25]{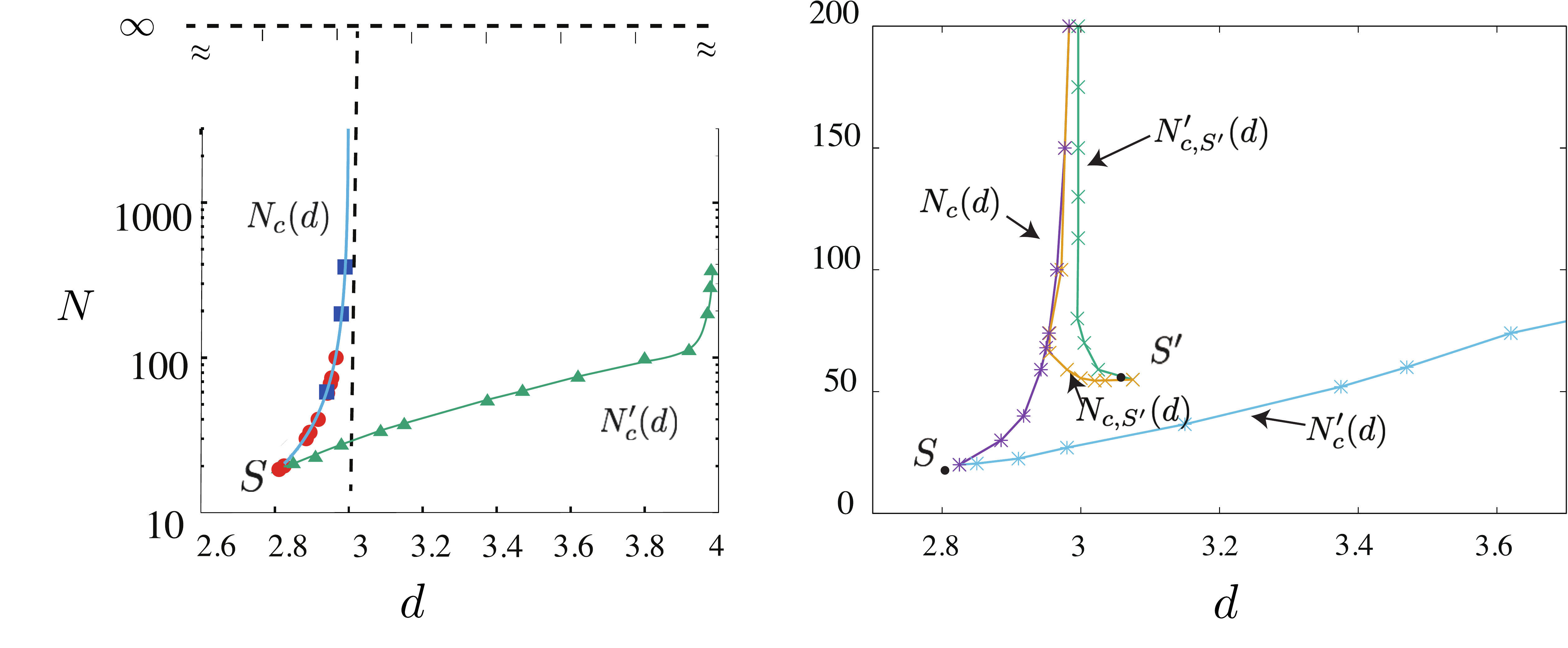}

\caption{Left: The two curves $N_c(d)$ and $N_c'(d)$ respectively defined by A$_2$=$\tilde{\textrm{A}}_3$ and SWF$_2$=SG$_3$ and the  curve $3.6/(3-d)$.
$N_c(d)$ is calculated with the LPA (red circles) and at order 2 of the derivative expansion (blue squares). Right: The four curves $N_{c}\left(d\right)$ (A$_2$=SG$_3$, violet stars), $N_{c}'\left(d\right)$ 
(SWF$_2$=SG$_3$, lightblue stars), $N'_{c,S'}\left(d\right)$ (SG$_3$=S$\tilde{\textrm{A}}_4$
green crosses) and $N_{c,S'}(d)$ (SA$_3$=S$\tilde{\textrm{A}}_4$, orange crosses).}
\label{fig:N_c}
\end{figure}

A summary of the critical lines for $d<3$ as well as their shapes is given in Figs. \ref{master_fig}, \ref{SASA} and \ref{fig:N_c}. \textcolor{black}{Finally, let us emphasize that the mere existence of the new FPs presented above necessarily changes the phase diagram of the O($N$) models because these FPs, which must each have a finite basin of attraction, drive the long-range physics of the systems described by coupling constants in these basins of attraction.}

 \section{ Nontrivial homotopy structures of the multicritical Fixed points}
 
 We have highlighted in our previous large $N$ analysis the existence of many new FPs and critical lines that were either not  known or fully acknowledged. However, at smaller values of $N$, say $N=1,2,3$, it is well known that all FPs have an upper-critical dimension where they become Gaussian. This is contrary to the nonperturbative FPs found above where $\tilde{\textrm{A}}_{3}$ (resp. SG$_3$) collapses with S$\tilde{\textrm{A}}_4$ (resp. SWF$_2$)  when the dimension is increased. How can these two situations be made compatible ? We show in the following that the answer lies in the existence of two special points of the $(d,N)$ plane that we call $S$ and $S'$ where critical lines meet. We show in detail below that these points have the interesting property that two FP solutions are swapped when they are followed by continuity in the $(d,N)$ plane along any path travelling around $S$ or $S'$.
 
 Let us then first recall that at the LPA a FP potential $\bar U$ is a solution  of Eq.~(\ref{flow-LPA-dimensionless-rescaled}) [or equivalently of Eq.~(\ref{flow-LPA-WP-essai})]. As such, it is a function of $d$ and $N$: $\bar U= \bar U(\bar\rho,d,N)$. Thus, these FPs can be followed smoothly  in the $(d,N)$ plane by varying continuously these parameters. Notice that when either $d$ or $N$ is varied, a FP can collide with another one for a given value of these parameters and disappear. Beyond this value it indeed does no longer exist as a physical FP but it still exists as a complex solution to the FP equation. From a mathematical point of view, it will be useful in the following to consider these complex FPs. 
 
Studying the behavior of all the FPs found above when they are smoothly followed along paths in the $(d,N)$ plane falls under homotopy theory or, more precisely in this case, monodromy theory, because a double valued structure of the FP potentials will be exhibited. We use in the following, somewhat abusively, the word homotopy instead of monodromy. We show below that the homotopy of the set of multicritical FPs of the O$(N)$ models considered as functions of $d$ and $N$ is nontrivial.

\begin{figure}[t]
\includegraphics[scale=0.27]{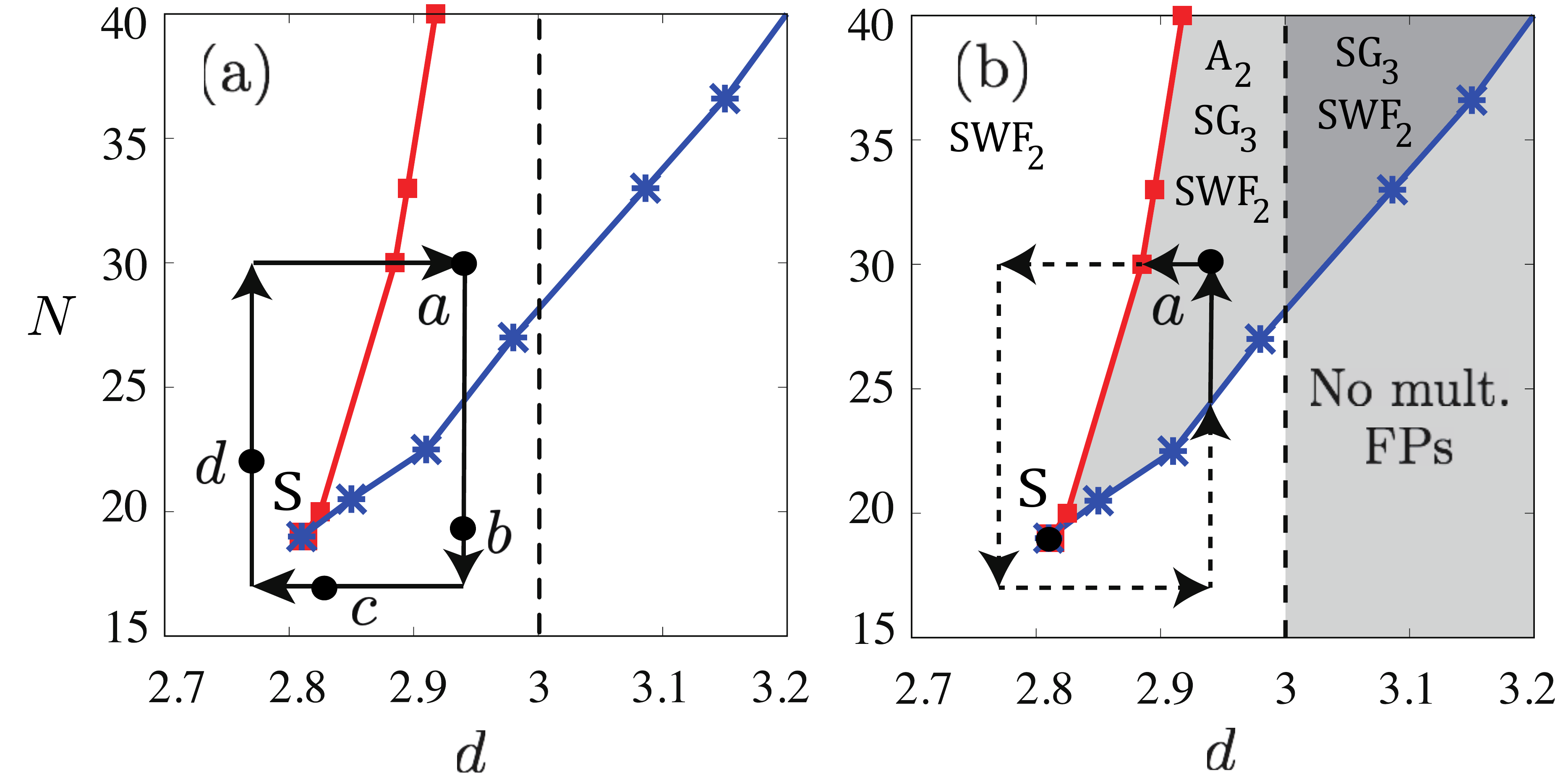}\caption{
The two lines $N_c(d)$ (red squares) and $N_c'(d)$ (blue stars) and their intersection point $S$.  The line $N_c'(d)$ crosses the $d=3$ axis for $N\simeq 28$.
Starting from $a$, the FP A$_2$ is followed along a clockwise (left) or anti-clockwise (right) closed path travelling around $S$.
On the clockwise path, A$_2$ becomes  SWF$_2$ after a full rotation. On the anti-clockwise path, A$_2$ collides with SG$_3$ on $N_c(d)$ and  becomes complex-valued. It remains so all along the dashed path. On $N_c'(d)$, it becomes  real again  but is now SWF$_2$. 
 In panel (b), we indicate which (real) multicritical FPs exist in each region. In the white region, there is only one multicritical FP with two directions of instability that is, for simplicity, denoted by SWF$_2$ and can be continuously followed from both A$_2$ and SWF$_2$ depending on the path followed. } 
\label{paths_around_S}
\end{figure}

\subsection{The first non trivial homotopy structure - The $\tilde{\textrm{A}}_3$ and SA$_3$ (or SG$_3$) FPs}

We have numerically found that by lowering the value of $N$, the lines $N_{c,S'}(d)$ and $N'_{c,S'}(d)$ intersects at a point $S'$ with, at LPA, $S'=(d_{S'}^{\text{LPA}}\simeq3.08,N_{S'}^{\text{LPA}}\simeq 55)$, see Fig. \ref{SASA}. Notice in particular that the critical lines where A$_2$=$\tilde{\textrm{A}}_3$ and SA$_3$=S$\tilde{\textrm{A}}_4$ split when $N$ is decreased whereas they were almost superimposed at large $N$. This is expected as the leading order of the boundary layer analysis only implies that the critical line on which SA$_3$ collapses with S$\tilde{\textrm{A}}_4$ converges to the critical line of A$_2$ and $\tilde{\textrm{A}}_3$ for sufficiently large $N$.

We may also notice that S$\tilde{\textrm{A}}_4$ collides with $\tilde{\textrm{A}}_3$ on $N'_{c,S'}(d)$ and with SA$_3$ on $N_{c,S'}(d)$ which implies that it ceases to exist as a real valued FP below the point $S'$. However, since $d_{S'}^{\text{LPA}}>3$, both $\tilde{\textrm{A}}_3$ and  S$\tilde{\textrm{A}}_4$ exist as physical FPs in $d=3$ at this order of approximations on a finite range of values of $N$,  between $N=55$ and $N=72$.

An interesting feature related to the existence of the point $S'$ is the existence a non trivial homotopy in the $\left(N,d\right)$ plane when  SA$_3$ is followed along a loop that travels around the point $S'$. This is a simple consequence of the topology in parameter space of what is known as a "cusp bifurcation". Further details on this bifurcation and its relationship with the RG and the point $S'$ are given in Appendix \ref{sec: bifurcation_section}. For now, it is sufficient to notice that along the clockwise path shown in Fig. \ref{SASA} and starting at the point $P$, SA$_3$  remains real after coming back to $P$ because along this path it does not collide with any other FP. However, back at $P$, the potential of the FP is no longer the initial potential. It becomes in fact the potential of $\tilde{\textrm{A}}_3$ as can be checked by continuously decreasing $d$ at fixed $N$ from the point $P$ down to the line $N_c(d)$ where it  collapses with A$_2$ and disappears if we go on decreasing $d$. Travelling twice around $S'$ we get back the SA$_3$ potential. 

Following instead an anti-clockwise  path,
SA$_3$ collides with S$\tilde{\textrm{A}}_4$ on $N_{c,S'}(d)$. It then becomes complex and remains so until the path crosses $N'_{c,S'}(d)$. It then becomes real again and, back to $P$, it is $\tilde{\textrm{A}}_3$. 

Thus, considered as functions of $N$ and $d$, the FP potentials of SA$_3$ and $\tilde{\textrm{A}}_3$ are bi-valued and can be interchanged by following by continuity one of these FPs along a path  that travels around $S'$. We conclude that starting with the FP SG$_3$ (=SA$_3$) in $d>3$ and decreasing $d$ along a path that travels above $S'$ it will collide with S$\tilde{\textrm{A}}_4$ on the line $N_{c,S'}(d)$ and then disappears. On the contrary, if the path travels below $S'$, it will collide with A$_2$ on $N_c(d)$.

\begin{figure}
\includegraphics[scale=0.27]{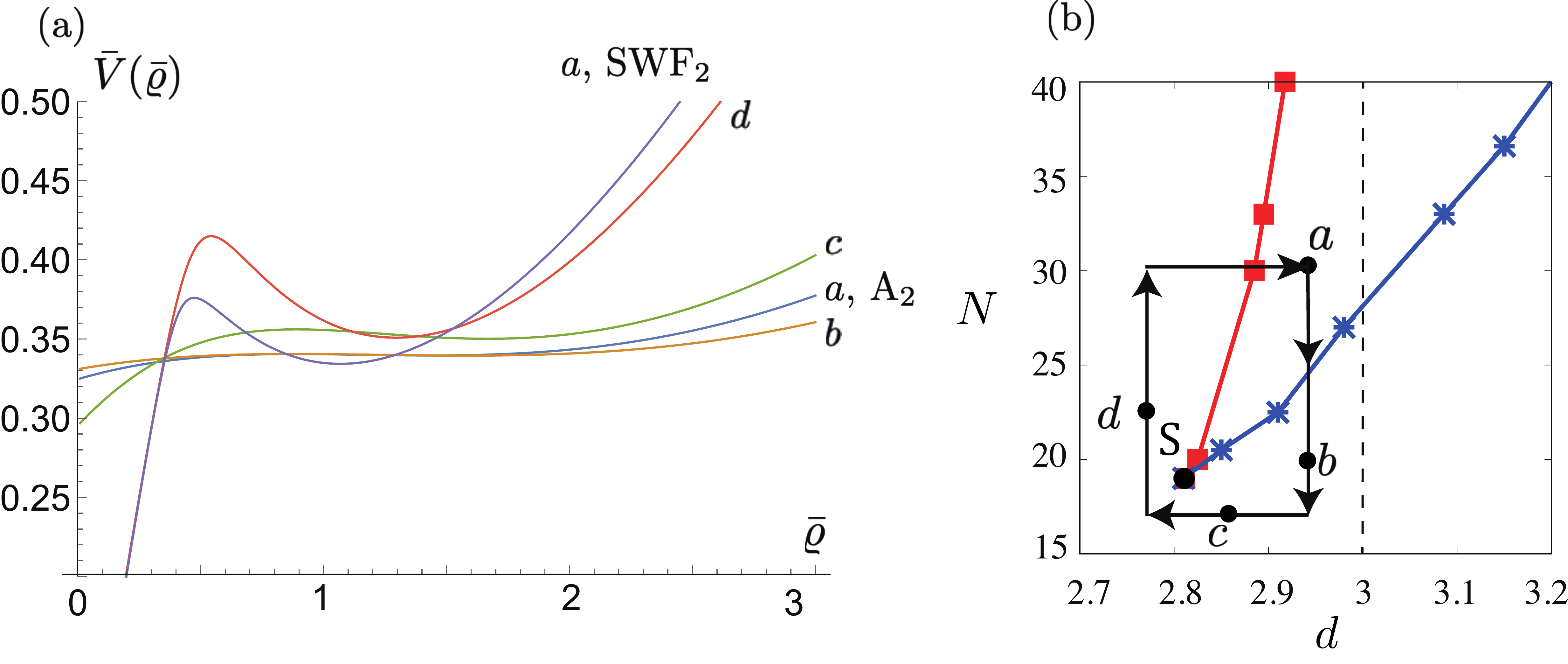}\caption{Evolution of the FP potential $\bar{V}(\bar{\varrho})$ in the W-P approach when  A$_2$ is followed from the point $a=(d=2.94, N=30)$ as a continuous function of $(N,d)$ along the clockwise closed path shown in the right panel (identical to the path shown in Fig. \ref{paths_around_S}(a)). In the left panel we show how the A$_2$ potential changes along the path $(a,b,c,d,a)$ shown in the right panel. In $a$, the A$_2$ potential is very flat because $a$ is close to the $d=3$ axis where it is the Gaussian FP. It remains so at point $b$ and deforms slightly in $c$. Then, it changes drastically between $c$ and $d$ which is the region where the double-valued structure plays an important role. Finally, it evolves slightly between $d$ and $a$ where it is clearly very different from the initial potential: It has become the SWF$_2$ FP.} 
\label{potential evolution}
\end{figure}

\subsection{The second non trivial homotopy structure - The  A$_2$ and SWF$_2$ FPs}

Let us now show that another homotopy structure is associated with A$_2$ and SWF$_2$.

We have shown in Fig. \ref{Ncddg3} that SWF$_2$ collides with SG$_3$ on the line $N_c'(d)$ and exists on the left of this curve. By building point by point this curve we find that it crosses the line $N_c(d)$ at a point $S$,  see Fig. \ref{fig:N_c}, in the same way $N_{c,S'}(d)$ crosses $N_{c,S'}'(d)$ at  $S'$. At LPA, we find $S=(d_S^{\text{LPA}}\simeq2.81$, $N_S^{\text{LPA}}\simeq19)$. Notice that the large $N$ equation of the $d_c(N)$ curve: $d_c(N)\simeq 3- 3.375/N$ remains a good approximation down to $S$.

At the point $S$, A$_2$=$\tilde{\textrm{A}}_3$=SWF$_2$=SG$_3$ which is possible because $N_S<N_S'$ and the FP $\tilde{\textrm{A}}_3$ is then identical to SG$_3$ around the point $S$. 

In Fig. \ref{paths_around_S}(b), we show which FPs exist in the different regions of the $(d,N)$ plane around the point $S$. It is very interesting to realize that contrary to common belief, there exists in $d=3$ for $N$ sufficiently large, that is,  $N>N_c'(d=3)\simeq 28$, a nontrivial twice unstable FP SWF$_2$ and a non trivial three times unstable FP SG$_3$ that have never been found previously \footnote{A non Wilson-Fisher FP has been mentioned in \cite{NonWF,NonWF2,NonWF3} but it is currently not clear whether it corresponds to one of the FP found in the present work.}. 

As with $S'$, a nontrivial homotopy structure is also associated  with  $S$ but this time for SWF$_2$ and A$_2$. Here again, following A$_2$ along either a clockwise or an anti-clockwise closed path travelling around $S$ the deformations of the FP potential considered as a function of $d$ and $N$ are different, see Figs. \ref{paths_around_S} and \ref{potential evolution}.

\begin{figure}[t]
\includegraphics[scale=0.65]{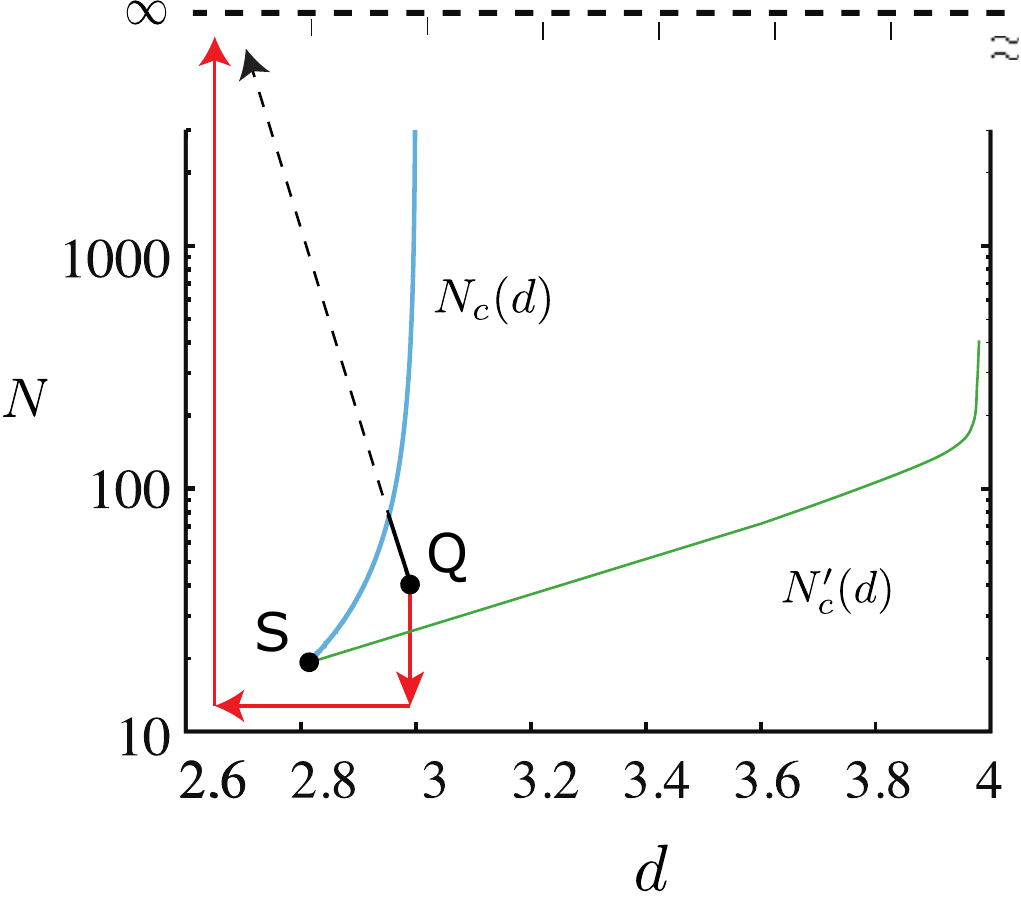}
\caption{Two paths that travel   below (shown in red) and above (shown in black) $S$. The FP A$_2$ is followed in the $(d,N)$ plane starting from $Q$ in $d=2.99$. On the path above $S$, A$_2$ vanishes by colliding with $\tilde{\rm A}_3$ on the line $N=N_c(d)$. On the path below $S$, A$_2$ exists everywhere but becomes indistinguishable with SWF$_2$ when it is away from the $d=3$ axis, and finally becomes cuspy when $N\to\infty$.}
\label{fig:2paths}
\end{figure}

  To conclude, when two FPs with the same degree of instability coexist in a region of the $(d,N)$ plane, below the point $S$ (resp. $S'$) it is largely arbitrary to call them A$_2$ or SWF$_2$ (resp. $\tilde{\textrm{A}}_3$  or SA$_3$) because they can be interchanged when they are continuously followed along paths travelling around $S$ (resp. $S'$). It is only in the infinitesimal neighborhood of some remarkable lines that it is possible to give them a name in an unambiguous way: In $d=3-\epsilon$ for instance, the FP whose potential is almost flat can be safely called A$_2$. The same holds true for SWF$_2$ on the $N_c(d)$ line. However, far from these remarkable lines, the only meaningful way to know which FP we are considering is to specify where it comes from and along which path.  This remark also holds for $\tilde{\textrm{A}}_3$, SA$_3$ and  SG$_3$ with now the first non trivial homotopy structure explained in the last section. For example, the name $\tilde{\textrm{A}}_3$ can specify a FP only unambiguously near the line $N_c(d)$. Let us also notice that the two homotopy structures described above ensure that the new multicritical FPs do not exist for   \textcolor{black}{lower values of $N$ such as} $N=1,2$ and 3 \textcolor{black}{since both $N_S$ and $N_S'$  are much larger than 3}.

\subsection{The solution of the paradox about the absence of a tricritical FP at $N=\infty$ and $d<3$}

Our analysis above solves the paradox raised in the introduction about the absence of any tricritical FP at $N=\infty$ and $d<3$. Let us indeed assume that we consider the perturbative tricritical FP A$_2$ in $d=3-\epsilon$ for an arbitrary value of $N$ and let us follow it up to $N=\infty$ in a dimension $d$ between 2 and 3. Our previous analysis shows that there are two kinds of non equivalent paths to reach $N=\infty$, those that travel below $S$  and those that travel above $S$ as shown in Fig. \ref{fig:2paths} in red and black. For a path that travels above $S$, A$_2$  collides with $\tilde{\textrm{A}}_3$ on $N_c(d)$ and then disappears. As a consequence, it cannot be found at $N=\infty$. For a path that travels below $S$, A$_2$ displays a clear boundary layer  when $d<d_S$ and $N>N_S$ and we may then call this FP SWF$_2$: It  becomes cuspy when $N\to\infty$. In this latter case it is not found in the usual large $N$ analysis that eliminates by construction these kinds of FPs. Thus, in both cases, no trace of a tricritical FP at $N=\infty$ can be found in the usual large $N$ approach.

\section{Conclusion}
We have solved in this article an old paradox of the O($N$) models: How can it be that a perturbative tricritical fixed point exists for all $N$ in dimension $d=3-\epsilon$ whereas no tricritical FP is found at $N=\infty$ in  $d<3$? It turns out that the solution to this paradox is incredibly intricate: It requires no less than the existence of four new fixed points, SWF$_2$, $\tilde{\textrm{A}}_3$, SG$_3$ and S$\tilde{\textrm{A}}_4$. These FPs  (i) never coincide with the gaussian FP whatever $d$ and $N$, (ii) appear by pairs when $d$ is decreased from two nontrivial lines $N_c'(d)$ and $N_{c,S'}'(d)$ and collide with other FPs on two other nontrivial lines $N_c(d)$ and $N_{c,S'}(d)$, see Fig. \ref{fig:N_c}, (iii)  are singular at $N=\infty$ except for $\tilde{\textrm{A}}_3$, (iv) are bi-valued in the $(d,N)$ plane (but for S$\tilde{\textrm{A}}_4$), that is, A$_2$ and SWF$_2$ (resp. $\tilde{\textrm{A}}_3$ and  SG$_3$) are interchanged when they are followed along paths travelling  around the point $S$ (resp. $S'$), see Figs. \ref{potential evolution} and  \ref{SASA}. 

Several lessons can be drawn from this study. 

First, the tricritical behavior of the O$(N)$ models was believed to be fully captured by the massless $(\boldsymbol\varphi^2)^3$ and therefore to become nontrivial only below $d=3$. We find on the contrary  that SWF$_2$ which is a twice unstable FP -- possibly tricritical -- exists and is highly nontrivial in $d=3$ for $N$ sufficiently large, that is, at LPA, for $N$ typically larger than 28, see Fig. \ref{paths_around_S}. This FP cannot be described by the perturbative massless $(\boldsymbol\varphi^2)^3$ theory and exists above $d=3$ up to the curve $N_c'(d)$, see Fig. \ref{fig:N_c}.  \textcolor{black}{Notice that barring miracles, the basin of attraction of SWF$_2$ cannot be empty. Since at finite $N$  the potential of SWF$_2$ shows all the properties of a well-defined theory, there should exist well-defined microscopic models having a RG flow terminating at SWF$_2$. Therefore, the long-distance physics of these models should be driven by SWF$_2$ and, as a result, the presence of this FP must change the phase diagram of the O($N$) models when it exists. This shows that at least for sufficiently large values of $N$, the multicritical behavior of the O($N$) models is {\it a priori} nontrivial in $d=3$.}
 Three other nontrivial FPs -- SG$_3$, $\tilde{\rm A}_3$ and S$\tilde{\textrm{A}}_4$ -- also exist in $d=3$ \textcolor{black}{and when they exist, they also change the phase diagram of the O($N$) models}. The FP S$\tilde{\textrm{A}}_4$ exists within the range $N\in [55,72]$ at LPA, see Fig. \ref{SASA}. Momentarily omitting the monodromy that requires changing the dimension, the FP SG$_3$ exists within the range $N\in [55,\infty[$ and the FP $\tilde{\rm A}_3$ within $N\in [28,72[$. If we then authorize dimensional changes, the non trivial monodromy around the point $S'$ leads us to identify SG$_3$ and $\tilde{\rm A}_3$ below $N=55$. This identification changes the domain of existence of SG$_3$ to the same as SWF$_2$, that is $N\in [28,\infty[$. It would of course be extremely interesting to confirm the existence of all these new FPs by other methods, the conformal bootstrap in particular. 

Second, neither of the nontrivial  FPs found above  bifurcate from the gaussian FP in any $d$. They therefore cannot be found in the $\epsilon=3-d$-expansion. The same holds true for the $1/N$ expansion for SG$_3$, SWF$_2$ and S$\tilde{\rm A}_4$ because at $N=\infty$ their effective FP potential is singular -- they show a cusp -- and  therefore cannot be found in the usual $N=\infty$ analysis and {\it a fortiori} in the standard $1/N$ expansion. The best that can be done perturbatively in a combined $\epsilon=3-d$ and large $N$-expansion is to find that the perturbative tricritical FP -- called A$_2$ here -- can exist only for $\epsilon N\le 36/\pi^2$, that is, on the right of the $N_c(d)$ line, see Eq. (\ref{g6}). A second FP, that we call here $\tilde{\textrm{A}}_3$, is in fact found perturbatively in a combined $1/N$ and $\epsilon$ expansion, see Eq. \eqref{beta-function1},  and it is found that it collapses with A$_2$ along the line $N_c(d)$ but within this approach it is difficult to determine on which interval in $d$ it exists. On the contrary, using the crudest approximation of the NPRG flow equations, that is, the LPA, we easily find the zoo of all new FPs and the domains where they exist, not only at large $N$ and small $\epsilon$ but for all $N$ and in all dimensions. Notice that whereas we have checked the stability of our approximations as for the existence of both the line $N_c(d)$ and of the point $S$ at order two of the derivative expansion, with moderate quantitative changes compared to LPA, a full study at order two remains to be done for all the other FPs and for the point $S'$. 

Third, however complicated the above picture may seem -- network of new FPs, nontrivial monodromies -- it is the simplest one that is consistent with all known results: \textcolor{black}{(i) Perturbation theory in $1/N$ and  $\epsilon$ makes doubtless that the usual tricritical FP A$_2$  ceases to exist for large $N$ as a real-valued FP above the critical line $N_c(d)$, (ii) therefore there must exist another FP -- called here $\tilde{\rm A}_3$ -- with which A$_2$ collides on this line, (iii) triviality implies that  $\tilde{\rm A}_3$  does not exist in dimensions $d>4$ which implies that it collides  with another FP in $d<4$ and thus does not exist in $d=4$, \footnote{\textcolor{black}{The author in \cite{Pisarski2} also mentioned this consistency requirement.}}  (iv)  the BMB line at $N=\infty$ is made of FPs called here A$(\tau)$ and $\tilde{\rm A}(\tau)$ parameterized by the $\phi^6$ coupling $\tau\in[0,\tau_{\rm BMB}]$, (v) all these A$(\tau)$ and $\tilde{\rm A}(\tau)$ FPs
 have a finite $N$ counterpart that are either A$_2(\alpha)$ or $\tilde{\rm A}_3(\alpha)$:  when A$_2(\alpha)$ or $\tilde{\rm A}_3(\alpha)$ are followed along the path $d=3 - \alpha(\tau)/N+O(1/N^2)$, see Eqs. \eqref{alpha-tau} and \eqref{beta-function1}, they reach A$(\tau)$ or $\tilde{\rm A}(\tau)$ on the BMB line, (vi) the BMB FP is the endpoint of the BMB line and is therefore $\tilde{\rm A}(\tau_{\rm BMB})$; its finite and large $N$ counterpart can be followed along the  line $d=3 - \alpha(\tau_{\rm BMB})/N+O(1/N^2)$ that we call above  $N_{c,S'}'(d)$, see Fig. \ref{fig:N_c}, (vii) since this finite $N$ counterpart of the BMB FP cannot exist on the right of the $N_{c,S'}'(d)$ line, it must collide with another FP -- that we call above S$\tilde{\text A}_4$ -- on this line, (viii) the S$\tilde{\text A}_4$ FPs are themselves the finite $N$ counterparts of FPs existing in $d=3$ and $N=\infty$: these FPs existing at  $N=\infty$ in $d=3$  are part of  the singular part of the BMB line, (ix) this singular part of the BMB line being also finite, S$\tilde{\text A}_4$ can only exist at finite and large $N$ on a finite interval of $d$: it collides with yet another FP -- called SA$_3$ -- on the line $N_{c,S'}(d)$ which is identical to $N_{c}(d)$ for $N\to\infty$, (x) once again, triviality implies that SA$_3$ should not exist above $d=4$, and indeed it collides in $d<4$ with another FP that we call SWF$_2$, (xi) the FPs A$_2$ and SWF$_2$ are exchanged when followed by continuity along a path that encircles the point $S$, (xii) $S$ is the intersection between the lines $N_{c}(d)$ and $N'_{c}(d)$, see Fig. \ref{fig:N_c}, (xiii) this nontrivial monodromy explains why there is only one tricritical FP for $N=1$, $d=2$ and no analytical tricritical FP at $N=\infty$ and $d<3$, (xiv) an analogous monodromy structure is found between SA$_3$ and $\tilde{\rm A}_3$ around the point $S'$, see Fig.\ref{SASA}.}

\bigskip

Fourth, even if we are only interested in the physics of the O($N$) models with integer values of $d$ and $N$, it is important to get a consistent picture of all the FPs found for all $d$ and $N$, that is, to understand in which portion of the $(d,N)$ plane they exist, what the flows between them are and what they are at $N=\infty$. It is then unavoidable to generalize the models to all real values of both $d$ and $N$. In doing so, we find that it is also unavoidable to cope with the nontrivial homotopy structures described above which is an intrinsically nonperturbative feature of these models. To the best of our knowledge, this is the first time that such a structure is found in the RG.

Fifth, one of the most intriguing feature of the new FPs found above is that their $N\to\infty$ limit is singular: Their potential shows a cusp, see Figs. \ref{c2cusp}, \ref{fig:ST}, \ref{C2FP}. This of course prevents the usual $N=\infty$ analysis which implicitly relies on analyticity properties. A nontrivial consequence of this nonanalyticity is that the LPA is no longer necessarily exact at $N=\infty$. It is thus important to understand if there exists an exact closure of the RG flow equations at $N=\infty$ allowing for nonanalytic behavior. It is also  important to devise a $1/N$ expansion for these FPs. We expect that the notion of boundary layer outlined above will be an important step in this direction. Finally, an important open question is the relationship between the nonanalyticities of the FP potentials at $N=\infty$ with some physical phenomena. A hint in this direction is the existence of a bound state in $d=3$ and $N=\infty$ right at the BMB FP \cite{David,david1985study}.

Sixth, our study generalizes the BMB phenomenon by showing that the usual BMB line of FPs must be supplemented by a line of singular FPs. It also shows how to extend it  at finite $N$ by revealing that the $N\to\infty$ limit must be taken simultaneously with the $d\to 3$ limit while keeping fixed the value of $(3-d)N$. We note that a similar situation where a large $N$ limit must be taken simultaneously with $\epsilon\to 0$  has also been found in other models in \cite{Gurau,Delporte}. We have also shown that the BMB phenomenon can no longer be considered as a curiosity occuring only at $d=3$ and $N=\infty$ because, together with its singular counterpart, its extension at finite $N$ is necessary for the consistency of the overall picture described above.

Let us now consider some open questions raised by our study. First, we could wonder whether the existence of nonperturbative and/or nonanalytic multicritical FPs is specific to the O($N$) models or whether it is likely to be generic. Two of us have shown in \cite{Yabunaka-Delamotte-PRL2017} that nonperturbative FPs also exist for the O($N$)$\otimes$O(2) model, although with some differences. We are therefore confident that what was shown here for the O($N$) models is rather generic. This lets open the possibility of having nontrivial tricritical behavior in $d=3$ for physical values of $N$ in some models. Second, the study that we have performed in this article focuses on the tricritical behavior of the O$(N)$ models. The same can be done for all $p$ times unstable  multicritical FPs around their critical dimension $d_c(p)=2+2/p$. Our preliminary investigations show that several features encountered with these FPs differ from what has been found in the tricritical case
\cite{defenu2020}. Third, the $d\to2$ limit of the O($N\geq2$)  multicritical FPs is still an open question. For example, while \cite{Codello2015,Codello2013} studies the critical exponents in the limit $d\rightarrow2$, they do not give the FP potentials. Our preliminary investigations show that here again nonanalytic FPs are found but now even for low values of $N$. It is of course an exciting question to study the $N\to2$ and $d\to2$ limit.

\begin{acknowledgements}
We thank J.-B. Zuber for many suggestions about the writing of this article. S. Y. was supported by Grant-in-Aid for Young Scientists (B) (15K17737 and 18K13516).
\end{acknowledgements}

\appendix
\section{Flow equation in the Wilson-Polchinski framework}\label{sec:WP}

Let us  give the definition  of the Wilsonian effective action in the Wilson-Polchinski version of the NPRG, following the notation of \cite{Ellwanger}. We introduce an IR-regulated propagator $P_k(q^2)$ that almost vanishes when $q\ll k$ or $\Lambda< q$ and coincides with $1/q^2$ for $k\ll q<\Lambda$. The partition function is given by 
\begin{equation}
 {\cal Z}_k[\boldsymbol{J}]= \int D\boldsymbol\varphi_i \exp(-S_\Lambda[\boldsymbol\varphi]-\Delta S_k[\boldsymbol\varphi]+ \boldsymbol{J}\cdot\boldsymbol\varphi)
\end{equation}
with the bare action $S_\Lambda[\boldsymbol\varphi]$, $\Delta S_k[\boldsymbol\varphi]=\frac{1}{2}\int_q Q_k(q^2) \boldsymbol\varphi_i(q)\boldsymbol\varphi_i(-q)$, where $Q_k(q^2)=(P_k(q^2))^{-1}$, and $\boldsymbol{J}\cdot\boldsymbol\varphi=\int_x J_i(x) \boldsymbol\varphi_i(x)$.  The wavenumber dependent Wilson effective action $S_{k}(\boldsymbol\phi)$ is defined as $ S_{k}(\boldsymbol\phi)=-\log{\cal Z}_k[\boldsymbol{J}]+\frac{1}{2}\int_{x,y}\boldsymbol\phi(x)\boldsymbol\phi(y)Q_k(x-y) $ for $\boldsymbol{J}(x)=\int_y Q_k(x-y) \boldsymbol\phi(y)$. 
The properties of the regulator explained above imply that :
\begin{equation}
   \begin{array}{llll}
  P_{k=\Lambda}(q)\simeq 0\ \  &\Rightarrow\ \ &S_{k=\Lambda}\simeq S_\Lambda[\boldsymbol\phi] \ \ &{\text{all fluctuations are frozen}}
\end{array}
\label{properties-Sk}
\end{equation} 
The Wilson-Polchinski exact RG equation reads
\begin{equation}
\partial_k  S_k[\boldsymbol\phi]=\frac{1}{2}\int_{x,y} \partial_k P_k(x-y)\left(-\frac{\delta^2 S_k}{\delta \boldsymbol\phi_{i}(x) {\delta \boldsymbol\phi_{i}(y)}}+\frac{\delta {S_k}}{\delta \boldsymbol\phi_{i}(x)}\frac{\delta S_k}{\delta \boldsymbol\phi_{i}(y)}\right).
\label{WP-flow}
\end{equation}
The local potential approximation (LPA) consists in approximating the effective  Wilson effective action $S_k[\boldsymbol\phi]$ as
\begin{equation}
 S_k[\boldsymbol\phi] = \int_{x}  V_k(\mu), 
\label{ansatz-order2}
\end{equation}

 where we defined $\mu=\phi_i \phi_i/2$. At the LPA, the RG equation for $V_k(\mu)$ is given by Eq.~(\ref{flow-LPA-WP-essai}) for any IR-regulated propagator  $P_k(q^2)$ after an appropriate rescaling \cite{Litim2018}.

\section{The  SWF$_2$ and SG$_3$ fixed points in the Ellwanger-Morris-Wetterich approach at large $N$}

Since we have now  obtained the SWF$_2$ and SG$_3$ FPs in the W-P version of the NPRG, both at large and infinite $N$, it is interesting to go back to the Ellwanger-Morris-Wetterich version of the NPRG  because it is the only one that can be studied in a controlled way beyond the LPA. We now show that the standard notion of boundary layer does not apply in this case which makes the construction of  SWF$_2$ and SG$_3$  much more difficult than in the W-P version of the RG.

\begin{figure}[t]
\includegraphics[scale=0.5]{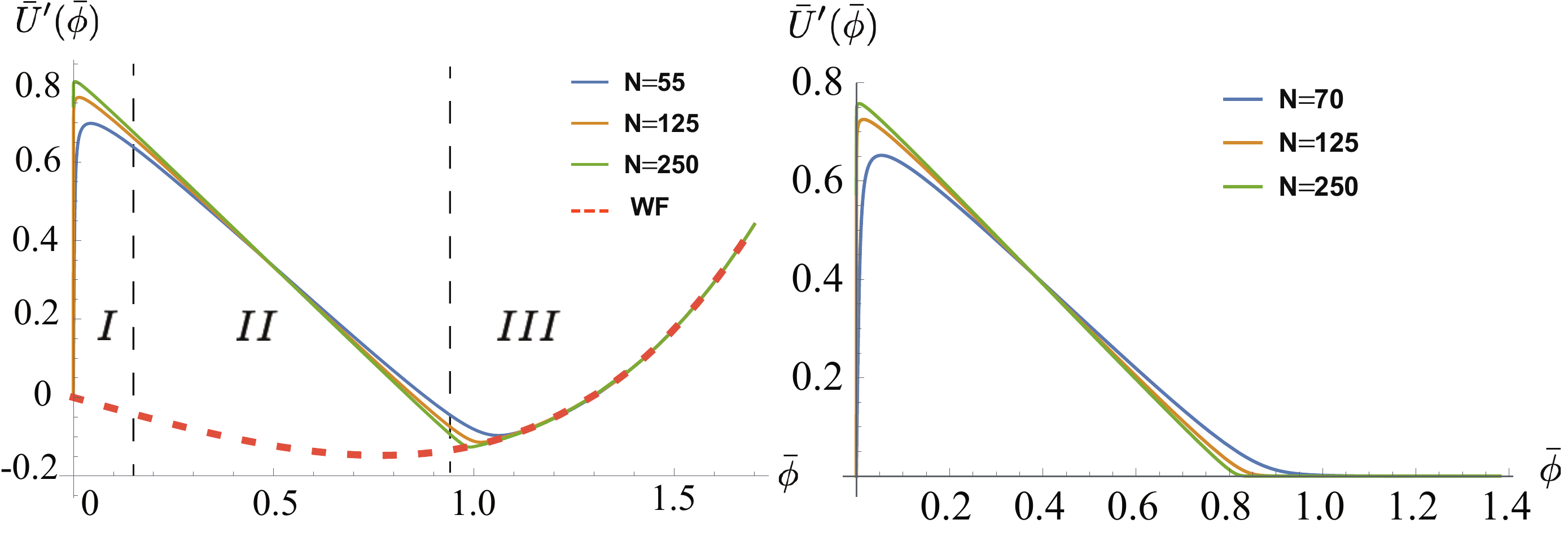}
 \caption{ $d=3.2$ and Ellwanger-Morris-Wetterich version of the RG. Left: $\bar{U}'(\bar{\phi})$ for  SWF$_{2}$  obtained by numerically integrating Eq.~(\ref{flow-LPA-dimensionless-rescaled}) for different values of $N$. The red dashed line corresponds to  the Wilson-Fisher FP for $N=\infty$. The regions I, II and III correspond to the regions with the same label in  Wilson-Polchinski in Fig. \ref{c2cusp}. Right: $\bar{U}'(\bar{\phi})$ for  SG$_{3}$ obtained by numerically integrating  Eq.~(\ref{flow-LPA-dimensionless-rescaled}) for different values of $N$.
 }
\label{C2FP}
\end{figure}

The translation from one version of the RG to the other can be made with the change of variables given in Eq.~(\ref{transformation-WP-Wetterich}). It is not completely straightforward and is best understood at finite  $N$ because it becomes singular at $N=\infty$. The LPA equation for the rescaled potential in Ellwanger-Morris-Wetterich version is given in Eq.~(\ref{flow-LPA-dimensionless-rescaled}).
We have numerically integrated the LPA FP equation following from Eq.~(\ref{flow-LPA-dimensionless-rescaled}) for several large values of $N$ and we show in Fig. \ref{C2FP} the derivative of the FP potential. 
The limiting shape at $N=\infty$ clearly shows up already at finite $N$.

Let us first analyze the shape of the  SWF$_2$ FP potential.

First, we, of course, retrieve that at large field the SWF$_2$ FP potential is (almost) identical to the WF potential for large values of $N$.

Second, the singularity that occurs at one isolated point $\bar\varrho_0$ in the W-P version now shows up on a finite interval in $\bar\phi$, shown as region II in Fig. \ref{C2FP}. We have indeed shown that the boundary layer in $\bar\varrho$ in W-P,  whose width goes to zero when $N\to\infty$, is mapped onto a finite interval of $\bar\phi$, even when $N=\infty$.  In this interval, $\bar{U}'(\bar\phi)$ has a slope that gets closer and closer to $-1$ as $N$ increases. This is what makes   the longitudinal propagator contribution -- the last term of the right hand side of Eq.~(\ref{flow-LPA-dimensionless-rescaled}) --   non negligible in the large $N$ limit. We have checked that this term diverges linearly in $N$ at large $N$ which makes it of the same order as the other terms in Eq.~(\ref{flow-LPA-dimensionless-rescaled}). 

Third, notice that at finite $N$ where $\bar{U}'(\bar\phi)$ is everywhere regular,  $\bar{U}'(0)=0$ because the O$(N)$ symmetry implies that $\bar{U}(\bar\phi)$ is a regular function of $\bar\phi^2$. However, the slope of $\bar{U}'(\bar\phi)$ at small field  and large $N$ is very large as can be seen on region I of Fig. \ref{C2FP}. 
We have \textcolor{black}{mentioned} that at $N\to\infty$, the region $0<\bar{\Phi}<\sqrt{2\bar{\varrho}_0}$ where $\bar V(\bar\varrho)=\bar\varrho$ in W-P version of the RG is mapped onto the point $\bar{\phi}=0^+$ in Ellwanger-Morris-Wetterich version.  In this limit, at  $\bar{\phi}=0^+$, $\bar{U}'(\bar{\phi})$ increases from $\bar{U}'(\bar{\phi}=0)=0$ up to a finite value  which can be obtained from the W-P solution and the change of variable: $\bar{\Phi}=\bar{\phi}+\bar{U}'(\bar{\phi})$. In $d=3.2$, we have found $\bar{U}'(\bar{\phi}=0^+)=\sqrt{2\bar{\varrho}_0}\simeq0.833$. This means that the slope of $\bar{U}'(\bar{\phi})$ at $\bar{\phi}=0$ goes to infinity when $N\to\infty$ and  $\bar{U}'(\bar{\phi}=0)$ is therefore undefined at $N=\infty$:  $\bar{U}'$ becomes discontinuous in this limit, that is, $\bar{U}'(0^+)=-\bar{U}'(0^-)\neq0$, and $\bar{U}(\bar{\phi})$ has a cusp at $\bar{\phi}=0$ at $N=\infty$, see Fig. \ref{C2FP}. The FP potential is therefore no longer a regular function of $\bar\phi^2$. We have numerically found the asymptotic behavior: $\bar{U}''(\bar{\phi}=0)\simeq 0.484 \exp(0.108 N)$ for large $N$ in $d=3.2$. 

At $N=\infty$, the extension of the region II where $\bar{U}'(\bar\phi)$ has a slope $-1$ can now be computed. In $d=3.2$ for instance, we draw a straight line of slope $-1$ starting at $(\bar\phi=0,\bar{U}'(0^+)= 0.833)$ and we find that it crosses the WF potential when $\bar\phi\simeq0.965$. At finite and large $N$, we have checked that indeed the interval $\bar\phi\in[0^+,0.965]$ is exactly mapped onto the (very narrow) boundary layer around $\bar\varrho_0$ in W-P. 

We conclude that the finite interval of $\bar\varrho$  where $\bar V(\bar\varrho)=\bar\varrho$ in the W-P version of the RG, that is, $0\le\bar\varrho\le\bar\varrho_0$, is mapped onto the point $\bar{\phi}=0^+$ in the Ellwanger-Morris-Wetterich version of the RG, region I of Fig. \ref{C2FP}, while the boundary layer in $\bar\varrho$ around $\bar\varrho_0$ in W-P is mapped onto the finite interval of $\bar\phi$ where \textcolor{black}{$\bar{U}''(\bar\phi)=-1$} in Ellwanger-Morris-Wetterich which is the region II of Fig. \ref{C2FP}. This is what explains why it would be much more difficult to study directly the singular FP potentials in the Ellwanger-Morris-Wetterich version of the RG: Instead of an isolated singularity at finite $\bar\varrho$ in W-P, there is a cusp at the origin followed by a finite interval  \textcolor{black}{$\bar{U}''(\bar\phi)=-1$}.

Let us now study the SG$_3$ FP. We show in Fig. \ref{C2FP} the derivative of its FP potential in Ellwanger-Morris-Wetterich parametrization  in $d=3.2$ for several large values of $N$. Here also, the limiting shape of the FP potential clearly shows up  when $N\to \infty$. At $N=\infty$, following the same arguments as above, we find that $\bar{U}'(\bar{\phi}=0)$ is undefined and that  $\bar{U}'(\bar{\phi}=0^+)=\sqrt{2/d}=0.791$. The slope of $\bar{U}'(\bar{\phi})$ is  $-1$ in the interval $[0^+,0.791]$. Then,  for $\bar{\phi}>0.791$, $\bar{U}'(\bar{\phi})=0$  which means that the SG$_3$ FP at $N= \infty$ is identical to the Gaussian FP at large field. 

To conclude, we can see from Fig. \ref{C2FP} that the {\it a priori} construction of SWF$_2$ and SG$_3$ is much less simple in the Ellwanger-Morris-Wetterich version of the RG than in the W-P one in particular because the contribution of the longitudinal propagator in Eq. \eqref{flow-LPA-dimensionless-rescaled-first} is singular on a finite field interval when $N\to\infty$, region II. 

\section{Flow diagram at $d=3$ and $N=\infty$}\label{sec:Phase diagram}

\begin{figure}[h]
\includegraphics[scale=0.5]{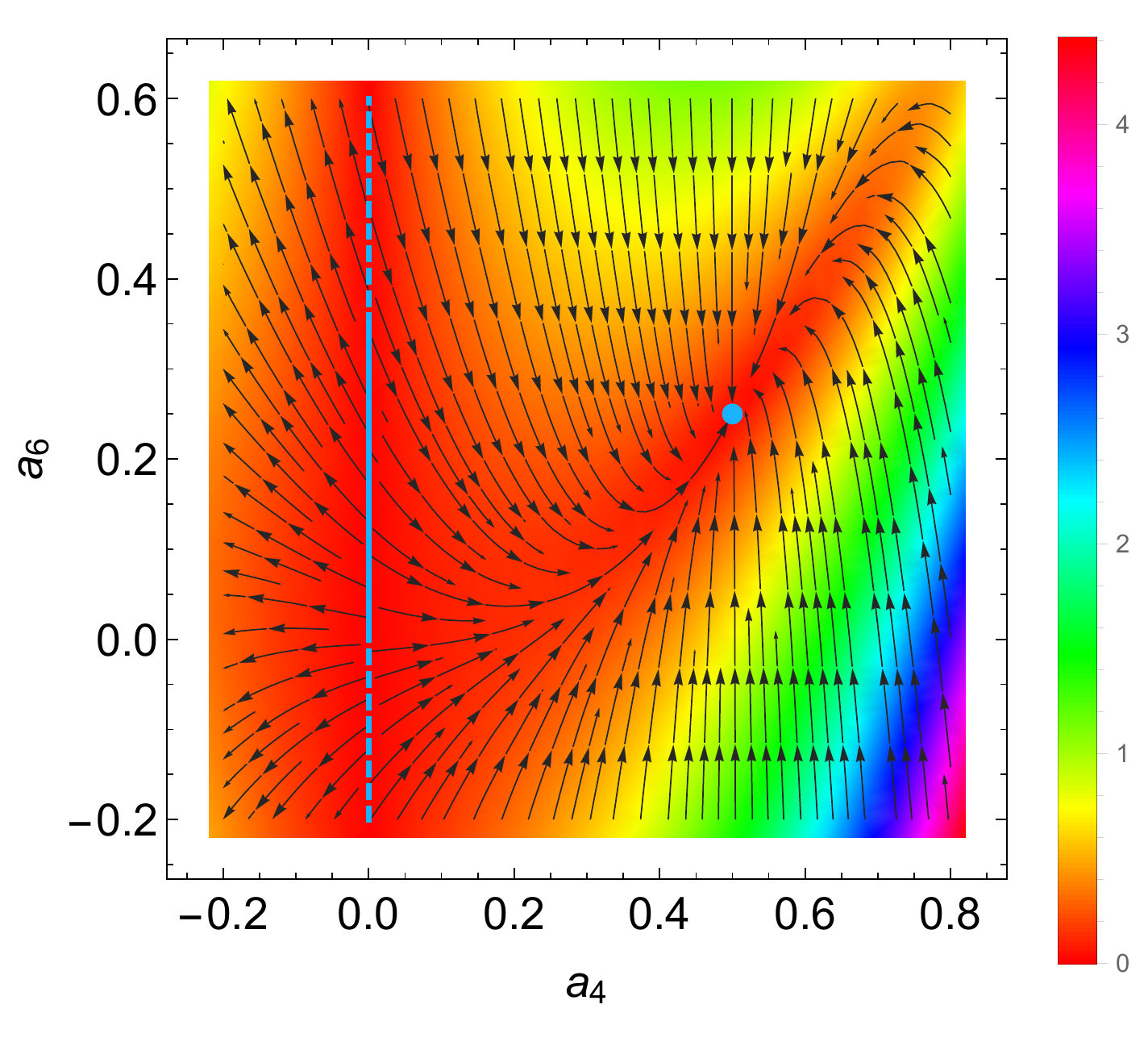}
\caption{Flow diagram at $\left(d=3,N=\infty\right)$
on the critical surface where $\text{\ensuremath{\kappa=1}}$. The light blue point represents the attractive infrared Wilson-Fisher fixed point. The light blue line represents the UV fixed points given by the BMB line $\left(a_{4}=0,0<a_{6}<a_{6}^{\rm{BMB}}\right)$. At the level of the expansion about the minimum at $\bar{\rho}=\kappa$ this line is infinite and given by the dashed line whereas it should be finite. The color code of the phase diagram represents the amplitude of the vector $\left(\beta_{a_{4}},\beta_{a_{6}}\right)$} 
\label{flow-diag}
\end{figure}

A simplified flow diagram can be obtained by expanding the effective potential $\bar{U}\left(\bar{\rho}\right)$
as $\bar{U}=\sum_{n}a_{2n}\left(\bar{\rho}-\kappa\right)^{n}/n!$ with $\kappa$ the (running) minimum of the effective potential. We then obtain the following system of flow equations,
valid at $N=\infty$ and $d=3$:
\begin{equation}
\begin{aligned}\frac{\textrm{d}}{\textrm{dt}}\kappa= & 1-\kappa\\
\frac{\textrm{d}}{\textrm{dt}}a_{4}= & -a_{4}\left(1-2a_{4}\right)\\
\frac{\textrm{d}}{\textrm{dt}}a_{6}= & -6a_{4}\left(a_{4}^{2}-a_{6}\right)\\
 & \cdots
\end{aligned}
\end{equation}
 from which can be computed the flow diagram and in particular the BMB line shown in Fig. \ref{flow-diag}.
 
\section{Analytical relation between $\alpha$ and $\tau$ on the BMB line }\label{sec:Analytical relation}

 We recall the LPA differential equation in the Wilson-Polchinski formulation:
\begin{equation}
1-d\bar{V}+\left(d-2\right)\,\bar{\varrho}\,\bar{V}'+2\,\bar{\varrho}\,\bar{V}'^{2}-\,\bar{V}'-\frac{2}{N}\,\bar{\varrho}\,\bar{V}''=0.\label{eq:V}
\end{equation}
Taking the derivative of this equation with respect to $\bar{\varrho}$ and writing $v=\bar{V}'$
and $1/N=\epsilon$ we obtain:
\begin{equation}
((d-2)\bar{\varrho}-2\epsilon-1)v'-2\bar{\varrho}\epsilon v''+2v\left(2\bar{\varrho} v'+v-1\right)=0.\label{eq:veq}
\end{equation}
We now present two methods to obtain the relation  between $\alpha$ and $\tau$ given in Eq. (\ref{alpha-tau}) of the main text. 
The first method is straightforward and requires expanding the potential in powers of $\epsilon$ and $\bar{\varrho}-1$ \footnote{One may also expand around the finite $N$ minimum $\bar{\varrho}_m$ of $v$. However, this point being a function of $\epsilon$, the expansion turns out to be in powers of $\epsilon^{1/2}$ which makes the calculations a little more difficult.}. Notice that a drawback of this method is that the expansion in $\bar{\varrho}$ hides the fact that the potential can be nonanalytic for another value of the field. This is the reason why we use in our second method a fully functional approach. This method has the advantage of yielding the potential at finite and large $N$ up to $1/N^2$ corrections and is therefore useful to get the behavior near $\bar{\varrho}=0$ where a divergence appears at $N=\infty$ and $\tau=\tau_{BMB}$.

The first method consists in Taylor expanding Eq.~(\ref{eq:veq}) about $\bar{\varrho}=1$ (the inflexion point)  with $v=\sum a_k (\bar{\varrho}-1)^k$. We moreover expand the couplings $a_k$ in powers of $\epsilon$ as $a_k=a_k^0+\epsilon a_k^1+O(\epsilon^2)$ where the $a_k^0$'s are the couplings at $N=\infty$ by Eq.~(9). The system of equations obtained by independently setting equal to 0  the coefficients of $\epsilon^n(\bar{\varrho}-1)^p$ yields the relation between $\alpha$ and $\tau$ given in Eq.~(\ref{alpha-tau}). Notice that in this method the $\tau$ dependence comes from the $a_k^0$'s.

The functional method consists in expanding $v$ as $v=v_{0}+\epsilon v_{1}+O(\epsilon^2)$ in Eq. (\ref{eq:veq}). At order $\epsilon$, this yields a differential equation on $v_1$ that depends on $\bar{\varrho}$, $v_0$, $v_0'$ and $v_0''$. Using Eq. (\ref{eq:veq}) and its derivative both evaluated at $\epsilon=0$, $v_0''$ and $v_0'$ can be eliminated in terms of $v_0$. This leads to:

\begin{equation}
\begin{aligned}32\bar{\varrho}^{2} v_0^{4}\left(\alpha\bar{\varrho}+4\bar{\varrho}\left(2\bar{\varrho}  v_1'+  v_1\right)-1\right)+16\bar{\varrho}^{2} v_0^{3}\left(-\alpha(\bar{\varrho}+1)+8(\bar{\varrho}-1)\left(2\bar{\varrho}  v_1'+  v_1\right)+1\right)+\\
2 v_0^{2}\left(\bar{\varrho}\left(-7\alpha\bar{\varrho}^{2}+6(\alpha+2)\bar{\varrho}+\alpha-6\right)+4(\bar{\varrho}-1)\bar{\varrho}\left(12(\bar{\varrho}-1)\bar{\varrho}  v_1'+(\bar{\varrho}-5)  v_1\right)+2\right)\\
2(\bar{\varrho}-1) v_0\left(\bar{\varrho}(\alpha(-\bar{\varrho})+\alpha-4)+8\bar{\varrho}(\bar{\varrho}-1)^{2}  v_1'+\left(-6\bar{\varrho}^{2}+4\bar{\varrho}+2\right)  v_1+2\right)+(\bar{\varrho}-1)^{3}\left((\bar{\varrho}-1)  v_1'-2  v_1\right)=0.
\end{aligned}\label{v1dif}
\end{equation}
We then assume that $v_1$ is analytic at $h=\bar{\varrho}-1=0$. The Taylor expansion of $v_0$
at $h=0$ is:

\begin{equation}
\begin{aligned}v_0= \frac{h^2 \tau }{2}-2 h^3 \tau ^2+\frac{5}{4} h^4 \tau ^2 (8 \tau -1)+h^5 (13-56 \tau ) \tau ^3+\\
\frac{7}{8} h^6 \tau ^3 (128 \tau  (3 \tau -1)+5)+h^7 \left(-2112 \tau ^6+912 \tau ^5-\frac{383 \tau ^4}{5}\right)+O\left(h^8\right).
\end{aligned}
\label{v0}
\end{equation}
Inserting Eq. (\ref{v0}) into Eq. (\ref{v1dif}), neglecting terms of order 5,  and dividing by $h^{3}$ 
yields:

\begin{equation}
-\tau h^2(\alpha +6 \tau  (12 \tau -5))+\left(8 h^2 \tau +h\right) v_1'+\left(6 h^2 \tau  (4 \tau -1)-8 h \tau -2\right) v_1-h \tau  (\alpha -12 \tau +4)-2 \tau=0.
\label{eqv1}
\end{equation}
Finally, replacing $v_1$  in Eq. (\ref{eqv1}) by its Taylor expansion: $v_1=v_1^{(0)}+v_1^{(1)}\,h+v_1^{(2)}\,h^{2}+O(h)^{3}$  leads to:

\begin{subequations}
\begin{align}v_1^{(0)} &=-\tau \label{systa}\\
v_1^{(1)} &=\tau  (-\alpha +20 \tau -4)\label{systb}\\
\alpha &= 36 \tau -96 \tau ^2.\label{systc}
\end{align}
\end{subequations}
Notice that it is because the $v_1^{(2)}$ term cancels in Eq. (\ref{systc}) that we can obtain a relation between $\alpha$ and $\tau$ only.

Let us finally notice that Eq. (\ref{systc}) can be retrieved in a functional way. The solution of Eq. (\ref{v1dif}) is:
\begin{equation}
\begin{aligned}v_1\left(\bar{\varrho}\right)=\exp\left(K\left(\bar{\varrho}\right)\right)\,\left(C-\int_{1}^{\bar{\varrho}}2e^{-K(\chi)}\left(\frac{16\alpha\chi^{3} v_0(\chi)^{4}-8\alpha\chi^{3} v_0(\chi)^{3}-7\alpha\chi^{3} v_0(\chi)^{2}-\alpha\chi^{3} v_0(\chi)-8\alpha\chi^{2} v_0(\chi)^{3}}{(4\chi v_0(\chi)+\chi-1)^{4}}\right.+\right.\\
\frac{6\alpha\chi^{2} v_0(\chi)^{2}+2\alpha\chi^{2} v_0(\chi)+\alpha\chi v_0(\chi)^{2}-\alpha\chi v_0(\chi)-16\chi^{2} v_0(\chi)^{4}}{(4\chi v_0(\chi)+\chi-1)^{4}}+\\
\left.\frac{\left.8\chi^{2} v_0(\chi)^{3}+12\chi^{2} v_0(\chi)^{2}-4\chi^{2} v_0(\chi)-6\chi v_0(\chi)^{2}+2 v_0(\chi)^{2}+6\chi v_0(\chi)-2 v_0(\chi)\right)}{(4\chi v_0(\chi)+\chi-1)^{4}}\,\right)d\chi
\end{aligned}\label{v1sol}
\end{equation}
where $C$ is an integration constant and
\begin{equation}
\begin{aligned}K\left(\bar{\varrho}\right)=\int_{1}^{\bar{\varrho}}-2\left(\frac{64\chi^{3} v_0(\chi)^{4}+64\chi^{3} v_0(\chi)^{3}+4\chi^{3} v_0(\chi)^{2}-6\chi^{3} v_0(\chi)-64\chi^{2} v_0(\chi)^{3}-24\chi^{2} v_0(\chi)^{2}}{(4\chi v_0(\chi)+\chi-1)^{4}}+\right.\\
\left.\frac{10\chi^{2} v_0(\chi)+20\chi v_0(\chi)^{2}-2\chi v_0(\chi)-2 v_0(\chi)-\chi^{3}+3\chi^{2}-3\chi+1}{(4\chi v_0(\chi)+\chi-1)^{4}}\,\right)d\chi.
\end{aligned}\label{K}
\end{equation}
Replacing $v_0$ in Eq. (\ref{v1sol})  by its Taylor expansion  (\ref{v0}) yields:

\begin{equation}
\begin{aligned}
v_1\left(\bar{\varrho}\right)=-\tau -h \tau  (\alpha -20 \tau +4)+h^2 \left(\tau ^2 (8 \alpha -156 \tau +37)+\tau  (\alpha +12 \tau  (8 \tau -3)) \log (h)+C \right)+O\left(h^3\log(h)\right).
\end{aligned}\label{log-term}
\end{equation}
The analyticity of $v_1$ implies that the  log term in Eq. (\ref{log-term}) is absent. This requires that its prefactor vanishes, that is, $\alpha= 36 \tau -96 \tau^2$ which is the same as Eq. (\ref{alpha-tau}) of the main text. To all orders checked (up
to 5th order) this also eliminates the following log terms. 

Notice that the expression (\ref{v1sol}) giving $v_1(\bar\varrho)$ is ill-conditioned for a numerical plot of this function because of the poles of the integrands of  $K$ in Eq.~(\ref{K}) and in $v_1$ in Eq.~(\ref{v1sol}). Although the final expression for $v_1$ is well-defined it is tricky to get rid of apparent divergencies showing up because of the poles within the integrands: This requires adding and subtracting divergencies and making some integration by parts. For this reason, it is simpler to numerically integrate Eq.~(\ref{v1dif}).

\section{Toy model of the boundary layer found in Section (\ref{BL})}
\label{sec:Toyboundarylayer}

\begin{figure}[h]
\begin{centering}
\includegraphics[width=7cm]{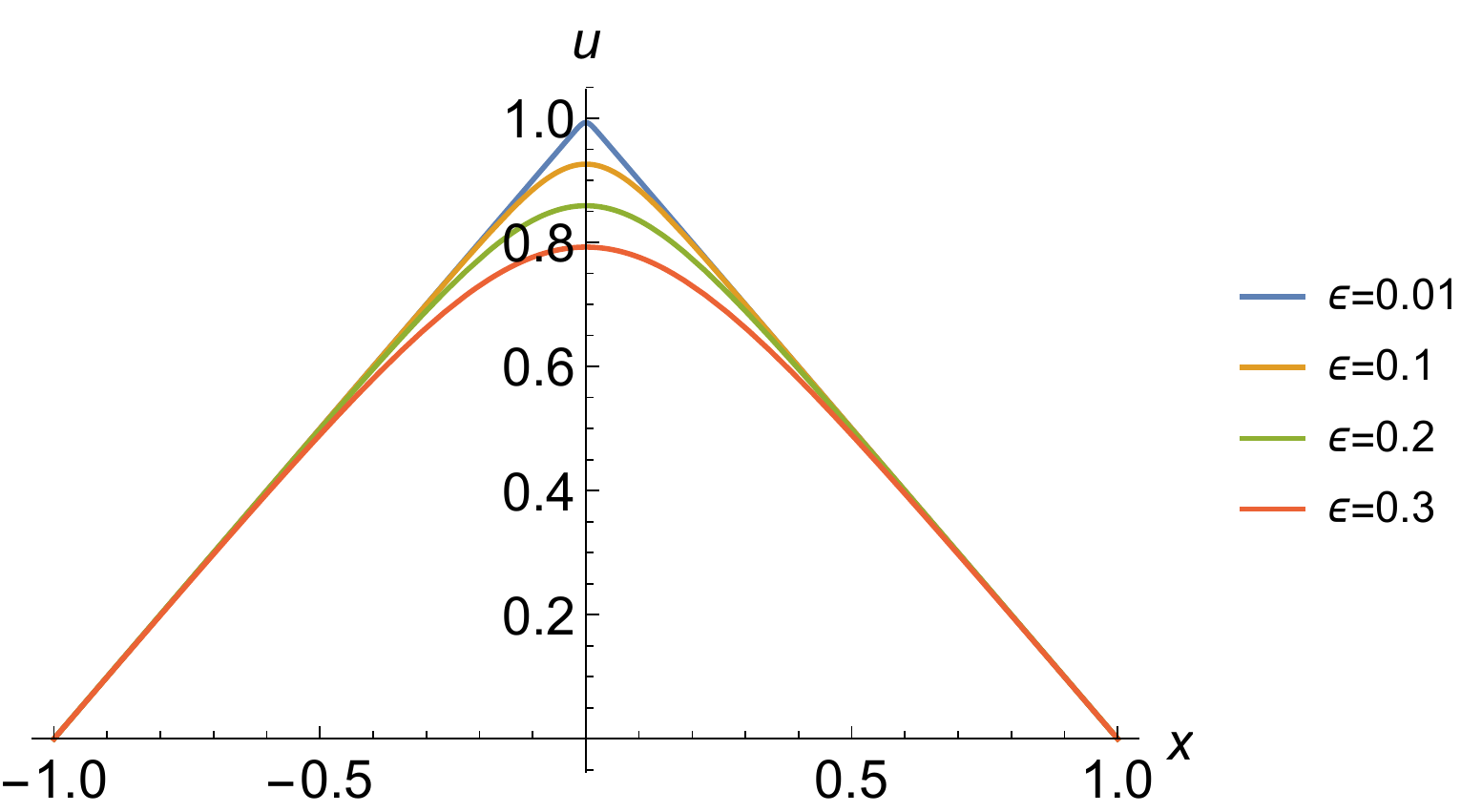}
\par\end{centering}
\caption{\label{fig:Boundary-layer-formation}Boundary layer formation of Eq. (\ref{eq:simple sing dif-1}). }
\end{figure}

Consider the following simple differential
equation that we now study as a toy model of the FP equation Eq.  (\ref{flow-LPA-WP-essai}):
\begin{subequations}
\begin{align}
&1-u'^{2}+\epsilon u''=0,\label{eq:simple sing dif-1}\\ 
&u\left(1\right)=u\left(-1\right)=0
\label{eq:simple sing dif-2}
\end{align}
\end{subequations}
where $\epsilon\ge0$ plays the role of $1/N$. 
The limit $\epsilon\rightarrow0^+$ is singular in the
sense that the very nature of the differential equation changes from
second to first order when $\epsilon$ is strictly vanishing. The exact
solution of Eqs. (\ref{eq:simple sing dif-1},\ref{eq:simple sing dif-2}) reads: 
\begin{equation}
u\left(x\right)=\epsilon\left(\text{\ensuremath{\log}}\left(\cosh\left(\frac{1}{\epsilon}\right)\right)-\text{\ensuremath{\log}}\left(\cosh\left(\frac{x}{\epsilon}\right)\right)\right)
\end{equation}
and is shown in Fig. \ref{fig:Boundary-layer-formation}. It will be convenient in the following to use the relation:
\begin{equation}
\epsilon\log\left(\cosh\left(\frac{x}{\epsilon}\right)\right)=  \epsilon\log\left(e^{\frac{x-|x|}{\epsilon}}+e^{-\frac{x+|x|}{\epsilon}}\right)+|x|-\epsilon\log\left(2\right).
\end{equation}
Thus, at fixed $x>0$ we find in the limit $\epsilon\to 0^+$:
\begin{equation}
u(x>0) \underset{\epsilon \to 0^+}{\sim} 1-x
\label{solx>0}
\end{equation}
and for $x<0$:
\begin{equation}
u(x<0) \underset{\epsilon \to 0^+}{\sim} 1+x
\label{solx<0}
\end{equation}
that are both solutions of Eq. (\ref{eq:simple sing dif-1}) with $\epsilon=0$. As for the boundary conditions, they are of course not satisfied simultaneously by the solutions above but $u(1)=0$ is satisfied by Eq. (\ref{solx>0}) and $u(-1)=0$ by Eq.  (\ref{solx<0}).

At small but finite $\epsilon$, there are two distinct regions of $x$ depending on the magnitude of $\vert x\vert/\epsilon$. For $|x|\gg\epsilon$:
\begin{equation}
    u(x)\sim C-\vert x\vert -\epsilon e^{-2\vert x\vert/\epsilon} 
\end{equation}
where $C=1+ O(\epsilon)$ is independent of $x$. Thus, for $\vert x\vert\gg\epsilon$, ${u'}^2\sim 1$ up to an exponentially small term in $\vert x\vert/\epsilon$ which means that $\epsilon u''(x)$ is exponentially small. Reciprocally, for $1\gg\epsilon\ge \vert x\vert$, $\epsilon u''(x)\sim O(1)$ and it cannot be neglected.

As a conclusion, with  equations as simple as (\ref{eq:simple sing dif-1}), it is relatively easy to build continuous solutions in the limit $\epsilon\to0$ by concatenating two  solutions obtained at $\epsilon=0$ at the price of having a corner (called cusp in our context). The very same mechanism is at work for the FP equation in W-P version of the RG where the problematic term is $\frac{2}{N}\bar{\varrho}\,\bar{V}''=O\left(1\right)$. Notice that this analysis is not as simple in the Ellwanger-Morris-Wetterich version.
\textcolor{black}{
\section{Details on the numerical methods}
\subsection{Shooting from vanishing fields and SpikePlot}
 Using the Ellwanger-Morris-Wetterich version of the FP equation written in terms of $\bar{\phi}$, we use the method of shooting from vanishing fields with initial conditions $\left(\bar{U}'(\bar{\phi}_{\textrm{min}})=s\bar{\phi}_{\textrm{min}}, \bar{U}''(\bar{\phi}_{\textrm{min}})=s\right) $ where  $\bar{\phi}_{\textrm{min}}$ is chosen to be sufficiently small that the Taylor expansion of $\bar{U}'$ about $\bar{\phi}=0$ can be stopped at linear order. Notice that we used $\bar{U}'(0)=0$ in the Taylor expansion which relies on the symmetry $\bar{U}(\bar{\phi})=\bar{U}(-\bar{\phi})$ of the potential and differentiability about $\bar{\phi}=0$. For each value of $s$, the LPA FP equation is integrated using Mathematica's NDSolve. Generically, starting at $\bar{\phi}_{\textrm{min}}$, the NDSolve program is unable to integrate the differential equation past $\bar{\phi}_{\textrm{max}}(s)$ which is the field value at which the potential blows up \cite{defenu2020, Hellwig, Morris_Derivative, Morris_truncations}. For generic $(d,N)$ there is a finite number of analytical FP solutions which then correspond to a finite number of values $s_l$ where $\bar{\phi}_{\textrm{max}}(s_l)=\infty$. These values $s_l$ appear as jumps or spikes in the numerical curve $\bar{\phi}_{\textrm{max}}(s)$ \cite{defenu2020, Hellwig, Morris_Derivative, Morris_truncations}. The critical lines where FPs collapse then correspond to the points $(d,N)$ where two spikes of the curve $\bar{\phi}_{\textrm{max}}(s)$ collide.
 }\textcolor{black}{
 \subsection{Finite difference algebraic method}
The second method consists in discretizing the differential equation using 5-point stencil finite differences. For this method we found it easier to use the Polchinski equation for $\bar{V}$.}

\textcolor{black}{
This finite difference approximation replaces the non linear LPA fixed-point differential equation by a system of non linear algebraic equations. The grid of points is taken within an interval $[0,\bar{\varrho}_{\textrm{max}}]$. The precision of the method then depends on $\bar{\varrho}_\textrm{max}$ and the density of points. Typically, we took a grid of 3000 points and chose the smallest $\bar{\varrho}_\textrm{max}$ such that the results would be independent of $\bar{\varrho}_\textrm{max}$.}

\textcolor{black}{The algebraic system is then solved using Mathematica's FindRoot where the initial conditions are chosen by continuity with respect to previous computations. More precisely, the FP solutions are followed by continuity taking sufficiently small steps in $d$ and $N$ and the initial conditions are updated at each step using the solution of the previous step. The very first initial condition can be obtained either starting from near the upper critical dimension and taking a small perturbation of the Gaussian FP or by using the first method above and progressively increasing the value of $\bar{\varrho}_\textrm{max}$ where at each step we use as initial condition the solution at a smaller value of $\bar{\varrho}_\textrm{max}$. }

\textcolor{black}{To obtain the eigenvalues and eigenfunctions we linearize the discretized system about the fixed point solution thereby leading to an ordinary matrix eigenvalue problem which we solve using Mathematica's Eigensystem.}

\textcolor{black}{As this method selects global solutions where the finite differences are sufficiently well bounded for all $\varrho$ such that the FindRoot method may converge, the solutions and eigenfunctions are indeed discretizations of analytical functions.}

 \section{Eigenfunctions of singular fixed points}
 \label{sec: eigenfunctions-sing-FP}
 In this section we explain why the singular counterpart of a regular FP has one extra relevant infrared eigendirection. 
 \\To compute eigenvalues one has to first find a solution to the FP
equation. Unfortunately, in the case of the singular FPs, we have not found an analytical approximation which works both on the inside and on the exterior of the boundary layer. The best global approximation we could find was to extend the boundary layer analysis of the SG$_3$ FP to all field values. This is a priori justified as for $\bar{\varrho}>\bar{\varrho}_0$, with $\bar{\varrho}_0$ the position of the cusp, the FP potential is a trivial constant. Yet, a comparison of this approximation with numerical solutions seems to indicate that this approximation does not yield a quantitative description of the finite $N$ corrections outside of the boundary layer. Nevertheless, it gives a good qualitative idea of the exponential corrections outside of the boundary layer and does converge to the infinite $N$ solution for all field values. Thus, purely for qualitative and pedagogical reasons, we shall first consider the eigenperturbations obtained by replacing the full SG$_3$ FP with its boundary layer approximation. The following discussion will then first focus on this FP but the paragraphs that follow will consider all singular FPs using general arguments.

We first display the approximation to the FP potential of SG$_3$ in $d=3$ :
\begin{equation}
\bar{V}^{*}\left(\bar{\varrho},\epsilon\right)=\frac{1}{2}\left(-2\epsilon\log\left(\text{sech}\left(\frac{1}{3\epsilon}\right)\cosh\left(\frac{1-3\text{\ensuremath{\bar{\varrho}}}}{6\epsilon}\right)\right)+\bar{\varrho}-1\right)+\frac{1}{3}
\label{eq:SG}
\end{equation}
where $\epsilon=1/N$. To find the eigenvalues associated to this
FP potential we consider a perturbation of this solution as $V=\bar{V}^{*}+\delta V\left(\bar{\varrho}\right)e^{\lambda t}$
which we insert into the time dependent LPA equation and linearize. This manipulation leads to:
\begin{equation}
-2\bar{\varrho}\epsilon\text{\ensuremath{\delta}V}''+2\bar{\varrho}\text{\ensuremath{\delta}V}'\left(\tanh\left(\frac{1-3\bar{\varrho}}{6\epsilon}\right)+1\right)+\bar{\varrho}\delta\text{V}'-\text{\ensuremath{\delta}V}'-\left(3+\lambda\right)\text{\ensuremath{\delta}V}=0.\label{eq:eigen diff equation}
\end{equation}
\begin{figure}
\begin{centering}
\includegraphics[width=12cm]{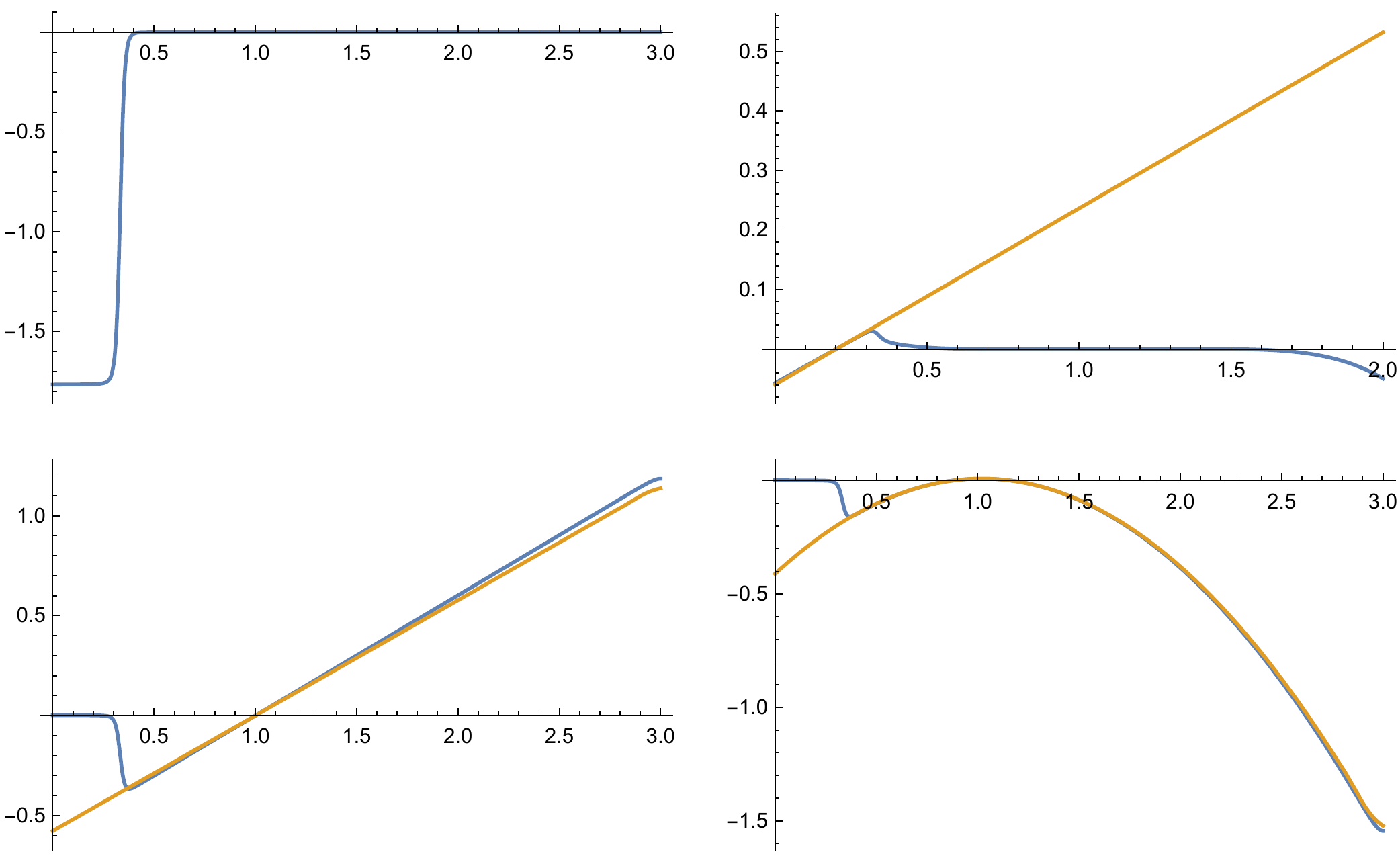}
\par\end{centering}
\caption{\label{fig:Eigenfunctions-of-the}Approximate eigenfunctions of the SG$_3$ FP at $\left(d=3,N=100\right)$.
Top: Eigenfunctions that become flat above the boundary layer. The
yellow curve on the top right corresponds to an eigenfunction of the
potential $\bar{V}=\bar{\varrho}$ which then shows that it is nearly
identical to the eigenfunction of SG$_3$ below the boundary layer.
Bottom: Eigenfunctions that become flat below the boundary layer.
The yellow curves correspond to eigenfunctions of the Gaussian potential
showing that it is nearly identical to the eigenfunction of SG
above the boundary layer.}
\end{figure}
Solving this equation numerically we find
Fig. \ref{fig:Eigenfunctions-of-the}. The upper left eigenfunction can be found as a tanh function to leading order in $1/N$ if we consider a boundary layer analysis of Eq.  (\ref{eq:eigen diff equation}) and we impose that the eigenfunction remains bounded. The plots of Fig. \ref{fig:Eigenfunctions-of-the} are similar to those obtained from the stability matrix using the full numerical SG$_3$ potential, solution of the LPA fixed point equation of the main text, rather than the approximation we took here. 

Notice that the eigenfunctions also exhibit boundary layers and that in the limit $\epsilon\rightarrow0$ they can be understood as piecewise eigenfunctions where the non trivial part is an eigenfunction of the regular counterpart of the potential, here the constant potential of the Gaussian FP, or of the high temperature FP $\bar{V}^*=\bar{\varrho}$. In the following paragraphs we detail more precisely this construction and explain why it occurs. This will allow us to understand why the singular counterpart of a regular potential has one extra relevant eigenvalue.

\begin{figure}
\begin{centering}
\includegraphics[width=18.5 cm]{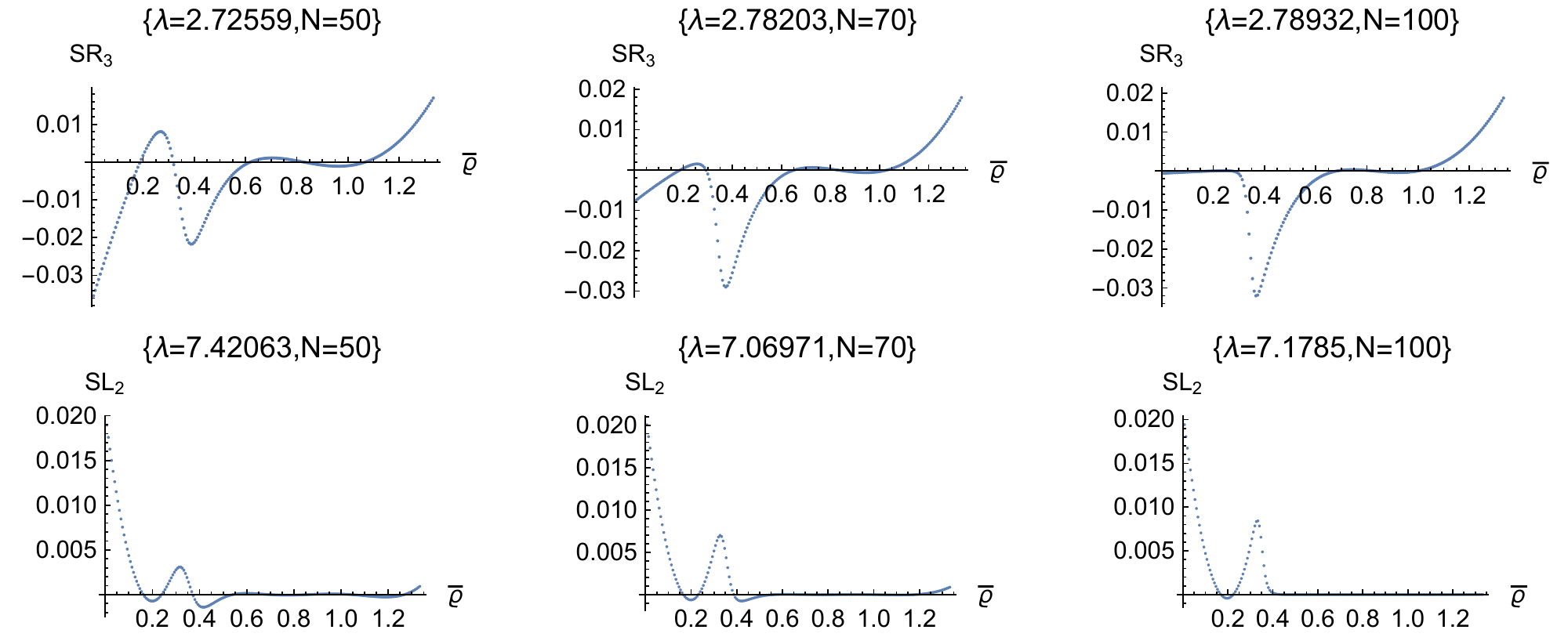}
\par\end{centering}
\caption{\label{fig:Eigenfunctions-at-} Plots of the singular eigenfunctions of the FP SWF$_2$ as a function of $\bar{\varrho}$
for $d=3.2$ and increasing $N$. We exhibit specifically the finite $N$ extension of the irrelevant eigenfunctions $SR_3$ and $SL_2$ defined in Eq. (\ref{SL}) and their respective eigenvalues $\lambda$. The indices correspond to the eigenvalues $\lambda_n=(d+2) n-d$ ( resp. $\lambda_m=2 m-d$) of the linear (resp. Wilson-Fisher) FP in the limit $N\rightarrow \infty$. The $SL_n$ (resp. $SR_m$) eigenfunctions become increasingly flat towards the right (resp. left) as $N$ increases while converging to a non trivial function with parity $(-1)^n$ (resp. $(-1)^m$) centered about $1/(d+2)\simeq 0.19$ (resp. $1/(d-2) \simeq 0.83$) where it vanishes as expected from the $SL$, $SR$ construction at $N\rightarrow\infty$. The eigenvalues also converge to that of $L_2$ (resp. $R_3$), that is $\lambda=7.2$ (resp. $\lambda=2.8$)}
\end{figure}

Consider first the eigenfunctions from the linear part of the singular FP for $\bar{\rho}<\bar{\varrho}_{0}$ where $\bar{\varrho}_{0}$ is the position of the cusp in the limit $N\rightarrow\infty$
which we denote as $L_{i}\left(\bar{\varrho}\right)$.
Consider also the regular part on the right, which we denote $R_{j}\left(\bar{\varrho}\right)$
for $\text{\ensuremath{\bar{\varrho}>\bar{\varrho}_{0}}}$. Then in the limit $N\rightarrow\infty$, the eigenfunctions of the FP SG$_3$ exhibit the
following construction:
\begin{equation}
\begin{aligned}SL_{m}\left(\bar{\varrho}\right)= & \begin{cases}
L_{m}\left(\bar{\varrho}\right) & \textrm{for \ensuremath{\bar{\varrho}<}\ensuremath{\bar{\varrho}_{0}}}\\
0 & \textrm{for \ensuremath{\bar{\varrho}>}\ensuremath{\bar{\varrho}_{0}}}
\end{cases}\\
SR_{m}\left(\bar{\varrho}\right)= & \begin{cases}
0 & \textrm{for \ensuremath{\bar{\varrho}<}\ensuremath{\bar{\varrho}_{0}}}\\
R_{m}\left(\bar{\varrho}\right) & \textrm{for \ensuremath{\bar{\varrho}>}\ensuremath{\bar{\varrho}_{0}}}
\end{cases}
\end{aligned}
.\label{SL}
\end{equation}
This construction is found again in Fig. \ref{fig:Eigenfunctions-at-} for the FP SWF$_2$ and similar constructions have been found for the other singular FPs. We now explain why this occurs.

First let us recall the Polchinski equation: 
\begin{equation}
\partial_{t}\bar{V}=1-d\,\bar{V}+(d-2)\bar{\varrho}\bar{V}'+2\bar{\varrho}\bar{V}'{}^{2}-\bar{V}'-\frac{2}{N}\bar\varrho\,\bar{V}''\label{eq:Polchinsk sing sectioni}.
\end{equation}
A FP solution verifies:
\begin{equation}
0=1-d\,\bar{V}^{*}+(d-2)\bar{\varrho}\bar{V}'^{*}+2\bar{\varrho}\bar{V}'^{*}{}^{2}-\bar{V}'^{*}-\frac{2}{N}\bar\varrho\,\bar{V}''^{*}.
\end{equation}
If we then insert $\bar{V}\left(d,N,\bar{\varrho},t\right)=\bar{V}^{*}\left(d,N,\bar{\varrho}\right)+\delta V\left(d,N,\bar{\varrho}\right)e^{\lambda(d,N) t}$
into Eq. (\ref{eq:Polchinsk sing sectioni}), where $\delta V$ is
a small perturbation, linearization leads to:
\begin{equation}
0=-\left(\lambda(d,N)+d\right)\,\delta V(d,N,\bar{\varrho})+(d-2)\bar{\varrho}\partial_{\bar{\varrho}}\delta V(d,N,\bar{\varrho})+4\bar{\varrho}\bar{V}'^{*}(d,N,\bar{\varrho})\partial_{\bar{\varrho}}\delta V(d,N,\bar{\varrho})-\partial_{\bar{\varrho}}\delta V(d,N,\bar{\varrho})-\frac{2}{N}\bar\varrho\,\partial_{\bar{\varrho},\bar{\varrho}}\delta V(d,N,\bar{\varrho}).\label{eq:eigenequation}
\end{equation}
For finite $N$, $\bar{V}^{*}(d,N,\bar{\varrho})$ and $\delta V(d,N,\bar{\varrho})$ are infinitely differentiable and $\lambda(d,N)$ is perfectly well defined and finite. The objective in the following will be to determine an approximation of $\lambda(d,N)$ for $N$ large based on its limit value $\lambda(d,\infty)$. Here we take the hypothesis that the $N$ dependence of $\lambda$ is sufficiently regular in $N$ that this limit exists. This is primarily motivated by numerical experiments which show convergence to finite values. In the following we will consider the existence of $\lambda(d,\infty)$ as a given and we will deduce the possible values it may take based on the structure of the singular potential $\bar{V}^{*}(d,\infty,\bar{\varrho})$.

\subsection{Spectrum of singular fixed points in the limit $N\rightarrow\infty$}

Consider field values $\bar{\varrho}>\bar{\varrho}_0$ above the position of the cusp. Taking the limit $N\rightarrow\infty$, $\bar{V}^{*}(d,N,\bar{\varrho})$ reduces to a regular potential and the eigenequation is identical to that of the usual regular potentials found in the limit $N\rightarrow\infty$. There is however one difference which is the restriction of the domain of study to $\bar{\varrho}>\bar{\varrho}_0$. In the following we will see that this restriction is irrelevant. 

In order to compute the value of $\lambda(d,\infty)$ there are two possibilities to consider. The first is that simply $\delta V(d,\infty,\bar{\varrho}>\bar{\varrho}_0)=0$ in which case the eigenequation is solved for any $\lambda(d,\infty)$ and there is thus no constraint on $\lambda(d,\infty)$ from the domain $\bar{\varrho}>\bar{\varrho}_0$. Usually one does not consider such a trivial case as there is often an implicit analyticity hypothesis on both the FP potential and the eigenfunctions which then implies that $\delta V(d,\infty,\bar{\varrho})$ is simply the null function. However, as the FP potential has a discontinuity, $\delta V(d,\infty,\bar{\varrho})$ can be non zero for $\bar{\varrho}<\bar{\varrho}_0$ as is the case of the large $N$ limit of the plots in Fig. \ref{fig:Eigenfunctions-of-the}. Whether or not the notion of a discontinuous eigenfunction is mathematically well defined is irrelevant for the present discussion as we are mainly interested in the large $N$ behavior of the eigenfunctions and eigenvalues where they are well defined at finite $N$. Discontinuous eigenfunctions at $N\rightarrow\infty$ can then be simply regarded as a mathematical abstraction of the functions themselves rather than their meaning in terms of eigenfunctions of an eigenequation. This is particularly true as our main goal is ultimately to obtain the leading order approximation of $\lambda(d,N)$ at large $N$ rather than the eigenfunctions themselves which are simply intermediate steps in our analysis.

The second possibility to consider is that $\delta V(d,\infty,\bar{\varrho}>\bar{\varrho}_0)$ is non zero and then the eigenvalues can be computed in the neighborhood of $\bar{\varrho}=1/(d-2)>\bar{\varrho}_0$ either from a Taylor expansion as was done for the flow diagram in Appendix \ref{sec:Phase diagram} or imposing analyticity in the neighborhood of $\bar{\varrho}=1/(d-2)$  as was done in \cite{Mati2017,Litim2018}. As $\bar{V}^{*}(d,\infty,\bar{\varrho}>\bar{\varrho}_0)$ is one of the usual regular potentials found in the limit $N\rightarrow\infty$, the eigenequation is the same as for those FP potentials and the eigenvalues found are also the same. Thus, for non zero $\delta V(d,\infty,\bar{\varrho}>\bar{\varrho}_0)$, $\lambda(d,\infty)$ belongs to the spectrum of $ \{ R_m(d,\bar{\varrho}), \, m \ge 0 \}\underset{\textrm{def}}{=}\{ R_m(d,\bar{\varrho})\}_m$. The eigenfunctions of the singular potential in the domain $\bar{\varrho}>\bar{\varrho}_0$ are also the same as for their regular counterpart in the neighborhood of $\bar{\varrho}=1/(d-2)$ as is clear by simply performing a Taylor expansion about this point. In this case the eigenfunction of the singular potential is the same as for the regular counterpart at least up to the radius of convergence of the Taylor expansion. As the limit $\bar{\varrho}\rightarrow\infty$ is included in the domain $\bar{\varrho}>\bar{\varrho}_0$, the asympotics at large fields should also be the same for the regular and singular FP potential. Thus, $\lambda$ being determined and the boundary conditions at $\bar{\varrho}=1/(d-2)$ and $\bar{\varrho}=\infty$ known, then via the uniqueness of the solution of the eigenequation given the eigenvalue and the boundary conditions, the eigenfunction of the singular potential in the domain $\bar{\varrho}>\bar{\varrho}_0$ is the same as the eigenfunction of its regular counterpart for that given eigenvalue. Thus the restriction of the interval of study to the domain $\bar{\varrho}>\bar{\varrho}_0$ is essentially irrelevant for the computation of the spectrum of the singular potential in the limit $N\rightarrow\infty$ when $\delta V(d,\infty,\bar{\varrho}>\bar{\varrho}_0)$ is non zero.

The situation is similar if we consider $\bar{\varrho}<\bar{\varrho}_0$ and then take the limit $N\rightarrow\infty$ where we have $\bar{V}^{*}(d,\infty,\bar{\varrho}<\bar{\varrho}_0)=\bar{\varrho}$. The first case where $\delta V(d,\infty,\bar{\varrho}>\bar{\varrho}_0)=0$ leads to no constraint on $\lambda(d,\infty)$. In the second case where the perturbation is non zero and the FP potential is $\bar{V}^*=\bar{\varrho}$, the eigenvalues can be computed from the neighborhood of $\bar{\varrho}=1/(d+2)<\bar{\varrho}_0$ as we will show below. The eigenfunctions of this FP potential are polynomial, as we will also show below, thus a finite Taylor expansion in the neighborhood of $\bar{\varrho}=1/(d+2)$ is sufficient to deduce the eigenfunctions for all $\bar{\varrho}$. The result is then similar to the case where $\bar{\varrho}>\bar{\varrho}_0$ as the eigenvalues are given by those of the FP potential $\bar{V}^*=\bar{\varrho}$ and the eigenfunctions of the singular potential for non zero $\delta V(d,\infty,\bar{\varrho}>\bar{\varrho}_0)$ is given by the eigenfunctions of the FP potential $\bar{V}^*=\bar{\varrho}$ in the domain $\bar{\varrho}<\bar{\varrho}_0$. Hence, the restriction to the interval $\bar{\varrho}<\bar{\varrho}_0$ is again irrelevant and the eigenfunctions and eigenvalues can be computed safely.   

Thus, for any non zero global perturbation, $\lambda(d,\infty)$ belongs either to the spectrum of $\{R_m\}_m$ or $\{L_n\}_n$. In other words, in the limit $N\rightarrow\infty$, the spectrum of eigenvalues of a singular potential is included within the union of the spectrum of eigenvalues that makes up its two parts. This does not necessarily imply that the inclusion holds in the other direction, that is that any eigenvalue of $L_n$ or $R_m$ is necessarily an eigenvalue of the singular potential. However, numerical experiments with the FP SG$_3$ and SWF$_2$ seem to infer that there is an equality between the two sets. In any case, numerically we have found that the relevant eigenvalues of a singular potential is equal to the union of relevant eigenvalues associated to its two parts. We now give arguments to explain this numerical result.

Let us remark that while eigenvalues are functions of $d$ and $N$, they are evidently not functions of $\bar{\varrho}$ so that a constraint in the domain $\bar{\varrho}>\bar{\varrho}_0$ (resp. $\bar{\varrho}<\bar{\varrho}_0$) applies globally for all $\bar{\varrho}>0$. As such, if $\lambda(d,\infty)$ belongs to the spectrum of the singular potential and of $R_m$ (resp. $L_n$) but not of $L_n$ (resp. $R_m$) then the global eigenfunction $\delta V(d,\infty,\bar{\varrho})$ is necessarily null for $\bar{\varrho}<\bar{\varrho}_0$ (resp. $\bar{\varrho}>\bar{\varrho}_0$) as $\delta V=0$ is the only "perturbation" that does not lead to a potentially contradictory constraint on the value of $\lambda$. As the intersection of the spectrum of eigenvalues of $L_n$ and $R_m$ is finite for irrational $d$ and the total dimension of eigenfunctions is infinite this implies that there is necessarily an infinite number of eigenvalues of $\delta V$ that do not belong to the intersection of the spectrum of both $L_n$ and $R_m$ for irrational $d$. For these eigenvalues the associated eigenfunctions are null either for $\bar{\varrho}>\bar{\varrho}_0$ or for $\bar{\varrho}<\bar{\varrho}_0$ as explained. If we then consider the continuity of these eigenfunctions with respect to $d$, we expect an infinite number of such eigenfunctions for general $d$. We mention this to note that singular eigenfunctions of the form $SL_n$ and $SR_m$ not only exist but are necessarily infinite within the spectrum of a singular FP in the limit $N\rightarrow\infty$. The finite $N$ extension of a subset of such eigenfunctions are illustrated for the FP SG$_3$ in Fig. \ref{fig:Eigenfunctions-of-the}.  For these eigenfunctions the associated eigenvalue is that of the non trivial part either $L_n$ or $R_m$. 

\subsection{High temperature fixed point spectrum}

In the following we show that the spectrum of the high-temperature FP has only one relevant eigenvalue corresponding to a trivial constant perturbation.

First, we consider the eigenvalues of the high temperature FP $\bar{V}=\bar{\varrho}$. In the limit $N\rightarrow\infty$ these eigenvalues are obtained from Eq. (\ref{eq:eigenequation}) by taking $\bar{V}^{*}=\bar{\varrho}$ and setting the $1/N$ term to zero:

\begin{equation}
0=-\left(d+\lambda\right)\delta V+(d+2)\bar{\varrho}\delta V'-\delta V'\label{eq:linear eigen}.
\end{equation}
This may be compared to the linearization of the Gaussian solution:
\begin{equation}
0=-\left(d+\lambda\right)\delta V+(d-2)\bar{\varrho}\delta V'-\delta V'\label{eq:gaussian eigen}
\end{equation}
whose eigenvalues are known to be :
\begin{equation}
\lambda_{n}=\left(d-2\right)n-d.
\end{equation}
Eq. (\ref{eq:gaussian eigen}) can be mapped to Eq. (\ref{eq:linear eigen})
by using the substitution $d\rightarrow d+4,\,\lambda\rightarrow\lambda-4$
which allows us to deduce the eigenvalues from Eq. (\ref{eq:linear eigen})
as:
\begin{equation}
\lambda_{n}=\left(d+2\right)n-d.
\end{equation}
Moreover, as the eigenvalues and eigenfunctions of the Gaussian FP can be obtained in the neighborhood of $\bar{\varrho}=1/(d-2)$ according to \cite{Mati2017,Litim2018}, then according to the map $d\rightarrow d+4$ between the high temperature FP and the Gaussian FP, we conclude that the neighborhood of $\bar{\varrho}=1/(d+2)$ is sufficient to compute the eigenvalues and eigenfunctions of the high temperature FP as was mentioned previously in this appendix section.  

\subsection{Boundary layer analysis of eigenfunctions}

Notice that the only relevant negative eigenvalue of the high temperature FP is obtained for $n=0$
which is $\lambda_{0}=-d$. Usually this eigenvalue is omitted
as it corresponds to perturbing the potential by an unphysical constant
however in the case of singular potentials in the limit $N\rightarrow\infty$ there are two eigenvalues $-d$ which correspond to adding
a constant to the left or to the right of the cusp. The only globally analytical eigenfunction of this type is a global constant perturbation. The other eigenfunction in the eigenspace of dimension two is then necessarily discontinuous. Any such piece-wise constant perturbation can be decomposed in terms of $SL_{0}$ and $SR_{0}$. $SR_{0}$ can be normalized as $\Theta\left(\bar{\varrho}-\bar{\varrho}_{0}\right)$
for $\bar{\varrho}>0$ where $\Theta$ is the Heaviside function.
In the same way we have $SL_{0}=\Theta\left(\bar{\varrho}_{0}-\bar{\varrho}\right)$. This is the limit $N\rightarrow\infty$ of the upper left eigenfunction of Fig. \ref{fig:Eigenfunctions-of-the}.
Notice then that  $SL_{0}+SR_{0}=1$ such that the analytical constant eigenperturbation can also be decomposed in terms of these two functions. As such, the set $\{SL_{0},SR_{0}\}$ constitutes a basis of the eigenspace associated to the eigenvalue $-d$. However, it is more convenient to consider the basis $\{SL_{0},1\}$ or $\{1,SR_{0}\}$ as one can then omit the trivial constant eigenperturbation. We then notice that in the case where the vector space is of dimension larger than one, the limit function of $\delta V$ is a priori unknown as it may correspond to $SL_{0}$ or $SR_{0}$. However, numerically we have found that the limit eigenfunction is always $SL_{0}$ as is visible for SG$_3$ in Fig. \ref{fig:Eigenfunctions-of-the}.

In the following we show that, within their boundary layers, all singular FP eigenfunctions are built from a $\tanh$ function. This $\tanh$ function simply marks the presence of the boundary layer in the eigenfunctions at finite $N$. Outside of this boundary layer the eigenfunctions are constrained to resemble those of either $L_n$ or $R_m$ according to the discussion above.  

In analogy with the boundary layer analysis of the singular FP potentials, we parametrize the perturbations as $\delta V=\delta V_0+\delta \tilde{V}/N$ and $\bar{\varrho}=\bar{\varrho}_0+\tilde{\rho}/N$. We also replace $V^*$ with its leading order approximation within the boundary layer. Then the leading order approximation of $\delta V$ within the boundary layer is:
\begin{equation}
    -(\lambda+d) \delta V_0+\delta \tilde{V}'(4 \bar{\varrho}_0 (V_1-V_2 \tanh (V_2 \tilde{\rho}))+(d-2) \bar{\varrho}_0-1)-2 \bar{\varrho}_0 \delta \tilde{V}''=0
\end{equation}
with $V_1=1/2+\bar{V}'(\bar{\varrho}_0^+)/2$ and $V_2=1/2-\bar{V}'(\bar{\varrho}_0^+)/2$. $\bar{V}'(\bar{\varrho}_0^+)$ and $\bar{V}'(\bar{\varrho}_0^-)$ can be found from the FP Polchinski equation in the limit $N\rightarrow\infty$ by considering the point $\bar{\varrho}$ where $\bar{V}(\bar{\varrho}_0)=\bar{\varrho}_0$. This is the point where two FP solutions of the singular potential intersect. Thus, replacing $\bar{V}(\bar{\varrho}_0)$ with $\bar{\varrho}_0$ we find :
\begin{equation}
    (d-2) \bar{\varrho}_0 \bar{V}'-d \bar{\varrho}_0+2 \bar{\varrho}_0 \bar{V}'^2-\bar{V}'+1=0.\label{eigen_boundary}
\end{equation}
Solving for $\bar{V}'$ leads to:
\begin{equation}
    \bar{V}'(\bar{\varrho}_0^-)=1,\quad \bar{V}'(\bar{\varrho}_0^+)=\frac{1-d \bar{\varrho}_0}{2 \bar{\varrho}_0}.  
\end{equation}
Inserting the expressions of $V_1$ and $V_2$ in terms of $\bar{\varrho}_0$ in Eq. (\ref{eigen_boundary}), solving the differential equation and then reverting back to the original variables $\bar{\varrho}$ and $\bar{V}$, we obtain:
\begin{equation}
   \delta V= \delta V_0+\frac{1}{N}A + \frac{1}{N}\frac{((\lambda+d) \delta V_0 N\left(\bar{\varrho}-\bar{\varrho}_0\right) - 4 \bar{\varrho}_0 B) \tanh{\frac{N\left(\bar{\varrho}-\bar{\varrho}_0\right) (1 - (2 + d) \bar{\varrho}_0)}{
   4 \bar{\varrho}_0}}}{-1 + (2 + d) \bar{\varrho}_0}
\end{equation}
where $A$ and $B$ are integration constants to be determined by boundary conditions and normalization of the eigenfunction. This is the general form of the eigenfunctions to leading order in $1/N$ within the boundary layer. This approximation can not be used outside of the boundary layer but at least qualitatively it represents a good approximation of the smoothed step function perturbation $SL_0$. We expect this step eigenfunction to have an eigenvalue of $-d$ as explained above.  Thus, considering this particular eigenfunction, the linear term that multiplies the tanh function is set to zero. 
We then obtain a pure tanh function which can represent a step function in the limit $N\rightarrow\infty$.

In conclusion, in the limit $N\rightarrow\infty$, the relevant eigenvalues of a singular FP is given by the relevant eigenvalues of the two FP potentials that form its construction. In particular, in the limit $N\rightarrow\infty$, there are two eigenvalues that equate to $-d$. Each eigenvalue $-d$ correspond to constant perturbations on either side of the cusp. For finite $N$ there is then one trivial eigenvalue corresponding to a global shift of the potential while the other is non trivial and receives finite $N$ corrections. This last eigenvalue has an eigenfunction that converges to a step function in the limit $N\rightarrow\infty$. 

\section{The curve $N'_c(d)$ at large $N$ and in the vicinity of $d=4$}\label{sec:curveN'c}

The curve $N_{c}'(d)$ is the location where SWF$_2$ = SG$_3$, see Fig. \ref{fig:N_c}. For large values of $N$, this line is asymptotic to the $d=4$ axis. We have numerically determined an excellent fit of this curve given by $N_c(d)\sim71.0-23.0\log(4-d)$ and which is valid typically for $4-d<10^{-2}$, see Fig. \ref{Nc'(d)}. 

The shape of this curve can be understood if we consider the results of Appendix \ref{sec: eigenfunctions-sing-FP}. Indeed, the point of collapse between the FPs SG$_3$ and SWF$_2$ can be characterized by the point at which an eigenvalue of SG$_3$ vanishes. The corresponding eigenvalue is the same along the critical curve and in particular at $(N=\infty,d=4)$. There the collapse can be understood instead in terms of the usual bifurcation of the WF FP from the Gaussian but only for field values above the position of the cusp. According to Appendix \ref{sec: eigenfunctions-sing-FP}, the relevant eigenvalues of SG$_3$ in the limit  $N\rightarrow \infty$ is the same as $G$ plus one extra eigenvalue $-d$. The eigenvalue of SG$_3$ that vanishes at $N=\infty,d=4$ is then the same eigenvalue as for the usual Gaussian FP at $d=4$. Numerically we have verified that this eigenvalue $\lambda^{\textrm{SG}}(d,N)$ differs from the eigenvalue of its regular counterpart $\lambda^{\textrm{G}}$ by an exponentially small correction for $N$ sufficiently large. The same can be found by replacing the FP potential of SG$_3$ by its leading order boundary layer approximation in Eq. (\ref{eq:SG}) where we have checked this for corrections of the order $10^{-11} \sim 10^{-13}$. A heuristic argument as to why this is the case is that the FP SG$_3$ differs from its regular counterpart by corrections that are exponentially small outside of the boundary layer, as is visible from Eq. (\ref{eq:SG}), and that at $N=\infty$ the eigenvalues are entirely determined by the neighborhood of $\bar{\varrho}=1/(d-2)$ which is also outside of the boundary layer. It is then natural to consider exponentially small corrections of the eigenvalues of SG$_3$ when compared to those of the Gaussian. However, the major weakness of the above argument is that it neglects the influence from the range of field values within the boundary layer where an expansion in $1/N$ can be done. It is then possible that the boundary layer also contributes a small correction of the form $A/N^m$ and that $A$ is sufficiently small such that, within the range of $N$ explored, the numerical correction exhibits only an exponentially suppressed form. In any case, in the following we will omit such power law corrections as we have not found any convincing evidence that they exist numerically \footnote{We mention however that non exponential corrections seem to be present for the eigenvalue which converges to $-d$ but this occurs at a level of accuracy where it is not clear whether these are artificial numerical corrections}. Thus, from this last consideration we have :

\begin{equation}
    \lambda^{\textrm{SG}}(d,N)=\lambda^{\textrm{G}}(d,N)\pm\exp\left(-\frac{N-C_2(d)}{C_1(d)}\right)+o\left(\exp\left(-\frac{N}{C_1(d)}\right)\right).
\end{equation}
For finite $N$ and along the critical line where SG$_3$ and SWF$_2$ collapse, we have $\lambda^{\textrm{SG}}(d'_c(N),N)=0$ which implies:

\begin{equation}
    \lambda^{\textrm{G}}(d'_c(N),N)=\pm\exp\left(-\frac{N-C_2(d'_c(N))}{C_1(d'_c(N))}\right)+o\left(\exp\left(-\frac{N}{C_1(d'_c(N))}\right)\right).
\end{equation}
Moreover, we have $\lambda^{\textrm{G}}=d-4$ which leads to:

\begin{equation}
    4-d'_c(N)=\pm\exp\left(-\frac{N-C_2(d'_c(N))}{C_1(d'_c(N))}\right)+o\left(\exp\left(-\frac{N}{C_1(d'_c(N))}\right)\right).
\end{equation}
 At this stage we use the fact that numerically $4-d'_c(N)>0$ which then allows us to choose the plus sign. Now taking the log of this last expression and neglecting the leading order corrections we arrive at: 

\begin{equation}
    N'_c(d)=-C_1(4)\log{(4-d)}+C_2(4).
\end{equation}

\begin{figure}[h]
\includegraphics[scale=0.24]{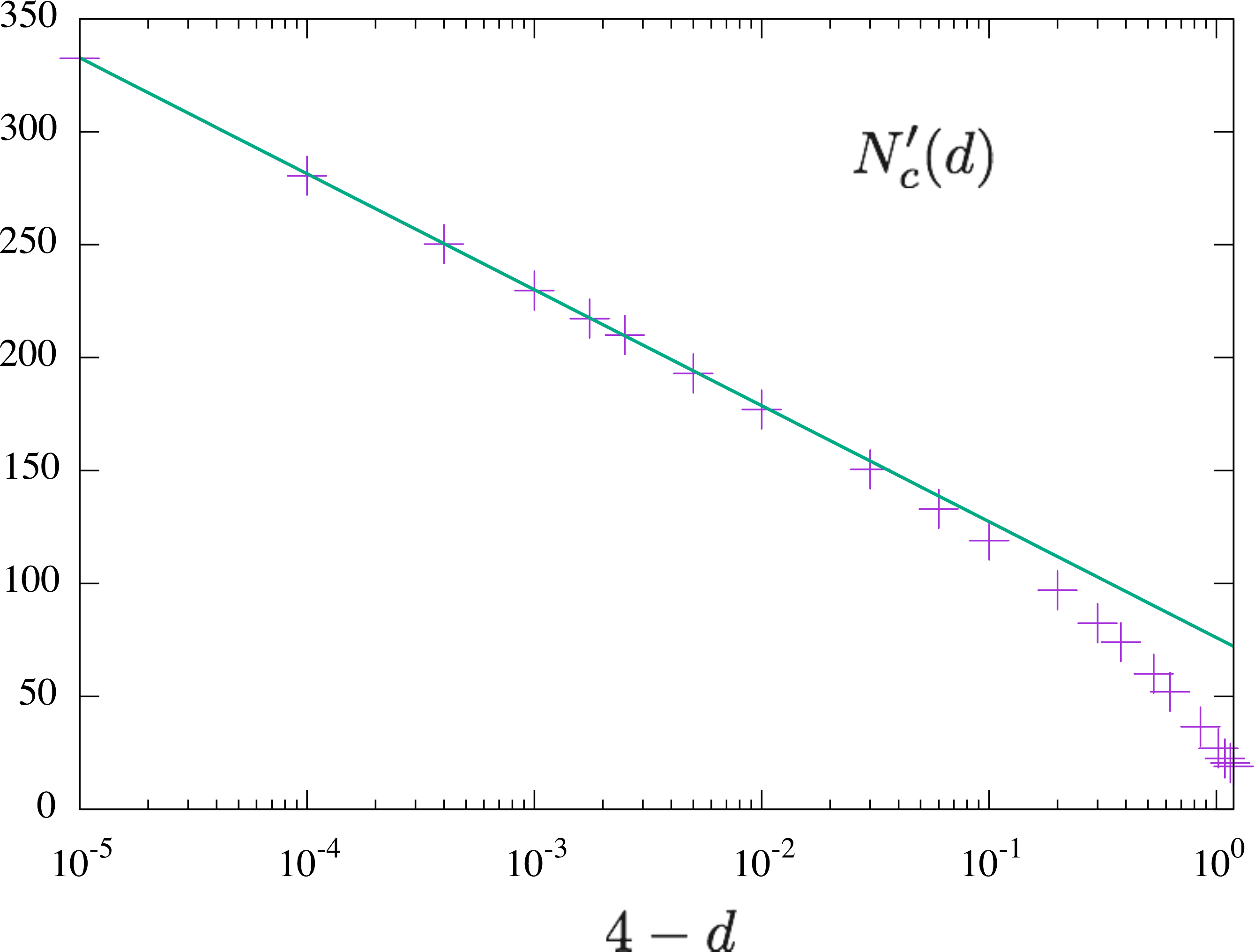}
\caption{The curve $N_{c}'(d)$ calculated at LPA  (purple crosses) and the curve $71.0-23.0\log(4-d)$ (green curve). }
\label{Nc'(d)}
\end{figure}

\section{A short review in bifurcation theory}\label{sec: bifurcation_section}

The following contains a short review of bifurcation theory whose purpose is mainly to clarify the bifurcations at the points $S$ and $S'$. In the process we will use notions of dynamical systems such as normal forms and center manifold reduction. We will also address the other types of bifurcations that take place in the main text, namely those that occur at  $d_c=2+2/p$ and on the new critical lines. This short review is based on \cite{bifurcation_1,bifurcation_2}.

Considering a set of coordinates for the effective action, the RG flow of the effective action can be understood as a an infinite dimensional dynamical system. When numerical calculations are required this infinite dimensional dynamical system is truncated to a finite system of flow equations where the usual theorems of dynamical systems apply. In particular, we will focus here on bifurcation theory as the bifurcations that take place at the points $S$ and $S'$ mentioned in the main text are not the usual ones found in RG systems. Nonetheless, it is interesting to note that the different types of bifurcations that take place in the main text, including those at $S$ and $S'$, are among the most common examples of bifurcations in dynamical systems. This is contrary to common belief which states that there is only one type of bifurcation in the O($N$) model, namely the so called transcritical bifurcations, which take place at $d_c=2+2/p$ when new multicritical FPs bifurcate from the Gaussian.

For pedagogical purposes we first review the two types of one dimensional bifurcations involving two FPs. We may exhibit these two types using the rather simple $\beta$ function of Eq. (\ref{beta-function1}):

\begin{equation}
    N \beta_{{\tilde g}_6}=-2\alpha\tilde{g}_6+12\tilde{g}_6^2-\pi^2\tilde{g}_6^3/2+O(1/N).
    \label{beta-function_bifurca}
\end{equation}
This $\beta$ function has three FPs: The Gaussian G with $\tilde{g}_6=0$, the tricritical FP A$_2$ with $\tilde{g}_6=\tilde{g}_{6,-}$ and the FP $\tilde{\textrm{A}}_3$ with $\tilde{g}_6=\tilde{g}_{6,+}$. For $\alpha=0$ (resp. $\alpha=\alpha_c$) , we have $\tilde{g}_{6,-}=0$ (resp.  $\tilde{g}_{6,-}=\tilde{g}_{6,+}$) and A$_2=$G (resp. A$_2=\tilde{\textrm{A}}_3$). Both $\alpha=0$ and $\alpha=\alpha_c$ correspond to bifurcations but they are not of the same type. In the case of the bifurcation at $\alpha=0$ both FPs remain real valued on either side of the bifurcation, that is for $\alpha>0$ and small or $\alpha<0$ and small. This is to be contrasted with the bifurcation at $\alpha=\alpha_c$ where for $\alpha>\alpha_c$, A$_2$ and $\tilde{\textrm{A}}_3$ become complex. This is most easily seen by expanding the $\beta$ function about the point of bifurcation $(\alpha_b,\tilde{g}_{6,b})$ to order 2 in the coupling. The reason for stopping the expansion at order two being that it is the minimal degree to incorporate both FP solutions involved in the bifurcation. In the case $\alpha=0$ this is the usual perturbative expansion where for $\alpha$ sufficiently small we may omit the cubic term. After re-scaling the coupling, in the neighborhood of $\alpha=0$, the flow of $\tilde{g}_{6}$ reduces to :

\begin{equation}
    \overset{.}{x}=-ax+x^2
    \label{transcritical_bifurcation}
\end{equation}
where the dot refers to a derivative with respect to RG time and $a$ is proportional to $\alpha$. Eq. (\ref{transcritical_bifurcation}) is called the normal form of a transcritical bifurcation. Normal forms are standard simplifications of dynamical systems that are particularly useful for studying bifurcations \footnote{These normal forms are said to be locally topologically equivalent to the original flow equations. The notion of topological equivalence here implies that there exists a local homeomorphism that maps the original dynamical system onto the new flow equations. This definition of normal forms is in slight contrast with that of \cite{Sethna}, for example, as we authorize local non analytical homeomorphisms given by the implicit function theorem. Restricting to analytical maps typically requires more terms in the normal form than what is needed to retrieve topological equivalence but allows for a more quantitative analysis such as computation of corrections to scaling \cite{Sethna}}. In essence, normal forms capture the characteristics of bifurcations such as the number of FPs that bifurcate and which FPs are real valued one either side of the bifurcation. The transcritical bifurcation is the typical bifurcation in the O($N$) model where multicritical FPs bifurcate from the Gaussian at $d_c=2+2/p$. As mentioned above, both FP roots of the polynomial on the right hand side of Eq. (\ref{transcritical_bifurcation}) remain real  on either side of the bifurcation. 

The second type of bifurcation involving two FPs appears at $\alpha=\alpha_c$. Considering a shift $\tilde{g}_6=\tilde{g}_{6,c}+\delta g$ where $\tilde{g}_{6,c}=\tilde{g}_{6,-}=\tilde{g}_{6,+}$, expanding about $\delta g$ for $\alpha-\alpha_c$ small and rescaling $\delta g$, the $\beta$ function has the form:

\begin{equation}
    \overset{.}{x}=-a-x^2
    \label{saddle_node_bifurcation}
\end{equation}
where $a\propto \alpha-\alpha_c$. In this case, for $a<0$, that is, $\alpha<\alpha_c$, there are two roots which correspond to A$_2$ and $\tilde{\textrm{A}}_3$ in the neighborhood of the bifurcation at $\alpha=\alpha_c$. If $a>0$ both roots are complex. In bifurcation theory, Eq. (\ref{saddle_node_bifurcation}) is called the normal form of a saddle-node bifurcation. Notice that for both normal forms the number of roots does not correspond to the total number of FP solutions of the full $\beta$ function in Eq. (\ref{beta-function_bifurca}). The normal form then essentially zooms into the region in $(a,x)$ space in which the bifurcation takes place thereby retaining the topology of the bifurcation while omitting irrelevant information. 

In the general case for one dimensional systems, the flow equation has the form 

\begin{equation}
    \overset{.}{x}=f(\bold{a},x)
    \label{general_bifurcation}
\end{equation}
where $\bold{a}=(a_1,a_2,\ldots, a_n)$ corresponds to the set of parameters of the system. A bifurcation involving two FPs takes place when two of the potentially numerous zeroes of $f$ coincide for some specific value of the parameter $\bold{a}$. When this occurs the normal form of $f$ in the neighborhood of the bifurcation takes on one of the above two forms, either Eq. \eqref{transcritical_bifurcation} or \eqref{saddle_node_bifurcation}. 

While this one dimensional bifurcation theory applies to Eq.  (\ref{beta-function_bifurca}), it is natural to wonder to what extent this one dimensional example is useful for the RG where in general there are many coupled flow equations. The answer to this question lies in what is called center manifold reduction which in some cases allows a simplification to the one dimensional case. The following paragraphs gives the general framework of this method.

We recall that although the RG is an infinite dimensional dynamical system, in practice, numerical implementations usually truncate to a finite but possibly large dynamical system. We write this system as

\begin{equation}
    \frac{d \vec{x}}{d t}=\vec{\beta}(\vec{x},s)
    \label{dynamical_bifurcari}
\end{equation}
\begin{figure}[h]
\includegraphics[scale=0.65]{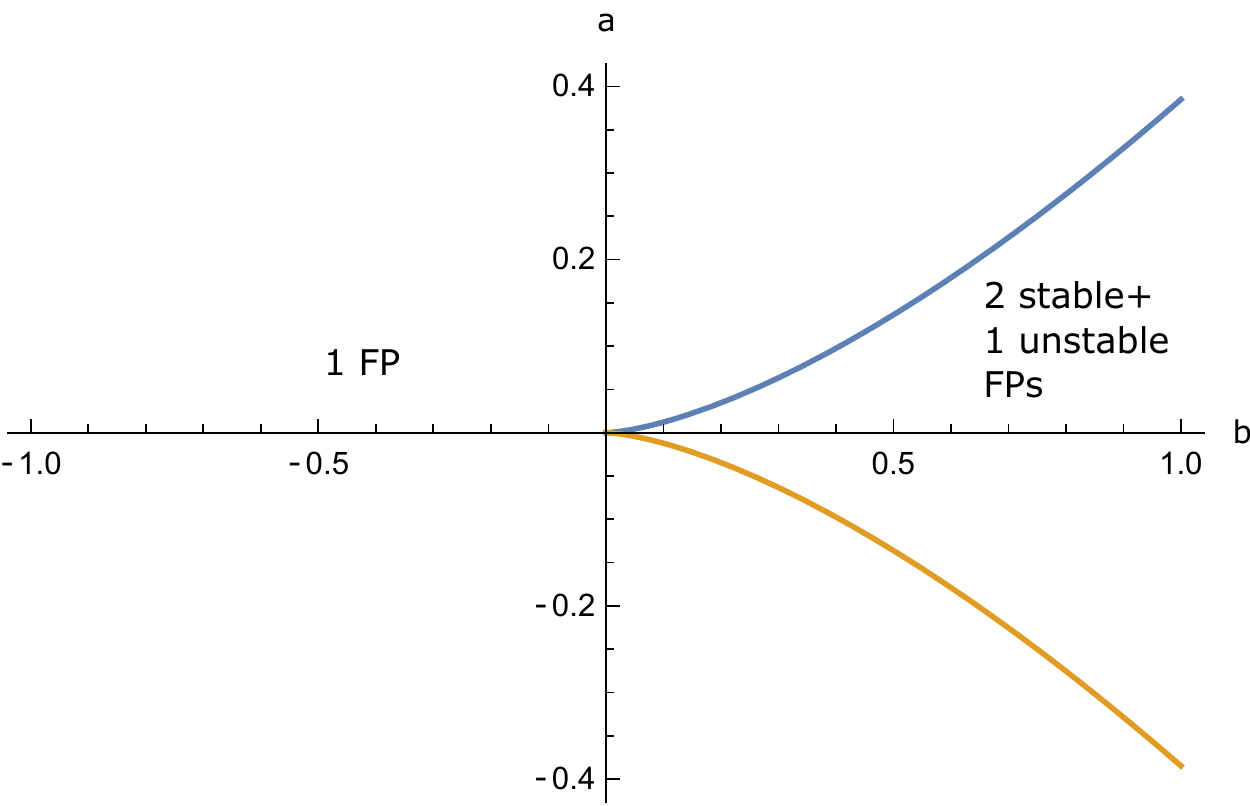}
\caption{Cusp bifurcation indicating the number of FP solutions of the dynamical system. The domain with 3 FPs is delimited by the blue and yellow curves where FPs collapse. Along the axis a=0, a pitchfork bifurcation occurs. }
\label{fig:cusp_bifurcation}
\end{figure}
where $s$ represents the sets of parameters of the system. Consider then a bifurcation that takes place at $s=s_0$ with FP solution $\vec{x}^*(s_0)$. Shifting the variable $\vec{x}$ as $\vec{x}=\vec{x}^*(s_0)+\vec{\chi}$, we may decompose $\vec{\beta}(\vec{x}^*(s_0)+\vec{\chi},s_0)$ as a linear and non linear part:

\begin{equation}
\frac{d \vec{\chi}}{d t}=L(s_0)\chi+R(\vec{\chi},s_0)
\end{equation}
where $L$ is a matrix and $R$ is at least quadratic in its variables. The eigenvalues of the matrix $L$ correspond to the RG eigenvalues of the FP $\vec{x}^*(s_0)$ in the neighborhood of the bifurcation point $s_0$. Consider now a change of variables $\chi\rightarrow\hat{\chi}$ which diagonalizes $L$ \footnote{Or puts it in  Jordan form in the case of logarithmic CFTs for example.}. For $s=s_0$, the dynamical system may then be written as

\begin{equation}
\begin{aligned} \frac{d \vec{\chi}_C}{d t}= & M_C(\vec{\chi}_C,\vec{\chi}_S,s_0)= C(s_0)\vec{\chi}_C+ R_C(\vec{\chi}_C,\vec{\chi}_S,s_0)\\
\frac{d \vec{\chi}_Y}{d t}= & M_Y(\vec{\chi}_C,\vec{\chi}_S,s_0)= Y(s_0)\vec{\chi}_Y+R_Y(\vec{\chi}_C,\vec{\chi}_S,s_0)
\end{aligned}
\end{equation}
where $C(s_0)$ and $Y(s_0)$ are both diagonal and $R_C(\vec{\chi}_C,\vec{\chi}_S,s_0)$ and $R_Y(\vec{\chi}_C,\vec{\chi}_S,s_0)$ are at least quadratic in their variables $\vec{\chi}_C,\vec{\chi}_S$. Moreover, all eigenvalues of $C(s_0)$ have zero real part while those of $Y(s_0)$ have non zero real part. The central manifold reduction then states that in the neighborhood of $s=s_0$, there is a local homeomorphism which maps  Eq. (\ref{dynamical_bifurcari}) to: 
\begin{figure}[t]
\includegraphics[scale=0.7]{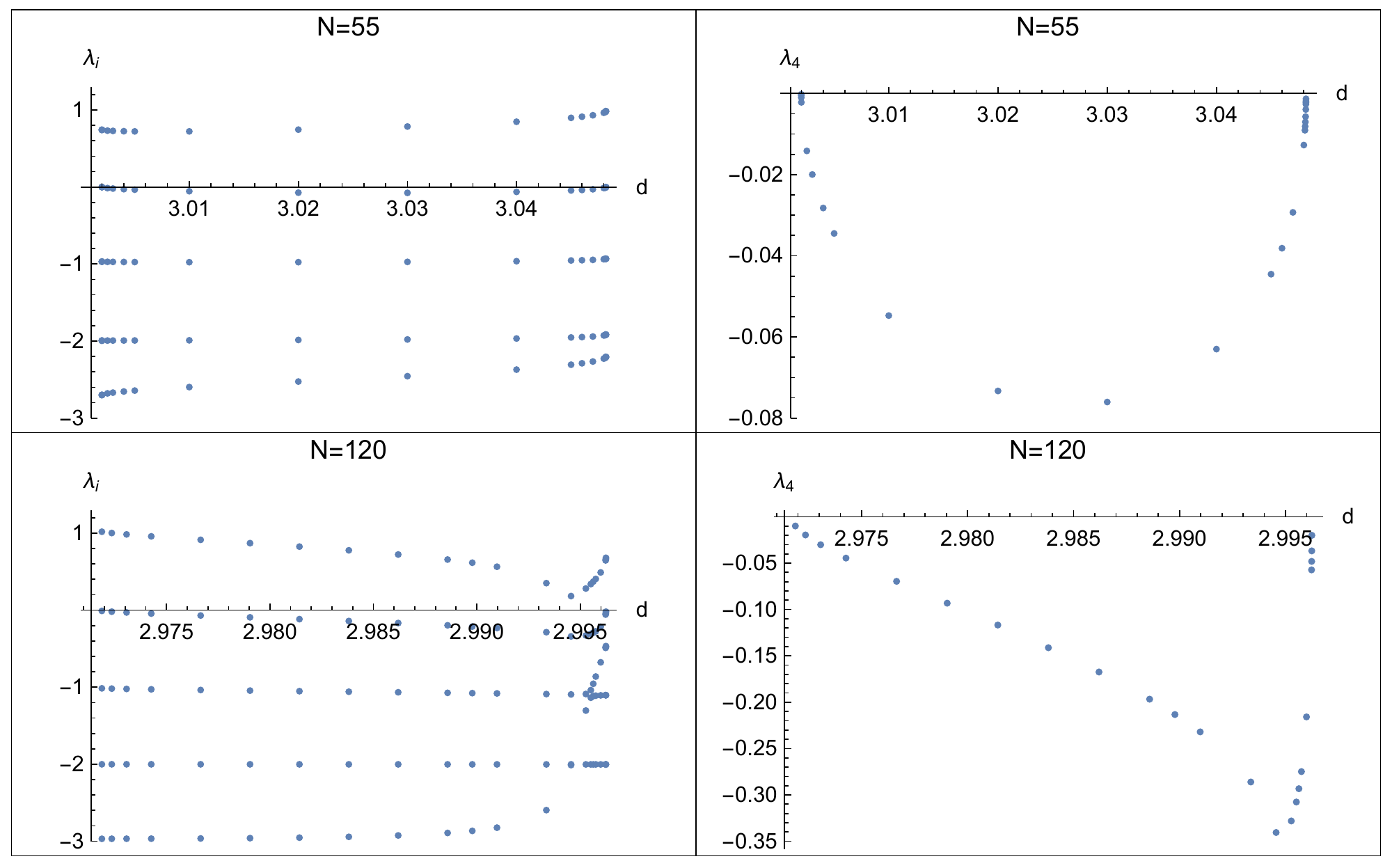}\caption{ Eigenvalues of the FP S$\tilde{\textrm{A}}_4$ as a function of $d$. The plots on the left show the evolution of the eigenvalues of S$\tilde{\textrm{A}}_4$ as functions of the dimension for $N=55$ (top) and $N=120$ (bottom). The plots on the right show a zoom of the eigenvalue which becomes marginal either by collapsing with SA$_3$ on the left of the axis labeled $d$ or with  $\tilde{\textrm{A}}_3$ towards the right of the same axis. Notice that the same eigenvalue becomes marginal on both sides of the collapse and in particular around the point $S'$. In the neighborhood of the bifurcation point $S'$, this allows a dimensional reduction of the RG system to a one dimensional system where the axis of the dimension is given by the associated nearly marginal eigendirection. Notice also that for $N=120$, the smallest relative eigenvalue, which is the one that is closest to $-d$ towards left of the axis, converges towards 0 but does not collapse before the other eigenvalue becomes marginal. As explained in Appendix \ref{sec: eigenfunctions-sing-FP}, a (non trivial) eigenvalue which is close to $-d$ is the signature of a singular FP. This eigenvalue seems to converge to zero due to the double "singular-regular" nature of the BMB point.} 
\label{bifurcations_of_SA4}
\end{figure}
\begin{equation}
\begin{aligned} \frac{d \vec{\chi_C}}{d t}= & M_C(\vec{\chi}_C,\vec{q}(\vec{\chi}_C),s)\\
\frac{d \vec{\chi_Y}}{d t}= & Y(s_0)\vec{\chi_Y}
\end{aligned}
\label{topo-dyn}
\end{equation}
where $\vec{\chi}_S$ has been replaced by a function $\vec{q}$ of $\vec{\chi}_C$. For the purpose here, we need not compute $\vec{q}$, it suffices to notice that the flow of $\vec{\chi}_C$ is now determined solely by $\vec{\chi}_C$. It is then said that Eq. (\ref{dynamical_bifurcari}) and Eq. (\ref{topo-dyn}) are topologically equivalent in the sense that the structure of the flow near the FPs involved in the bifurcation is the same for both flows. The important point in the above equations is that the variables whose eigenvalues vanish during the bifurcation decouple from those whose eigenvalues remain finite after having performed the local change of coordinates. Thus, Eq. (\ref{topo-dyn}) implies that it is sufficient to reduce the dimensionality of the system to the dimension of the variables whose eigenvalues vanish at the bifurcation. This is clearly a notable simplification as all bifurcations in this paper take place where a single eigenvalue becomes marginal. In particular, this  applies for the critical line $N'_c(d)$ where the FPs SG$_3$ and SWF$_2$ collapse in a saddle-node bifurcation. Notice that this also aligns with the fact that for all critical lines in this paper, the difference in the number of relevant eigenvalues of the FPs involved in the collapse is equal to one in absolute value. 

It is then interesting to note that among the typical examples of bifurcation there is one more which is called a pitchfork bifurcation with normal form: 

\begin{equation}
    \overset{.}{x}=z x-x^3.
\end{equation}
Notice that in this case the polynomial is of degree three and at $z=0$ the right hand side has a triple root at $x=0$. This triple root then corresponds to the bifurcation of three FPs. This is to be contrasted with the double root of transcritical and saddle-node bifurcations as the existence of two FPs requires expanding the dynamical system to at least quadratic order in the dynamical variables. The pitchfork bifurcation is a specialized case of the more general "imperfect" or "cusp" bifurcation with normal form:
\begin{figure}[h]
\includegraphics[scale=0.7]{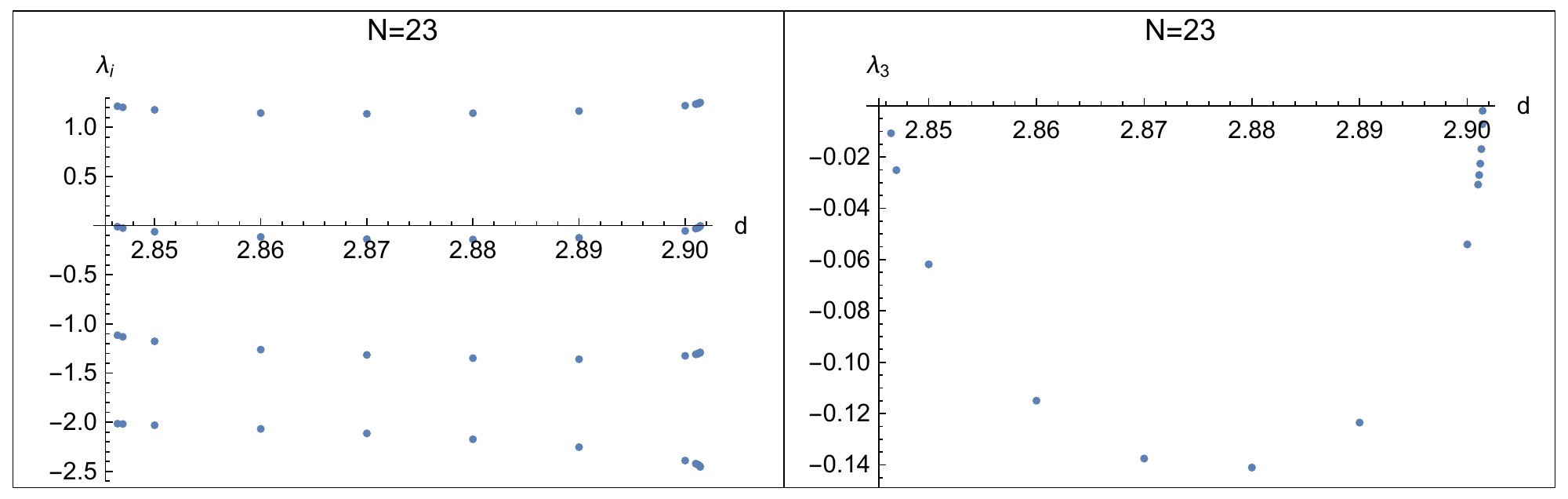}\caption{ Eigenvalues of the FP SG$_3\sim \tilde{\textrm{A}}_3$ around the point $S$ as a function of $d$. The plot on the left shows the evolution of the eigenvalues of SG$_3\sim \tilde{\textrm{A}}_3$ for $N=23$. The plot on the right shows a zoom of the eigenvalue which becomes marginal either by collapsing with A$_2$ on the left of the axis labeled $d$ or with  SWF$_2$ towards the right of the same axis.} 
\label{bifurcations_of_SG}
\end{figure}

  \begin{figure}[h]
\begin{centering}
\includegraphics[width=8cm]{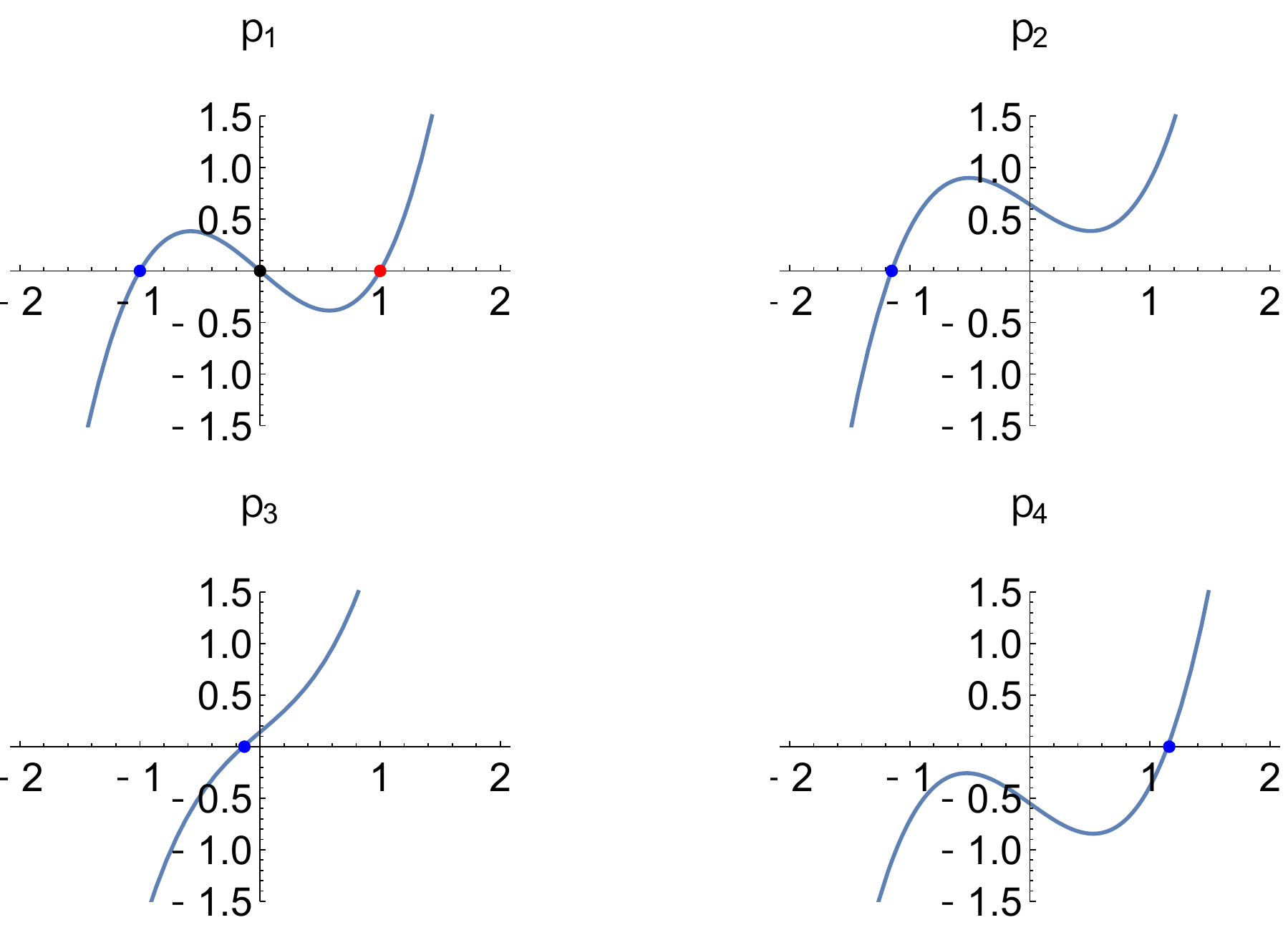}\includegraphics[width=8cm]{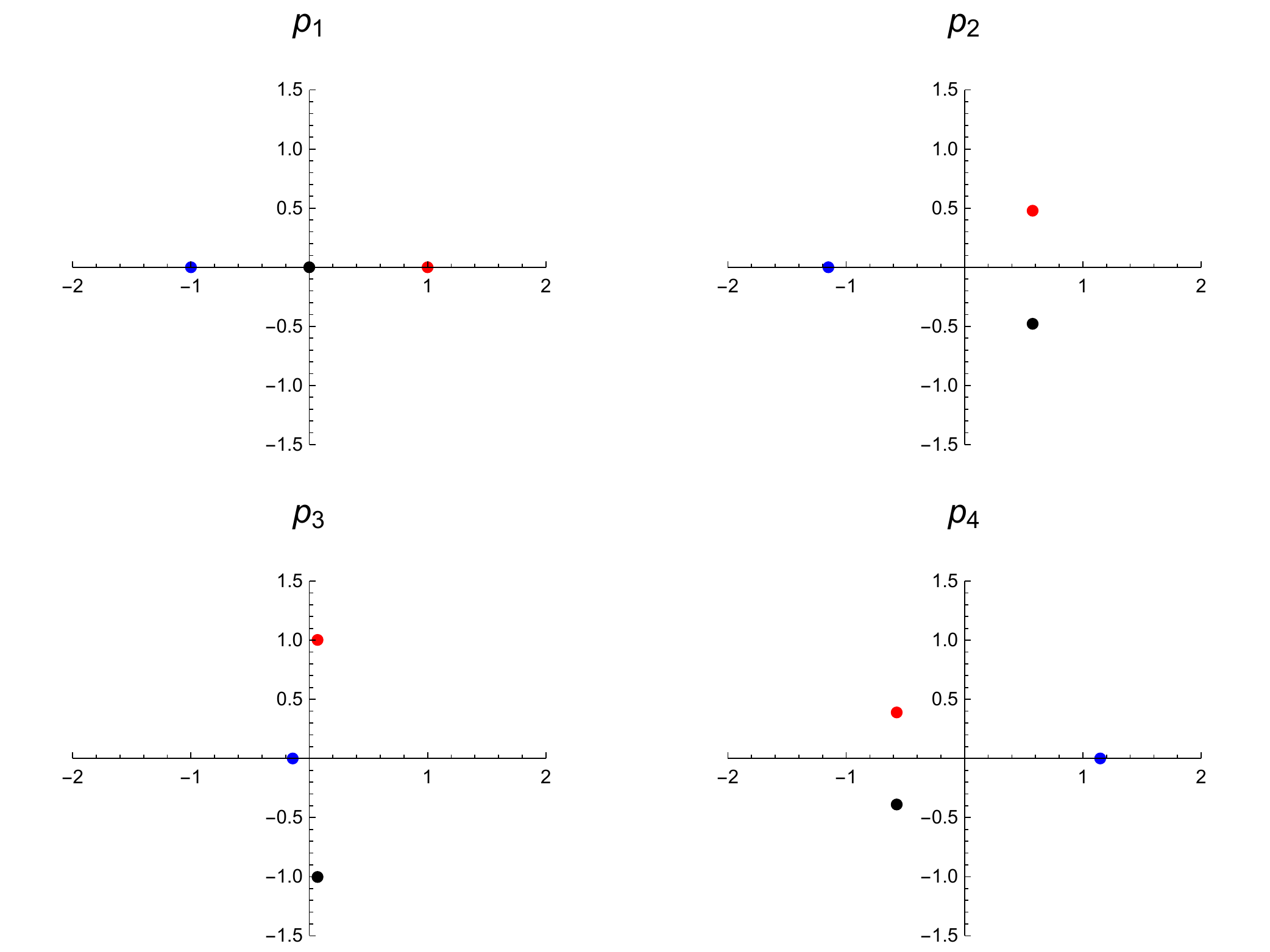}
\par\end{centering}
\caption{\label{fig:Left:-Polynomial-function}Left: Polynomial function $P_{\theta}\left(x\right)$
as a function of $x$ for different values of $\theta$. The leftmost root in $p_1$ is followed by continuity when $\theta$ is increased from 0 to $2\pi$ and becomes the red one at $\theta=2\pi$. Right: positions of all the roots, real and complex, when $\theta$ is varied. }
\end{figure}
\begin{equation}
    \overset{.}{x}=a+bx-x^3.
\end{equation}
The topology in parameter space of this bifurcation is given in Fig. \ref{fig:cusp_bifurcation}. The same topology occurs for the points $S$ and $S'$. Of course, it is still necessary to verify that the bifurcation is indeed one dimensional by verifying the number of eigenvalues that vanish. In Fig. \ref{bifurcations_of_SA4} we show that only one eigenvalue of the FP S$\tilde{\textrm{A}}_4$ vanishes as this FP approaches the point $S'$. The same is true for the point $S$ in Fig. \ref{bifurcations_of_SG}. The monodromy structure around these points can be explained by choosing a closed path in $(a,b)$ parameter space that encircles the point $(a=0,b=0)$. Considering a unit circle gives a particularly simple choice of coordinates:
\begin{equation}
\overset{.}{x}=P_{\theta}\left(x\right)=-\sin\theta+x \cos\theta-x^{3}.
\end{equation}
In Fig. \ref{fig:Left:-Polynomial-function}, the  roots of $P_\theta$ are shown at four values of $\theta$. The leftmost root, shown in blue in panel $p_1$, is followed by continuity when $\theta$ increases from 0 to $2\pi$. When $\theta=2\pi$, the three real roots are of course identical to those shown in $p_1$ but the blue root is now at the place of the red one.

\end{document}